\documentclass[a4paper,12pt]{article}
\usepackage{multicol}
\usepackage[left=1.5cm,right=1.5cm,top=2cm,bottom=2cm]{geometry}

\usepackage{cmap}					
\usepackage{mathtext} 				
\usepackage[T2A]{fontenc}			
\usepackage[utf8]{inputenc}			
\usepackage[english,russian]{babel}	
\usepackage{ccaption} 
\captiondelim{. } 
\usepackage{cite}

\usepackage{amsfonts,amssymb,amsthm,mathtools} 
\usepackage{amsmath}
\usepackage{icomma} 

\usepackage{euscript}	 
\usepackage{mathrsfs} 

\usepackage{graphicx}  
\usepackage{subfig}
\usepackage{float}
\graphicspath{{images/}}  
\usepackage{wrapfig} 

\usepackage{array,tabularx,tabulary,booktabs} 
\usepackage{longtable}  
\usepackage{multirow} 

\title{RESONANT SPONTANEOUS BREMSSTRAHLUNG EFFECT IN THE SCATTERING OF ULTRARELATIVISTIC ELECTRONS ON NUCLEI IN A STRONG LASER FIELD}
\author{S. P. Roshchupkin$^1$, A.V. Dubov$^1$, V. V. Dubov$^1$,  S. S. Starodub$^2$\\\\
$^1$Peter the Great St. Petersburg Polytechnic University,\\
195251, St-Petersburg, Russian Federation, Russia\\
$^2$Institute of Applied Physics, National Academy of Sciences of Ukraine,\\
 40000, Sumy, Ukraine}
\date{}

\begin{document} 

\maketitle

\selectlanguage{english}

\begin{abstract}
The process of resonant spontaneous bremsstrahlung radiation during the scattering of ultrarelativistic electrons with energies of the order $\sim 100\ \text{GeV}$ by the nuclei in strong laser fields with intensities up to $I\sim {{10}^{24}}\ \text{Wc}{{\text{m}}^{\text{-2}}}$ is theoretically studied. Under resonant conditions, an intermediate electron in the wave field enters the mass shell. As a result, the initial second-order process by the fine structure constant is effectively reduced to two first-order processes: laser-stimulated Compton effect and laser-assisted Mott process. The resonant kinematics for two reaction channels (A and B) is studied in detail. It is shown that in the resonant case there is a characteristic parameter that determines a significant number of absorbed laser photons in the laser-stimulated Compton effect. This parameter is determined by the parameters of the laser installation, the energy of the initial electrons and is proportional to the intensity of the laser wave. An analytical resonant differential cross-section with simultaneous registration of the frequency and the outgoing angle of a spontaneous gamma-quantum is obtained. It is shown that the resonant differential cross-section has the largest value in the region of average laser fields ($I\sim {{10}^{18}}\ \text{Wc}{{\text{m}}^{\text{-2}}}$) and can be of the order of $\sim {{10}^{18}}$ in units  ${{Z}^{2}}\alpha r_{e}^{2}$. With an increase in the intensity of the laser wave, the value of the resonant differential cross-section decreases and for the intensity $I\sim {{10}^{24}}\ \text{Wc}{{\text{m}}^{\text{-2}}}$ is of the order of $\sim {{10}^{4}}$ in units  ${{Z}^{2}}\alpha r_{e}^{2}$. The obtained results reveal new features of spontaneous emission of ultrarelativistic electrons on nuclei in strong laser fields and can be tested at international laser installations.
\end{abstract}

\section{Introduction}

Theoretical study of fundamental processes of quantum electrodynamics (QED) in strong laser fields (see, for example, articles \cite{1,2,3,4,5,6,7,8,9,10,11,12,13,14,15,16,17,18,19,20,21,22,23,24,25,26,27,28,29,30,31,32,33,34}, reviews \cite{35,36,37,38,39,40} and monographs \cite{41,42,43,44}) is connected not only with the development of QED, but also with the creation of powerful laser radiation sources \cite{45,46,47,48,49}. It is important to emphasize that higher-order QED processes with respect to the fine structure constant in a laser field can proceed both in a resonant and non-resonant way. Here, the so-called Oleinik resonances may occur \cite{2,3}, due to the fact that lower-order processes are allowed in the laser field by the fine structure constant. It is important to note that the resonant differential cross-section can significantly exceed the corresponding non-resonant one \cite{36,37,38,39,40}.

Spontaneous bremsstrahlung during electron scattering on the nucleus in the field of a light wave has been studied for a long time (see, for example,\cite{1,5,6,7,8,9,10,11,12,13,14,15,16,17,18,19,20,21,22,23,24,25,26,27,28,29,36,37,38,39,40,41,42,43,44}). At the same time, Oleinik resonances were studied in \cite{8,9,10,11,12,13,17,18,19,20,21,22,23,24,25,26,27,28,29,36,37,38,39,42,43,44}. In recent articles \cite{27,28,29}, the process of spontaneous emission of ultrarelativistic electrons on nuclei in the field of a weak electromagnetic wave was considered.

In this article, unlike the previous ones, we will study the resonant spontaneous emission of ultrarelativistic electrons on nuclei in medium and strong laser fields in the Born approximation (${v}/{c>>{Z}/{137}}$, $v$ is the velocity of the electron, $c$ is the speed of light in a vacuum, $Z$ is the charge of the nucleus). We will consider resonant processes for high-energy particles when the main parameter is the classical relativistic-invariant parameter

\begin{equation} \label{eq1}
\eta =\frac{{eF\mathchar'26\mkern-10mu\lambda  }}{{m{c^2}}}
\end{equation}
numerically equal to the ratio of the field work at the wavelength to the rest energy of the electron ($e$ and $m$ are the charge and mass of the electron, $F$ and $\mathchar'26\mkern-10mu\lambda={c}/{\omega }\;$  are the electric field strength and wavelength, $\omega $ is the frequency of the wave). 

We will use the relativistic system of units: $\hbar=c=1$.

\section{The Amplitude SB of an Electron on a Nucleus in a Strong Light Field}

\begin{figure}[h]
	\begin{minipage}[t]{0.5\textwidth}
		\centering
		\includegraphics[width=0.5\textwidth]{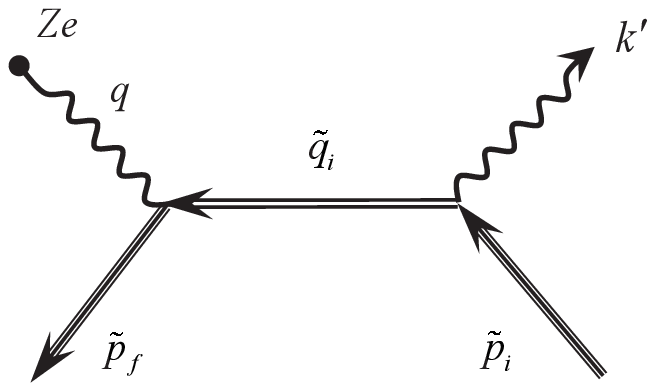} \\ A
	\end{minipage}
	\hfill
	\hfill
	\begin{minipage}[t]{0.5\textwidth}
		\centering
		\includegraphics[width=0.5\textwidth]{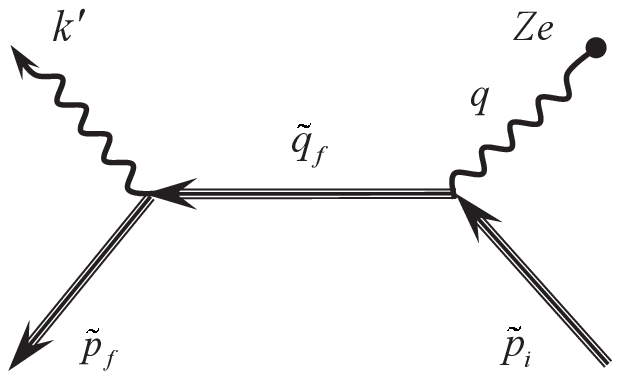} \\ B
	\end{minipage}
	\caption{Feynman diagrams of the electron-nucleus process in the field of a plane electromagnetic wave. The double incoming and outgoing lines correspond to the Volkov functions of the electron in the initial and final states, the inner line corresponds to the Green function of the electron in the plane wave field~ \eqref{eq3}. The wavy lines correspond to the 4-momenta of the spontaneous gamma-quantum  and the \textquotedblleft pseudophoton\textquotedblright of the recoil of the nucleus.}
	\label{<figure1>}
\end{figure}

We choose the 4-potential of a plane monochromatic circularly polarized electromagnetic wave propagating along the axis $z$ in the following form:

\begin{equation} \label{eq3}
	A\left( \phi  \right)=\frac{F}{\omega }\left( e_x \cos \phi +\delta \cdot e_y \sin \phi  \right),\quad \phi =kx=\omega \left( t-z \right).
	\end{equation}
Here $\delta =\pm 1$, ${{e}_{x,y}}=\left( 0,{{\mathbf{e}}_{x,y}} \right)$ and $k=\left( \omega ,\mathbf{k} \right)$ are 4-polarization vectors and the 4-momentum of the photon of the external field, and: ${{k}^{2}}=0,\ e_{x,y}^{2}=-1,\ {{e}_{x,y}}k=0$. This is a second-order process with respect to the fine structure constant and is described by two Feynman diagrams (see Figure \ref{<figure1>}). The wave functions of the electron are determined by the Volkov functions \cite{50}, the intermediate states of the electron are given by the Green function in the field of a plane light wave (\ref{eq3}) \cite{51,52}. The amplitude of such a process after simple calculations can be represented in the following form (see, for example, \cite{27,28,29}):
\begin{equation} \label{eq4}
S_{fi}=\sum_{l=-\infty }^{\infty } S_l,
\end{equation}
where the partial amplitude with the emission and absorption of $\vert l \vert$-photons of the wave has the following form:

\begin{equation} \label{eq5}
S_l=-i\cdot \frac{8{{\pi }^{{5}/{2}\;}}\cdot Z{{e}^{3}}}{\sqrt{2{\omega }'{{{\tilde{E}}}_{i}}{{{\tilde{E}}}_{f}}}}\cdot \exp \left( i{{\varphi }_{fi}} \right)\cdot \left[ {{{\bar{u}}}_{f}}{{M}_{l}}{{u}_{i}} \right]\cdot \frac{\delta \left(q_0\right)}{\mathbf{q}^2}.
\end{equation}
Here it is indicated:
\begin{equation}\label{eq6}
	{{M}_{l}}=\sum\limits_{r=-\infty }^{\infty }{\left[ {{M}_{r-l}}\left( {{{\tilde{p}}}_{f}},{{{\tilde{q}}}_{i}} \right)\cdot \left( \frac{{{{\hat{{q}}}}_{i}}+{{m}}}{\tilde{q}_{i}^{2}-m_{*}^{2}} \right)\cdot {{F}_{-r}}\left( {{{\tilde{q}}}_{i}},{{{\tilde{p}}}_{i}} \right)+{{F}_{-r}}\left( {{{\tilde{p}}}_f},{{{\tilde{q}}}_{f}} \right)\cdot \left( \frac{{{{\hat{{q}}}}_{f}}+{{m}}}{\tilde{q}_{f}^{2}-m_{*}^{2}} \right)\cdot {{M}_{r-l}}\left( {{{\tilde{q}}}_{f}},{{{\tilde{p}}}_{i}} \right) \right]}
\end{equation}
In Exps. \eqref{eq5}-\eqref{eq6} $\varphi_{fi}$ is the phase independent of the summation indices, $u_i$,$\bar{u}_f$ are Dirac bispinors, ${{\tilde{p}}_{i}}$ and ${{\tilde{p}}_{f}}$ are the 4-quasimomenta of the initial and final electrons, ${{\tilde{q}}_{i}}$ and ${{\tilde{q}}_{f}}$ are the 4-quasimomenta of intermediate electrons for channels A and B, ${{m}_{*}}$ is the effective mass of the electron in the plane wave field, $q$ is the 4-transmitted momentum:

\begin{equation}\label{eq7}
	{{\tilde{q}}_{i}}={{\tilde{p}}_{i}}-{k}'+rk,\text{  }{{\tilde{q}}_{f}}={{\tilde{p}}_{f}}+{k}'-rk.\text{ }
\end{equation}	

\begin{equation}\label{eq8}
q={{\tilde{p}}_{f}}-{{\tilde{p}}_{i}}+{k}'-lk.
\end{equation}

\begin{equation}\label{eq9}
{{\tilde{p}}_{j}}={{p}_{j}}+{{\eta }^{2}}\frac{{{m}^{2}}}{2\left( k{{p}_{j}} \right)}k,\text{  }{{\tilde{q}}_{j}}={{q}_{j}}+{{\eta }^{2}}\frac{{{m}^{2}}}{2\left( k{{q}_{j}} \right)}k,\quad j=i,f
\end{equation}

\begin{equation}\label{eq10}
\tilde{p}_{i,f}^{2}=m_{*}^{2},\text{  }{{m}_{*}}=m\sqrt{1+{{\eta }^{2}}}.
\end{equation}
Here ${k}'={\omega }'\left( 1,\mathbf{{n}'} \right)$ is the 4-momentum of the spontaneous gamma-quantum, ${{p}_{i,f}}=\left( {{E}_{i,f}},{{\mathbf{p}}_{i,f}} \right)$ is the 4-momentum of the initial and final electrons. Expressions with a cap in the relation \eqref{eq6} and further mean the scalar product of the corresponding 4-vector with the Dirac gamma matrices: ${{\gamma }^{\mu }}=\left( {{\gamma }^{0}},\mathbf{\gamma } \right)$, $\mu =0,1,2,3$. For example, ${{\hat{\tilde{q}}}_{i}}={{\tilde{q}}_{i\mu }}{{\gamma }^{\mu }}={{\tilde{q}}_{i0}}{{\gamma }^{0}}-\mathbf{\tilde q_{i}} \boldsymbol {\gamma}$. Amplitudes ${{M}_{r-l}}$ and ${{F}_{-r}}$ (see Fig. \ref{<figure1>}) in the relation \eqref{eq6} have the form:

\begin{equation}\label{eq11}
	{{M}_{r-l}}\left( {{{\tilde{p}}}_{2}},{{{\tilde{p}}}_{1}} \right)={{a}^{0}}\cdot {{L}_{r-l}}\left( {{{\tilde{p}}}_{2}},{{{\tilde{p}}}_{1}} \right)+b_{-}^{0}\cdot {{L}_{r-l-1}}+b_{+}^{0}\cdot {{L}_{r-l+1}},
\end{equation}

\begin{equation}\label{eq12}
{{F}_{-r}}\left( {{{\tilde{p}}}_{2}},{{{\tilde{p}}}_{1}} \right)=\left( a{{\varepsilon }^{*}} \right)\cdot {{L}_{-r}}\left( {{{\tilde{p}}}_{2}},{{{\tilde{p}}}_{1}} \right)+\left( {{b}_{-}}{{\varepsilon }^{*}} \right)\cdot {{L}_{-r-1}}+\left( {{b}_{+}}{{\varepsilon }^{*}} \right)\cdot {{L}_{-r+1}}.
\end{equation}
In these expressions $\varepsilon _{\mu }^{*}$ is the 4-polarization vector of the spontaneous gamma-quantum, and the matrices ${{a}^{\mu }},b_{\pm }^{\mu }$ are determined by the relations

\begin{equation}\label{eq13}
{{a}^{\mu }}={{\tilde{\gamma }}^{\mu }}+{{\eta }^{2}}\frac{{{m}^{2}}}{2\left( k{{{\tilde{p}}}_{1}} \right)\left( k{{{\tilde{p}}}_{2}} \right)}{{k}^{\mu }}\hat{k},
\end{equation}

\begin{equation}\label{eq14}
b_{\pm }^{\mu }=\frac{1}{4}\eta m\cdot \left[ \frac{{{{\hat{\varepsilon }}}_{\pm }}\hat{k}{{\gamma }^{\mu }}}{\left( k{{{\tilde{p}}}_{2}} \right)}+\frac{{{\gamma }^{\mu }}\hat{k}{{{\hat{\varepsilon }}}_{\pm }}}{\left( k{{{\tilde{p}}}_{1}} \right)} \right],\text{   }{{\hat{\varepsilon }}_{\pm }}={{\hat{e}}_{x}}\pm i\delta \cdot {{\hat{e}}_{y}}.
\end{equation}
Special functions ${{L}_{r-l}}$ and $\text{ }{{L}_{-r}}$, and their arguments are given by expressions \cite{31}:

\begin{equation}\label{eq15}
L_{r'}\left( {\tilde p}_2,{\tilde p}_1 \right)=\exp \left( -ir' \chi_{{\tilde p}_2 {\tilde p}_1} \right)\cdot J_{r'}\left( {{\gamma }_{{{{\tilde{p}}}_{2}}{{{\tilde{p}}}_{1}}}} \right),
\end{equation}

\begin{equation}\label{eq16}
tg{{\chi }_{{{{\tilde{p}}}_{2}}{{{\tilde{p}}}_{1}}}}=\delta \cdot \frac{\left( {{e}_{y}}{{Q}_{{{{\tilde{p}}}_{2}}{{{\tilde{p}}}_{1}}}} \right)}{\left( {{e}_{x}}{{Q}_{{{{\tilde{p}}}_{2}}{{{\tilde{p}}}_{1}}}} \right)},\quad {{Q}_{{{{\tilde{p}}}_{2}}{{{\tilde{p}}}_{1}}}}=\frac{{{{\tilde{p}}}_{2}}}{\left( k{{{\tilde{p}}}_{2}} \right)}-\frac{{{{\tilde{p}}}_{1}}}{\left( k{\tilde{p}_{1}} \right)},
\end{equation}

\begin{equation}\label{eq17}
{{\gamma }_{{{{\tilde{p}}}_{2}}{{{\tilde{p}}}_{1}}}}=\eta m\sqrt{-Q_{{{{\tilde{p}}}_{2}}{{{\tilde{p}}}_{1}}}^{2}}.
\end{equation}
Note that, $\left( k{{{\tilde{p}}}_{1,2}} \right)=\left( k{{p}_{1,2}} \right)$ and also in the case of amplitudes ${{M}_{r-l}}\left( {{{\tilde{p}}}_{f}},{{{\tilde{q}}}_{i}} \right)$ and ${{M}_{r-l}}\left( {{{\tilde{q}}}_{f}},{{{\tilde{p}}}_{i}} \right)$ in Exps. \eqref{eq11}, \eqref{eq13}-\eqref{eq14} it is necessary to put ${{\tilde{p}}_{1}}\to {{\tilde{q}}_{i}},\text{  }{{\tilde{p}}_{2}}\to {{\tilde{p}}_{f}}$ and ${{\tilde{p}}_{1}}\to {{\tilde{p}}_{i}},\text{  }{{\tilde{p}}_{2}}\to {{\tilde{q}}_{f}}$, and for the amplitudes ${{F}_{-r}}\left( {{{\tilde{q}}}_{i}},{{{\tilde{p}}}_{i}} \right)$ and ${{F}_{-r}}\left( {{{\tilde{p}}}_{f}},{{{\tilde{q}}}_{f}} \right)$in the Exps. \eqref{eq12}-\eqref{eq14}: ${{\tilde{p}}_{1}}\to {{\tilde{p}}_{i}},\text{  }{{\tilde{p}}_{2}}\to {{\tilde{q}}_{i}}$ and ${{\tilde{p}}_{1}}\to {{\tilde{q}}_{f}},\text{  }{{\tilde{p}}_{2}}\to {{\tilde{p}}_{f}}.$We should note the obtained amplitude of the SB process \eqref{eq4}-\eqref{eq17} is valid for arbitrary frequencies and intensities of a circularly polarized electromagnetic wave.

\section{Poles of the SB Amplitude in a Strong Field}

The resonant behavior of the amplitude \eqref{eq5}-\eqref{eq6} is due to the quasi-discrete structure of the system: an electron + a plane electromagnetic wave, as a result of which the 4-quasimomentum of the intermediate electron, due to the implementation of the laws of conservation of energy-momentum in the components of the process, lies on the mass shell \cite{27,28,29,34}. Because of this, for channels A and B, we get:

\begin{equation}\label{eq18}
\tilde{q}_{i}^{2}=m_{*}^{2},
\end{equation}

\begin{equation}\label{eq19}
\tilde{q}_{f}^{2}=m_{*}^{2}.
\end{equation}
We will study the most interesting case of ultrarelativistic electron energies, when the spontaneous gamma-quantum and the final electron fly out in a narrow cone along the momentum of the initial electron. In this case, the direction of propagation of the wave lies far from the given narrow cone of particles (if the direction of propagation of the wave lies inside the narrow cone of particles, then the resonances disappear \cite{27,28,29,34}).

\begin{equation}\label{eq20}
{{E}_{i,f}}>>m,
\end{equation}

\begin{equation}\label{eq21}
{{{\theta }'}_{i,f}}=\measuredangle \left( \mathbf{{k}'},{{\mathbf{p}}_{i,f}} \right)<<1,\quad {{\theta }_{if}}=\measuredangle \left( {{\mathbf{p}}_{i}},{{\mathbf{p}}_{f}} \right)<<1,	
\end{equation}

\begin{equation}\label{eq22}
\quad {\theta }'=\measuredangle \left( \mathbf{{k}'},\mathbf{k} \right)\sim1,\quad {{\theta }_{i,f}}=\measuredangle \left( \mathbf{k},{{\mathbf{p}}_{i,f}} \right)\sim 1.	
\end{equation}
It should be noted that in strong laser fields, when the classical parameter $\eta \gtrsim 1$, instead of the mass and energy of the electron, it is necessary to use the effective mass and quasi-energy \cite{34}. Therefore, the condition \eqref{eq20} must be replaced with the following one:

\begin{equation}\label{eq23}
	\frac{{{{\tilde{E}}}_{i,f}}}{{{m}_{*}}}\approx \frac{{{E}_{i,f}}}{m\sqrt{1+{{\eta }^{2}}}}\left[ 1+{{\left( \eta \frac{m}{2{{E}_{i,f}}} \right)}^{2}}\frac{1}{{{\sin }^{2}}\left( {{{\theta }_{i,f}}}/{2}\; \right)} \right]\sim \begin{cases}
		 E_{i,f}/m>>1,\quad  & \text{if}\quad \eta <<1  \\
		 E_{i,f}/{\left( \eta m \right)}>>1,\quad & \text{if}\quad \eta \gtrsim 1  \\
	\end{cases}.
	\end{equation}
From the second condition of the relation \eqref{eq23}, we obtain a restriction on the maximum intensity of the laser wave:

\begin{equation}\label{eq24}
\eta <<{{\eta }_{\max }}=\frac{{{E}_{i}}}{m}>>1.
\end{equation}
In this paper, we will consider the energies of the initial electrons ${{E}_{i}}\lesssim 100\ \text{GeV}$. We estimate the maximum electric field strength of the laser wave in this case. From Exp. \eqref{eq24} we get: $\eta << {\eta_{\max}} \sim 10^5$ or for the intensity of the laser wave $F<<{F_{\max}} \sim 10^{15} {\text{V}}/{\text{cm}}$;  $\left( I<<{{I}_{\max}} \sim 10^{28} {{\text{W}}/{\text{c} \text{m}^{\text{2}}}} \right)$. Thus, further consideration of the resonant SB process will be valid for sufficiently large wave intensities. However, these fields are at least two orders of magnitude smaller than the critical Schwinger field ${{F}_{*}}\approx 1.3\cdot {{10}^{16}}\ {\text{V}}/{\text{cm}}\;$.

\begin{figure}[h]
		\centering
		\includegraphics[width=0.8\textwidth]{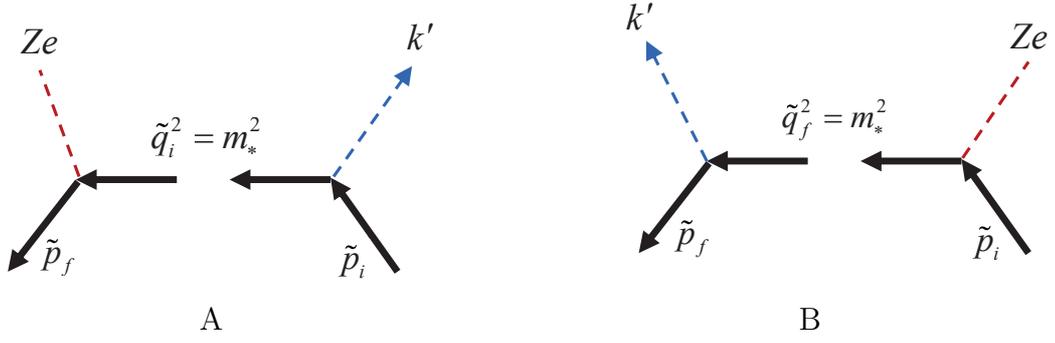} \\ A \qquad \qquad \qquad \qquad \qquad \qquad \qquad \qquad \qquad B
	\caption{Resonant spontaneous bremsstrahlung of an electron in the field of a nucleus and a plane electromagnetic wave.}
	\label{<figure2>}
\end{figure}
Expressions defining the 4-quasimomenta of the intermediate electrons ${{\tilde{q}}_{i}}$ and ${{\tilde{q}}_{f}}$ \eqref{eq7}, as well as the transmitted 4-momenta $q$ \eqref{eq8}, for channels A and B (see Fig. \ref{<figure2>}) in resonance, it is convenient to write as:

\begin{equation}\label{eq25}
{{\tilde{p}}_{i}}+rk={{\tilde{q}}_{i}}+{k}',
\end{equation}

\begin{equation}\label{eq26}
q={{\tilde{p}}_{f}}-{{\tilde{q}}_{i}}+\left( r-l \right)k
\end{equation}
and

\begin{equation}\label{eq27}
{{\tilde{q}}_{f}}+rk={{\tilde{p}}_{f}}+{k}',
\end{equation}

\begin{equation}\label{eq28}
q={{\tilde{q}}_{f}}-{{\tilde{p}}_{i}}+\left( r-l \right)k.
\end{equation}
Since both the equalities $\tilde{p}_{i}^{2}=\tilde{q}_{i}^{2}=m_{*}^{2}$ and $\tilde{p}_{f}^{2}=\tilde{q}_{f}^{2}=m_{*}^{2},\text{  }{{k}^{2}}={{{k}'}^{2}}=0,$ then equations \eqref{eq25} and \eqref{eq27} are valid only for $r\ge 1$. Hence and from the form of the amplitude \eqref{eq5}-\eqref{eq6} it follows (see also Fig. \ref{<figure2>}) that ${{F}_{-r}}$ \eqref{eq12} represents the amplitude of the process of emission of a gamma-quantum with the 4-momentum ${k}'$  by an electron due to the absorption of r-photons of the wave (laser-stimulated Compton effect) \cite{35}. Taking into account the value of the transmitted 4-momenta $q$ \eqref{eq26} or \eqref{eq28}, the value ${{M}_{r-l}}$ \eqref{eq11} is the amplitude of the scattering of an electron on the nucleus in the field of a wave with the absorption or emission of $\left| r-l \right|$-photons of the wave. This process was studied in the nonrelativistic limit by Bunkin and Fedorov \cite{1,41}, and in the general relativistic case by Denisov and Fedorov \cite{41}. Hence, in the absence of interference of the amplitudes A and B (see Fig. \ref{<figure2>}), the resonant SB process of an electron on the nucleus in the wave field, is effectively reduced to two consecutive processes of the first order according to the fine structure constant: the scattering of an electron by the nucleus in the wave field and the laser-stimulated Compton effect (see Fig. \ref{<figure2>}). It is easy to verify that when the directions of momenta of the spontaneous gamma-quantum and the photons of the laser wave coincide, the simultaneous fulfillment of the resonant conditions \eqref{eq18} or \eqref{eq19} with the laws of conservation of the 4-momentum \eqref{eq25} or \eqref{eq27} is impossible. Therefore, resonances can only occur when these photons move in non-parallel motion.

We determine the resonant frequency $\left( {{{{\omega }'}}_{\eta i\left( r \right)}} \right)$ of the spontaneous gamma-quantum for channel A. We take into account the relations \eqref{eq20}-\eqref{eq21} in the resonant condition \eqref{eq25}. After simple calculations \cite{41}, we get

\begin{equation}\label{eq29}
{x}'_{\eta i\left( r \right)}=\left[ 1+\left(1+{{\delta}'^2_{\eta i}} \right)\frac{r_{\eta}}{r} \right]^{-1},\quad x'_{\eta i\left( r \right)}=\frac{\omega'_{\eta i\left( r \right)}}{E_i}.
\end{equation}
Here it is indicated:

\begin{equation}\label{eq30}
{{r}_{\eta }}=\frac{{{m}^{2}}\left( 1+{{\eta }^{2}} \right)}{4{{E}_{i}}\omega {{\sin }^{2}}\left( {{{\theta }_{i}}}/{2}\; \right)},
\end{equation}

\begin{equation}\label{eq31}
\delta'_{\eta i}=\frac{{{E}_{i}}{{{{\theta }'}}_{i}}}{{{m}_{*}}}=\frac{{{E}_{i}}{{{{\theta }'}}_{i}}}{m\sqrt{1+{{\eta }^{2}}}}.
\end{equation}

Note that in the ultrarelativistic parameter \eqref{eq31}, we put ${{\tilde{E}}_{i}}\approx {{E}_{i}}$ by virtue of condition \eqref{eq24}. In formula \eqref{eq29} $r=1,2,3,\ldots $ is the resonance number (the number of photons of the wave absorbed by the initial electron in the laser-stimulated Compton effect); ${{r}_{\eta }}$ is a characteristic parameter of the process that determines the number of photons of the wave that make the main contribution to the laser-stimulated Compton effect. We estimate the ${{r}_{\eta }}$ value  \eqref{eq30}  for a specific experimental setup. So, for the oncoming motion of the electron flow and the laser wave $\left( {{\theta }_{i}}=\pi  \right)$, $\omega =1\ \text{eV}$ and the energy of the initial electrons ${{E}_{i}}=62.5\ \text{GeV}$ from formula \eqref{eq30}, we obtain:

\begin{equation}\label{eq32}
{{r}_{\eta }}=\left( 1+{{\eta }^{2}} \right).
\end{equation}
Note that in strong fields $\left( \eta >>1 \right)$ with a small number of absorbed photons of the wave $\left( r<<{{r}_{\eta }}\approx {{\eta }^{2}}>>1 \right)$, the resonant frequencies of the spontaneous photon are small compared to the energy of the initial electron

\begin{equation}\label{eq33}
x'_{\eta i\left( r \right)}\approx \frac{1}{\left( 1+{{\delta'}^2_{\eta i}} \right)}\frac{r}{{{r}_{\eta }}}<<1\quad \left( r<<{{r}_{\eta }} \right).
\end{equation}

\begin{figure}[!h]
\centering
\subfloat[]
{\includegraphics[width=0.5\textwidth]{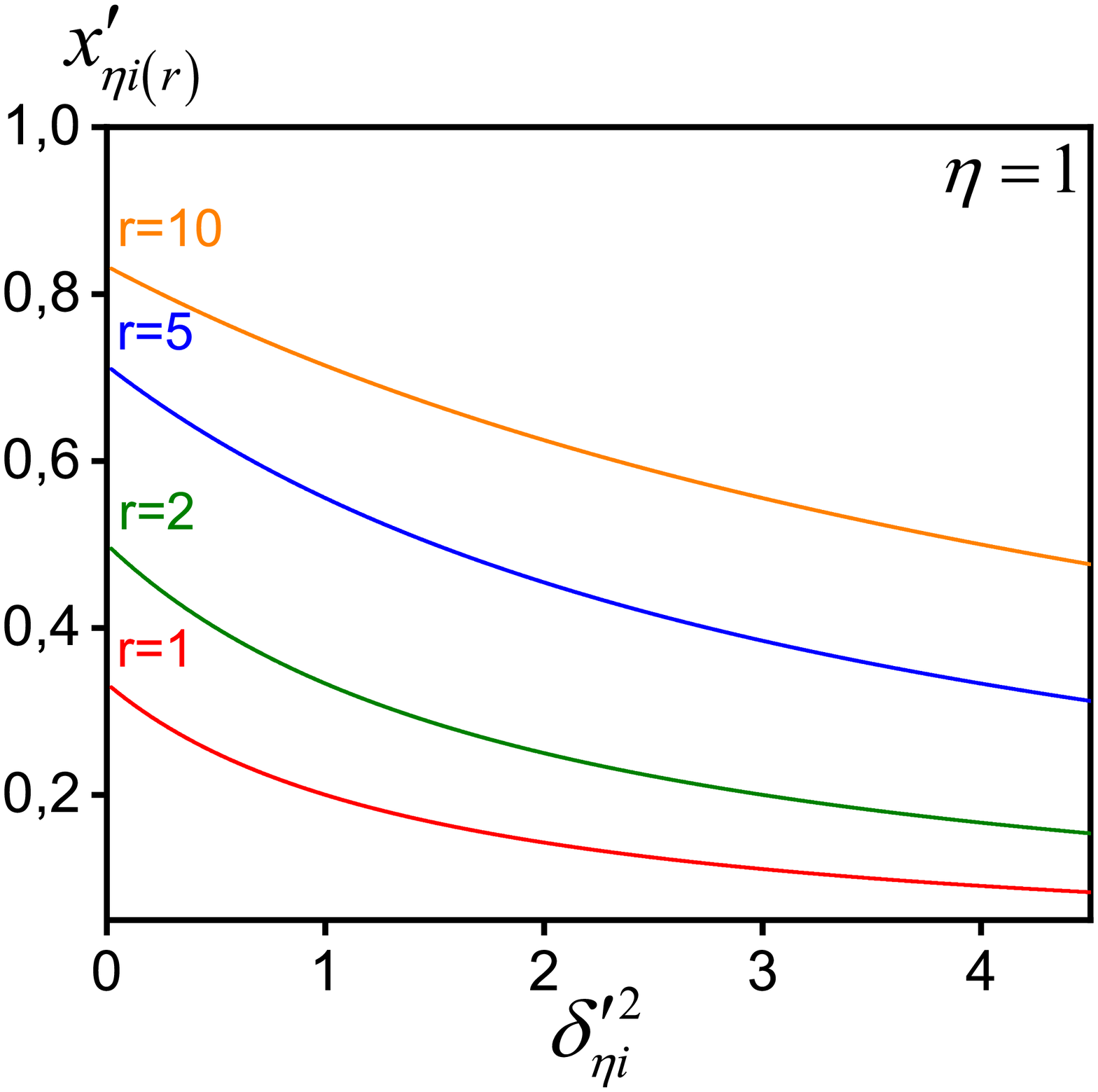}\label{<figure3a>}}
\subfloat[]
{\includegraphics[width=0.5\textwidth]{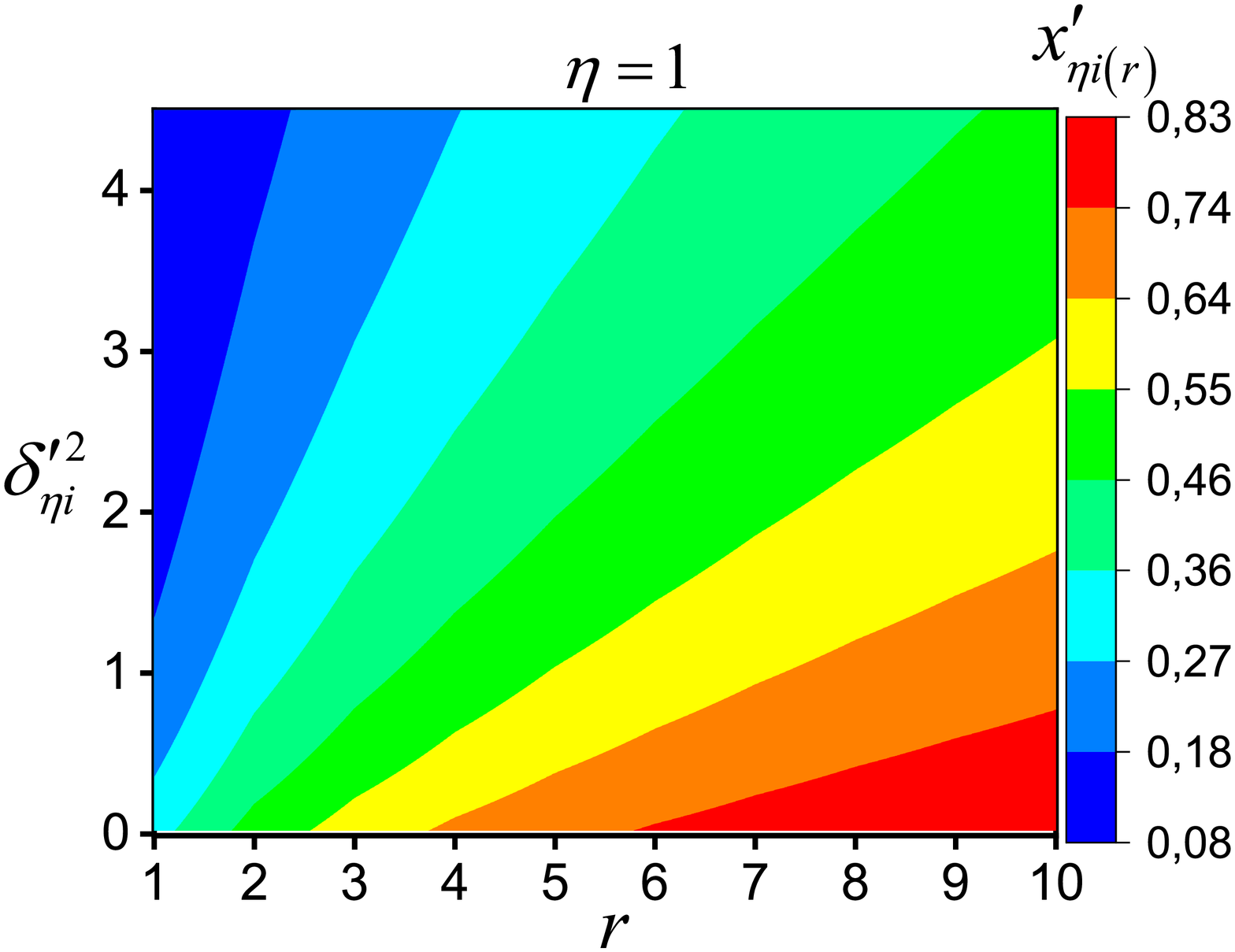}\label{<figure3b>}}
	\caption{Resonant frequency of the spontaneous photon for channel A \eqref{eq29} (in units of energy of the initial electron).  Fig.\ref{<figure3a>}  shows the dependence of the resonant frequency ${{{x}'}_{\eta i\left( r \right)}}$ \eqref{eq29} on the square of its outgoing angle relative to the momentum of the initial electron for a fixed number of absorbed photons of the wave. Fig.\ref{<figure3b>} shows the dependence of the resonant frequency ${{{x}'}_{\eta i\left( r \right)}}$ \eqref{eq29} on the square of its outgoing angle relative to the momentum of the initial electron and also for different number of absorbed photons of the wave. The energy of the initial electrons ${{E}_{i}}=62.5\ \text{GeV}$, the angle between the momenta of the electrons and the laser wave ${{\theta }_{i}}=\pi $, the frequency of the wave $\omega =1\ \text{eV}$, the intensity of the laser wave $I\approx 1.861\cdot {{10}^{18}}\ \text{Wc}{{\text{m}}^{-2}}$.}
	\label{<figure3>}
\end{figure}

\begin{figure}[!h]
\centering
\subfloat[]
{\includegraphics[width=0.5\textwidth]{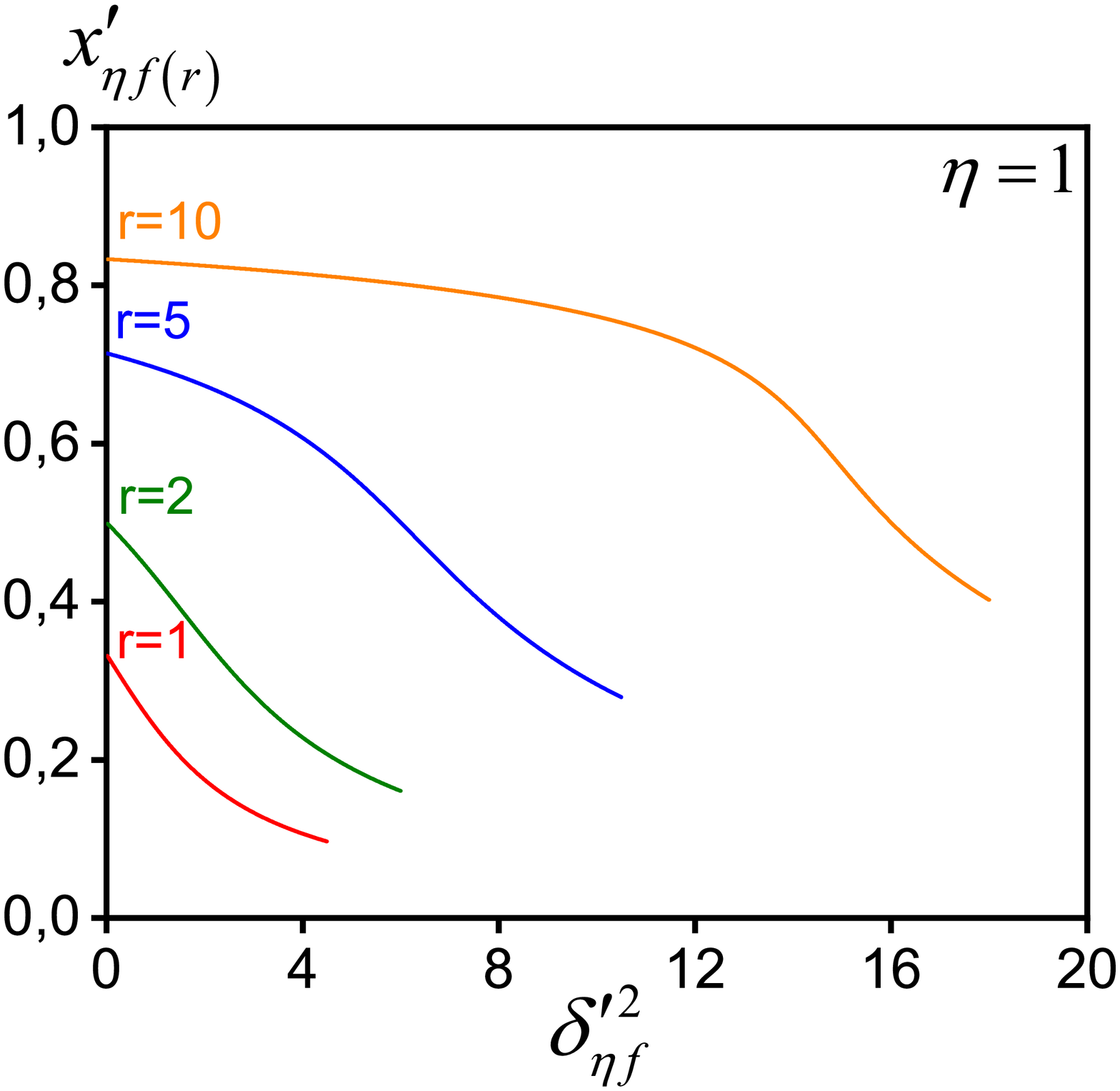}\label{<figure4a>}}
\subfloat[]
{\includegraphics[width=0.5\textwidth]{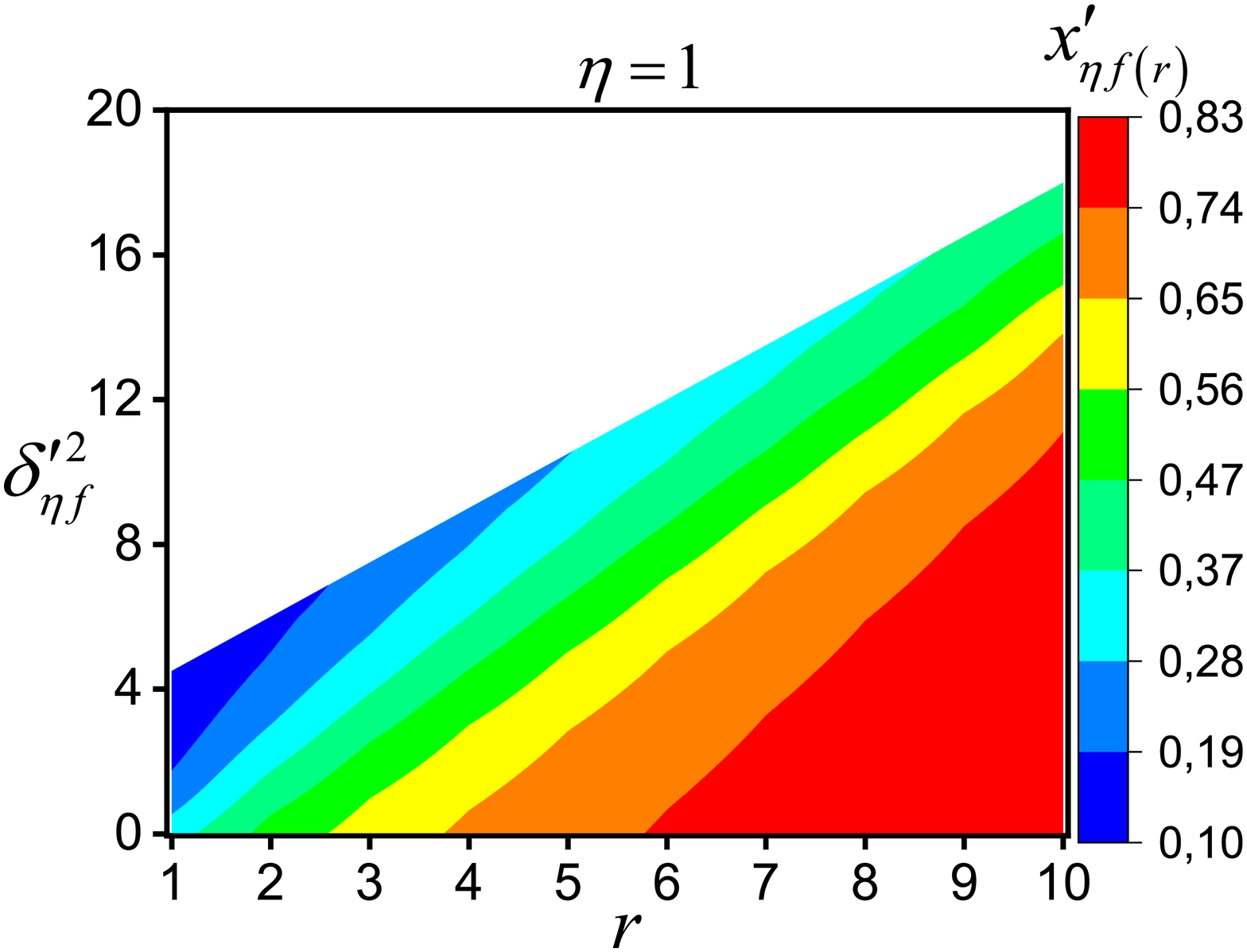}\label{<figure4b>}}
	\caption{ Resonant frequency of the spontaneous photon for channel B \eqref{eq38}, \eqref{eq39} (in units of energy of the initial electron).  Fig.\ref{<figure4a>} shows the dependence of the resonant frequency ${{{x}'}_{\eta f\left( r \right)}}$ \eqref{eq39} on the square of its outgoing angle relative to the momentum of the final electron for a fixed number of absorbed photons of the wave. Fig.\ref{<figure4b>} shows the dependence of the resonant frequency ${{{x}'}_{\eta f\left( r \right)}}$ \eqref{eq39} on the square of its outgoing angle relative to the momentum of the final electron and also for different number of absorbed photons of the wave. The energy of the initial electrons ${{E}_{i}}=62.5\ \text{GeV}$, the angle between the momenta of the electrons and the laser wave ${{\theta }_{i}}=\pi $, the frequency of the wave $\omega =1\ \text{eV}$, the intensity of the laser wave $I\approx 1.861\cdot {{10}^{18}}\ \text{Wc}{{\text{m}}^{-2}}$.}
	\label{<figure4>}
\end{figure}

\begin{figure}[!h]
	\centering
	\subfloat[]
	{\includegraphics[width=0.5\textwidth]{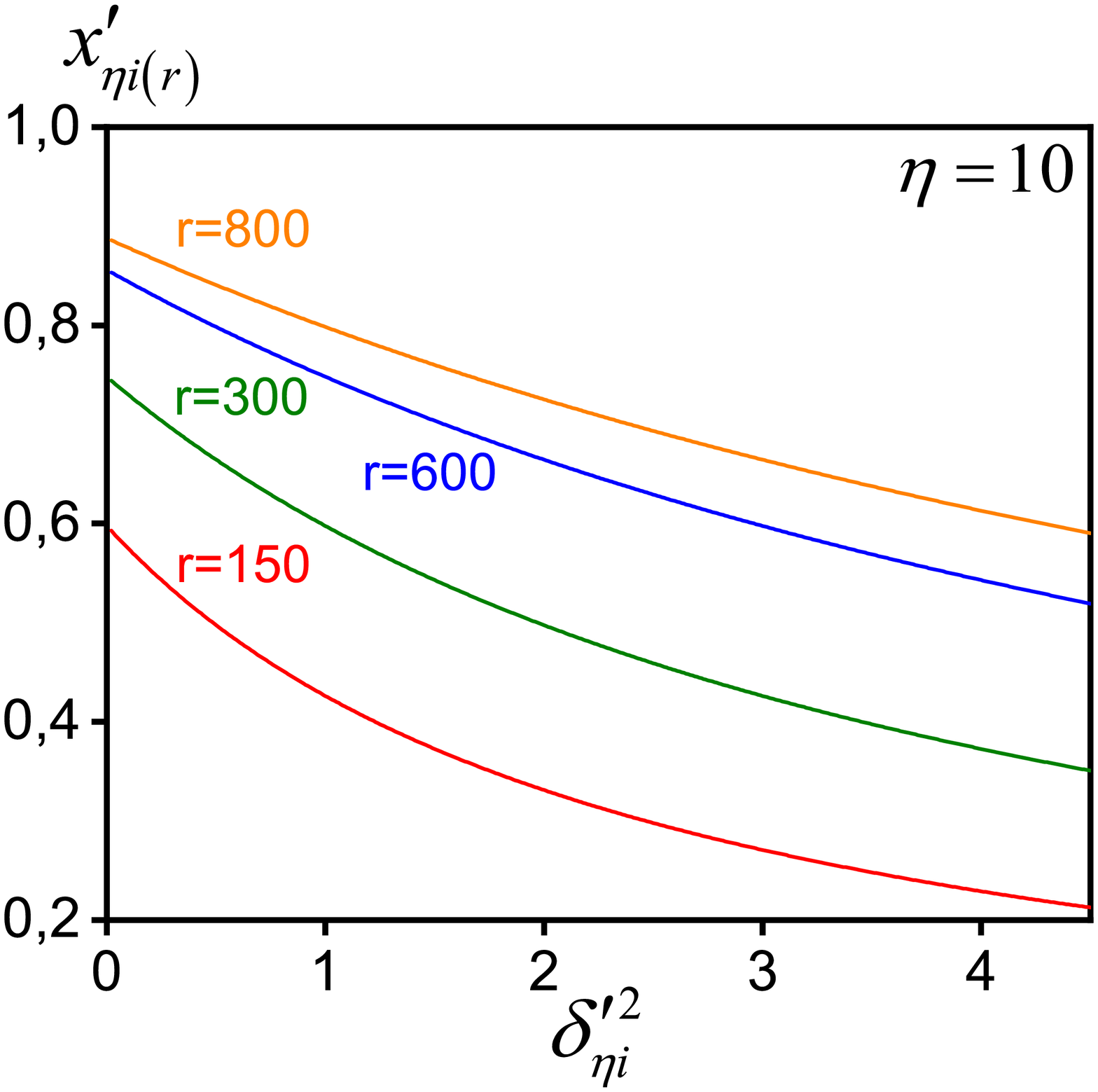}\label{<figure5a>}}
	\subfloat[]
	{\includegraphics[width=0.5\textwidth]{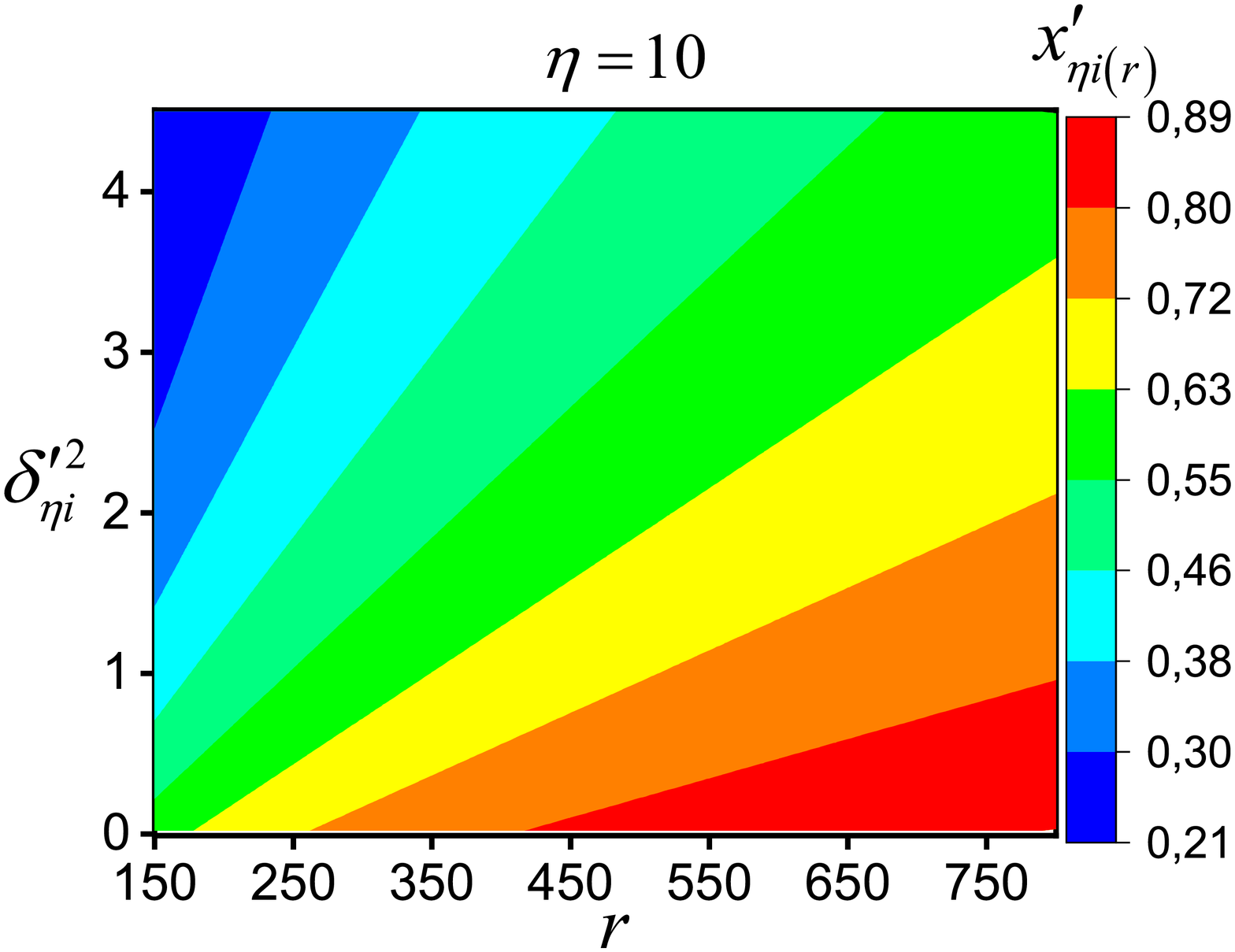}\label{<figure5b>}}
	\caption{Resonant frequency of the spontaneous photon for channel A \eqref{eq29} (in units of energy of the initial electron).  Fig.\ref{<figure5a>} shows the dependence of the resonant frequency ${{{x}'}_{\eta i\left( r \right)}}$ \eqref{eq29} on the square of its outgoing angle relative to the momentum of the initial electron for a fixed number of absorbed photons of the wave. Fig.\ref{<figure5b>} shows the dependence of the resonant frequency ${{{x}'}_{\eta i\left( r \right)}}$ \eqref{eq29} on the square of its outgoing angle relative to the momentum of the initial electron and also for different number of absorbed photons of the wave. The energy of the initial electrons ${{E}_{i}}=62.5\ \text{GeV}$, the angle between the momenta of the electrons and the laser wave ${{\theta }_{i}}=\pi $, the frequency of the wave $\omega =1\ \text{eV}$, the intensity of the laser wave $I\approx 1.861\cdot {{10}^{20}}\ \text{Wc}{{\text{m}}^{-2}}$.}
	\label{<figure5>}
\end{figure}

\begin{figure}[!h]
\centering
\subfloat[]
{\includegraphics[width=0.5\textwidth]{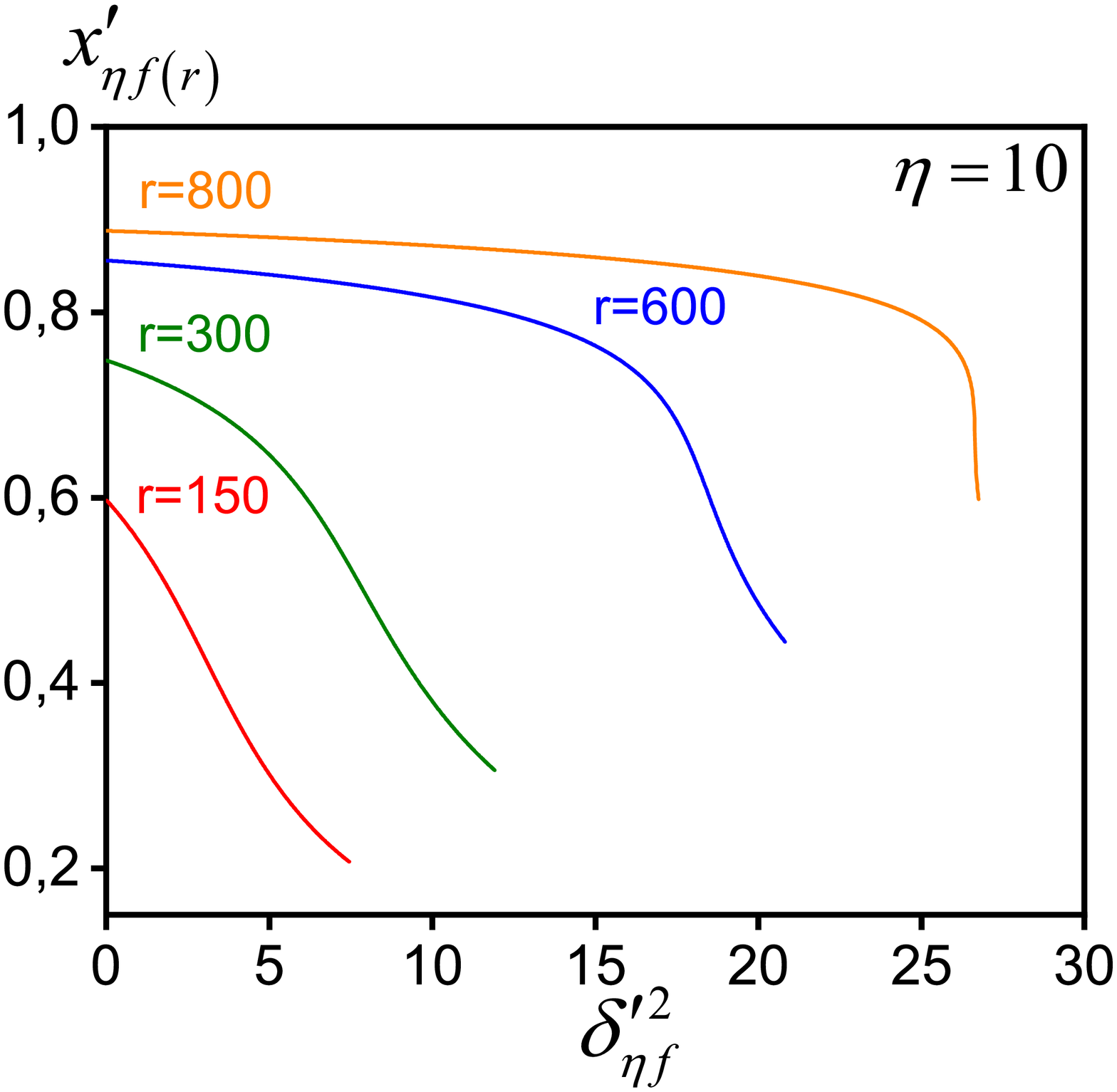}\label{<figure6a>}}
\subfloat[]
{\includegraphics[width=0.5\textwidth]{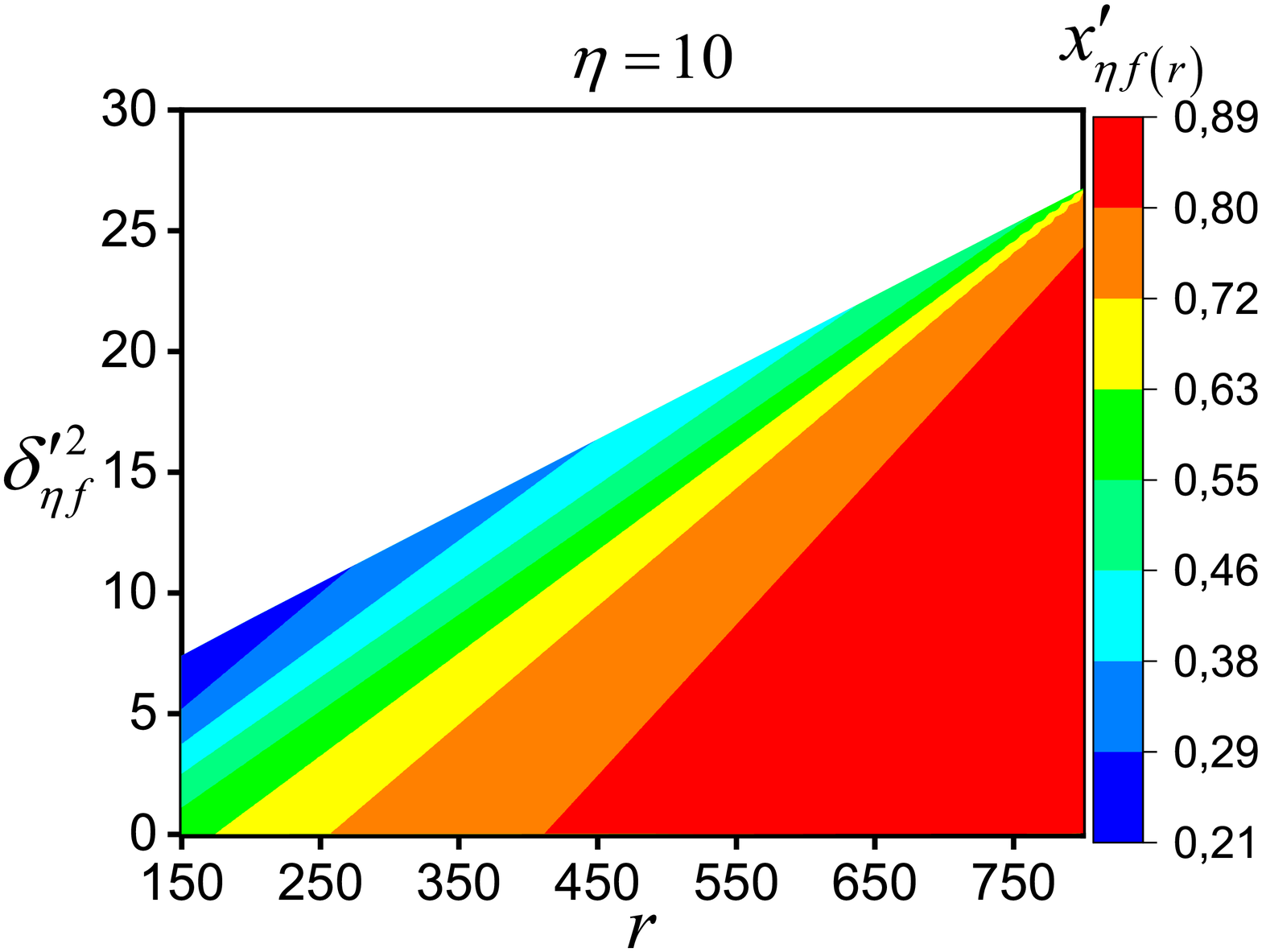}\label{<figure6b>}}
	\caption{Resonant frequency of the spontaneous photon for channel B \eqref{eq38}, \eqref{eq39} (in units of energy of the initial electron).  Fig.\ref{<figure6a>} shows the dependence of the resonant frequency ${{{x}'}_{\eta f\left( r \right)}}$ \eqref{eq39} on the square of its outgoing angle relative to the momentum of the final electron for a fixed number of absorbed photons of the wave. Fig.\ref{<figure6b>} shows the dependence of the resonant frequency ${{{x}'}_{\eta f\left( r \right)}}$ \eqref{eq39} on the square of its outgoing angle relative to the momentum of the final electron and also for different number of absorbed photons of the wave. The energy of the initial electrons ${{E}_{i}}=62.5\ \text{GeV}$, the angle between the momenta of the electrons and the laser wave ${{\theta }_{i}}=\pi $, the frequency of the wave $\omega =1\ \text{eV}$, the intensity of the laser wave $I\approx 1.861\cdot {{10}^{20}}\ \text{Wc}{{\text{m}}^{-2}}$.}
	\label{<figure6>}
\end{figure}

In this paper, we will be interested in resonant frequencies comparable to the energy of the initial electron. Therefore, we will assume that the number of absorbed photons of the wave is comparable to or greater than the characteristic parameter

\begin{equation}\label{eq34}
r\gtrsim {{r}_{\eta }}.
\end{equation}
The equation \eqref{eq30}  shows that $r_\eta \sim \eta^2\sim I \left( \text{Wcm}^{-2} \right)$. Therefore, the characteristic parameter $r_\eta$, as well as the number of absorbed photons of the wave in the resonant SB process, increase in proportion to the wave intensity. It follows from expression \eqref{eq29} that if a spontaneous gamma-quantum  is emitted along the momentum of the initial electron $\left( {\delta'}^2_{\eta i}=0 \right)$, then the resonant frequency takes a maximum value equal to (see the curves in Fig. \ref{<figure3a>} – Fig. \ref{<figure6a>})

\begin{equation}\label{eq35}
x'_{\eta i\left(r\right)} \left(0\right)=x'_{\eta \left( r \right)\max }=\left[1+\frac{r_\eta}{r} \right]^{-1}.
\end{equation}
With an increase in the outgoing angle of the spontaneous gamma-quantum , the resonant frequency \eqref{eq29} decreases and tends to zero as $\sim {{{\delta }'}^{-2}_{\eta i}}<<1$. Note that for weak fields, when $\eta <<1$ the expression for the resonant frequency of channel A \eqref{eq29} passes into the well-studied case (see \cite{27}).

We obtain the equation for the resonant frequency $\left( {{{{\omega }'}}_{\eta f\left( r \right)}} \right)$ of the spontaneous gamma-quantum in the case of channel B (see Fig. \ref{<figure2>}). Given the kinematics Eqs. \eqref{eq20}-\eqref{eq21}, from the expressions \eqref{eq19}, \eqref{eq27} we get:

\begin{equation}\label{eq36}
{\delta'}^2_{\eta f} {x'}^3_{\eta f\left(r\right)}-2{\delta'}^2_{\eta f} {x'}^2_{\eta f\left(r \right)}+\left( 1+{\delta'}^2_{\eta f}+\frac{r}{{{r}_{\eta }}} \right)x'_{\eta f\left(r\right)}-\frac{r}{r_\eta}=0,\quad x'_{\eta f\left(r\right)}=\frac{\omega'_{\eta f\left(r\right)}}{E_i}.
\end{equation}
Here it is indicated

\begin{equation}\label{eq37}
\delta'_{\eta f}=\frac{{E_i}{\theta'_f}}{m_*}=\frac{{E_i}{\theta'_f}}{m\sqrt{1+{{\eta }^{2}}}}.
\end{equation}
We will be interested in the real roots of Eq. \eqref{eq36} in the interval $0<x'_{\eta f\left( r \right)}<1$. It is easy to see that Eq.\eqref{eq36} has one real root when the parameter ${{\delta'}^{2}_{\eta f}}=0$, i.e. , the spontaneous gamma-quantum is emitted along the direction of the momentum of the final electron. In this case, the resonant frequency of the spontaneous gamma-quantum in channel B takes the maximum value $x'_{\eta f\left( r \right)}=x'_{\eta \left( r \right)\max }$ that coincides with the corresponding value for channel A (see Exp. \eqref{eq35}). For further analysis of equation \eqref{eq36}, we assume ${{\delta'}^{2}_{\eta f}}\ne 0$. A simple analysis of the cubic equation \eqref{eq36} in this case shows that there are two intervals for the number of absorbed photons of the wave, in which the resonant frequencies and the outgoing angles of the spontaneous gamma-quantum qualitatively changes. So, in the range of values

\begin{equation}\label{eq38}
{{r}_{\eta }}\lesssim r\le 8{{r}_{\eta }}
\end{equation}
there is a single valid solution to equation \eqref{eq36} for the resonant frequency of a spontaneous gamma-quantum:

\begin{equation}\label{eq39}
x'_{\eta f\left( r \right)}=\frac{2}{3}+\left( {{\alpha }_{\eta \left( r \right)+}}+{{\alpha }_{\eta \left( r \right)-}} \right),
\end{equation}
where
\begin{equation}\label{eq40}
	{{\alpha }_{\eta \left( r \right)\pm }}={{\left[ -\frac{{{b}_{\eta \left( r \right)}}}{2}\pm \sqrt{{{Q}_{\eta \left( r \right)}}} \right]}^{{1}/{3}\;}},\quad {{Q}_{\eta \left( r \right)}}={{\left( \frac{{{a}_{\eta \left( r \right)}}}{3} \right)}^{3}}+{{\left( \frac{{{b}_{\eta \left( r \right)}}}{2} \right)}^{2}},
\end{equation}

\begin{equation}\label{eq41}
	{{a}_{\eta \left( r \right)}}=\frac{1}{3{{{\delta }'}^{2}_{\eta f}}}\left[ 3\left( 1+\frac{r}{{{r}_{\eta }}} \right)-{{\delta'}^2_{\eta f}} \right],\quad {{b}_{r}}=\frac{1}{27{{{\delta }'}^{2}_{\eta f}}}\left[ 18+2{{{\delta }'}^{2}_{\eta f}}-\frac{9r}{{{r}_{\eta }}} \right]	.
\end{equation}
At the same time, the outgoing angle of the spontaneous gamma-quantum relative to the final electron varies from values close to zero to a certain maximum value

\begin{equation}\label{eq42}
0<{{{\delta }'}^{2}_{\eta f}}\le {{{\delta }'}^{2}_{\eta f\max }},\quad {{{\delta }'}^{2}_{\eta f\max }}=3\left( 1+\frac{r}{{{r}_{\eta }}} \right).
\end{equation}
In this case, the resonant frequency of the spontaneous gamma-quantum  \eqref{eq39} changes from the maximum value $x'_{\eta \left( r \right)\max }$ \eqref{eq35} at ${\delta'_{\eta f}}=0$ to the minimum value equal to

\begin{equation}\label{eq43}
x'_{\eta f\left( r \right)\min }=\frac{1}{3}\left\{ 2-{{\left[ \frac{8-\left( {r}/{{{r}_{\eta }}}\; \right)}{1+\left( {r}/{{{r}_{\eta }}}\; \right)} \right]}^{{1}/{3}\;}} \right\}
\end{equation}
at the maximum value of the outgoing angle ${{\delta'}^2_{\eta f\max }}$ \eqref{eq42}. For the outgoing angles ${{\delta'}^2_{\eta f}}>{{\delta'}^2_{\eta f\max }}$, there is no resonant spontaneous emission (see the curves in Fig. \ref{<figure3b>} – Fig. \ref{<figure6b>}).

\begin{figure}[!h]
	\centering
	\subfloat[]
	{\includegraphics[width=0.5\textwidth]{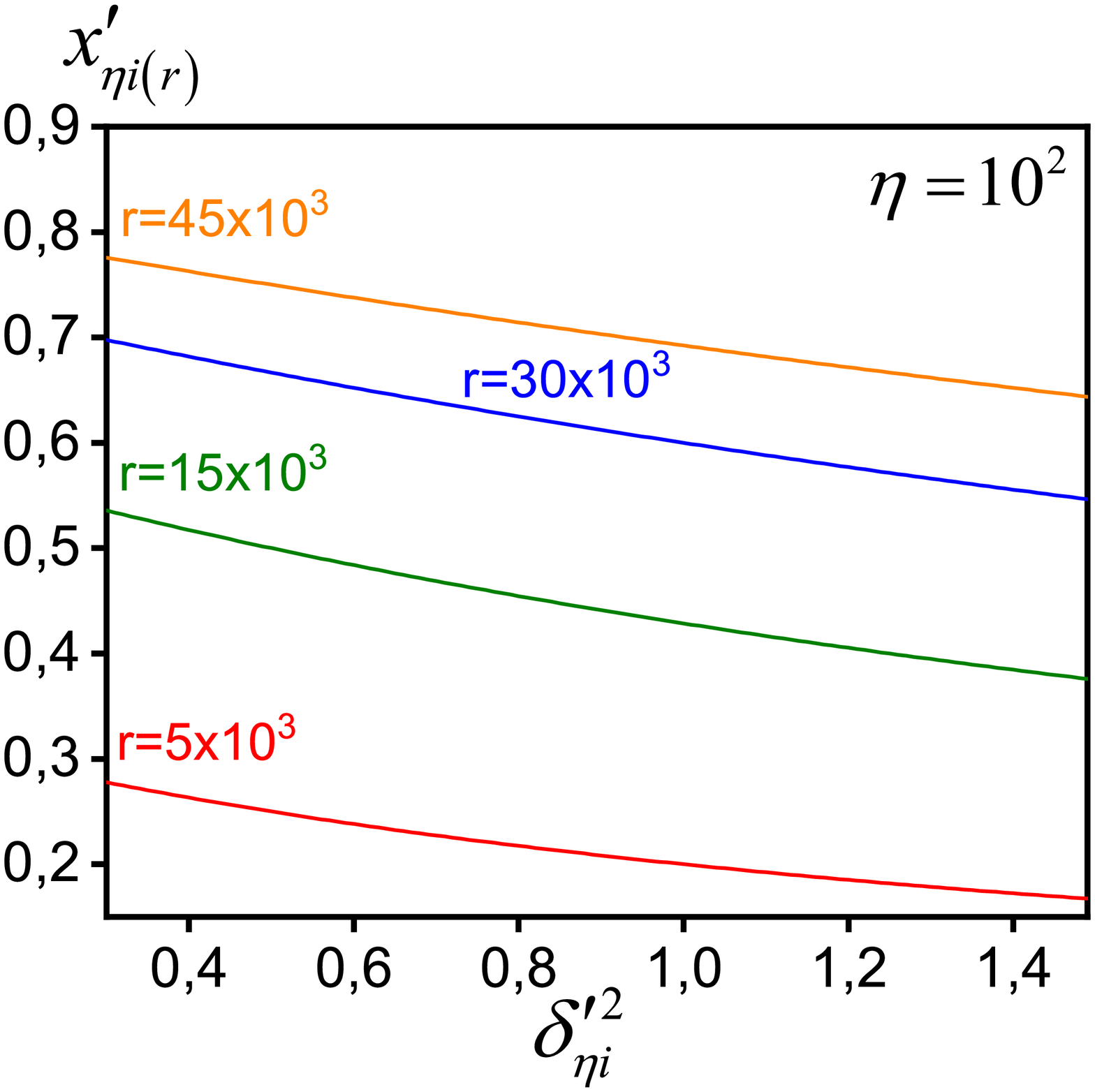}\label{<figure7a>}}
	\subfloat[]
	{\includegraphics[width=0.5\textwidth]{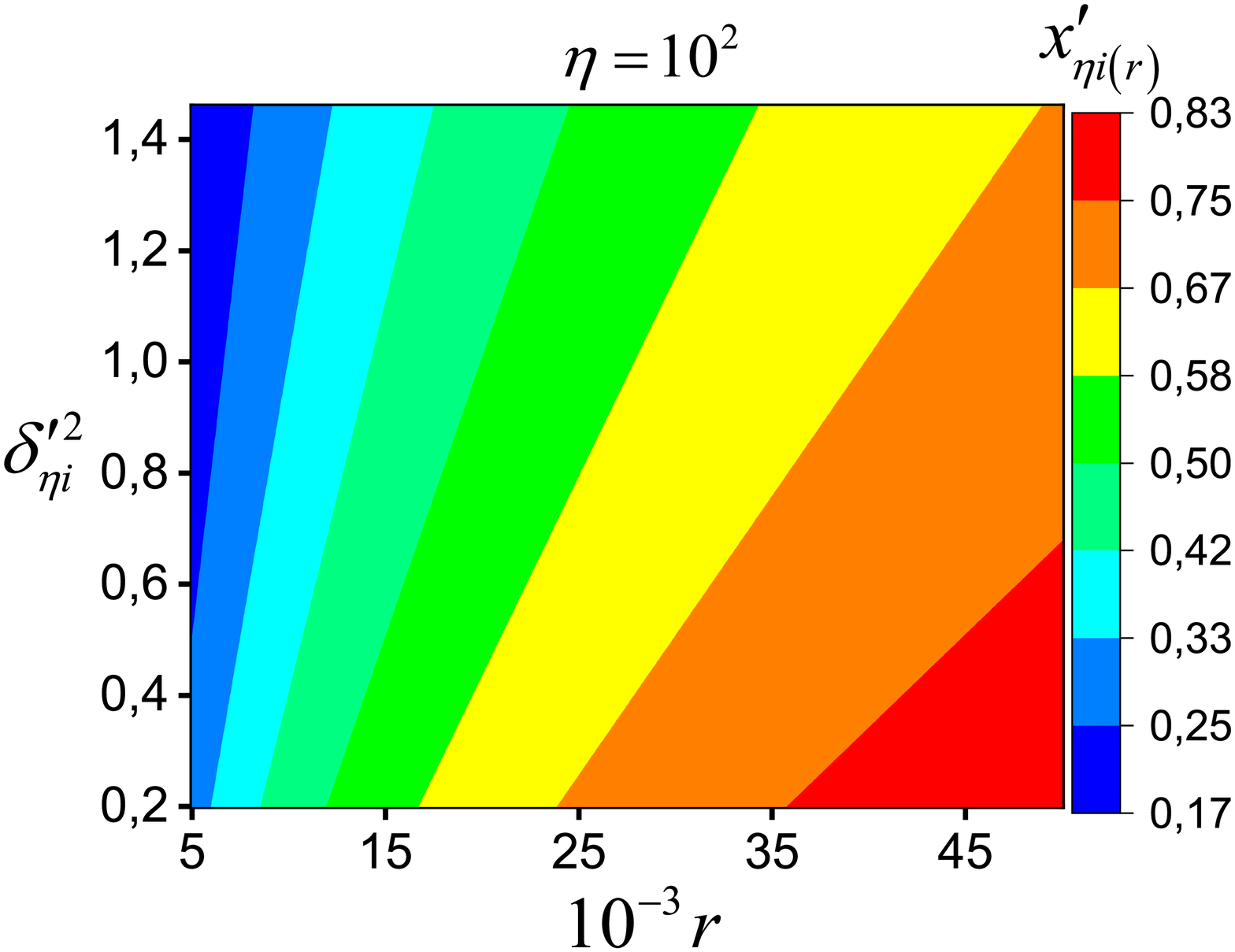}\label{<figure7b>}}
	\caption{Resonant frequency of the spontaneous photon for channel A \eqref{eq29} (in units of energy of the initial electron).  Fig.\ref{<figure7a>} shows the dependence of the resonant frequency ${{{x}'}_{\eta i\left( r \right)}}$ \eqref{eq29} on the square of its outgoing angle relative to the momentum of the initial electron for a fixed number of absorbed photons of the wave. Fig.\ref{<figure7b>} shows the dependence of the resonant frequency ${{{x}'}_{\eta i\left( r \right)}}$ \eqref{eq29} on the square of its outgoing angle relative to the momentum of the initial electron and also for different number of absorbed photons of the wave. The energy of the initial electrons ${{E}_{i}}=62.5\ \text{GeV}$, the angle between the momenta of the electrons and the laser wave ${{\theta }_{i}}=\pi $, the frequency of the wave $\omega =1\ \text{eV}$, the intensity of the laser wave $I\approx 1.861\cdot {{10}^{22}}\ \text{Wc}{{\text{m}}^{-2}}$.}
	\label{<figure7>}
\end{figure}

\begin{figure}[!h]
	\centering
	\subfloat[]
	{\includegraphics[width=0.5\textwidth]{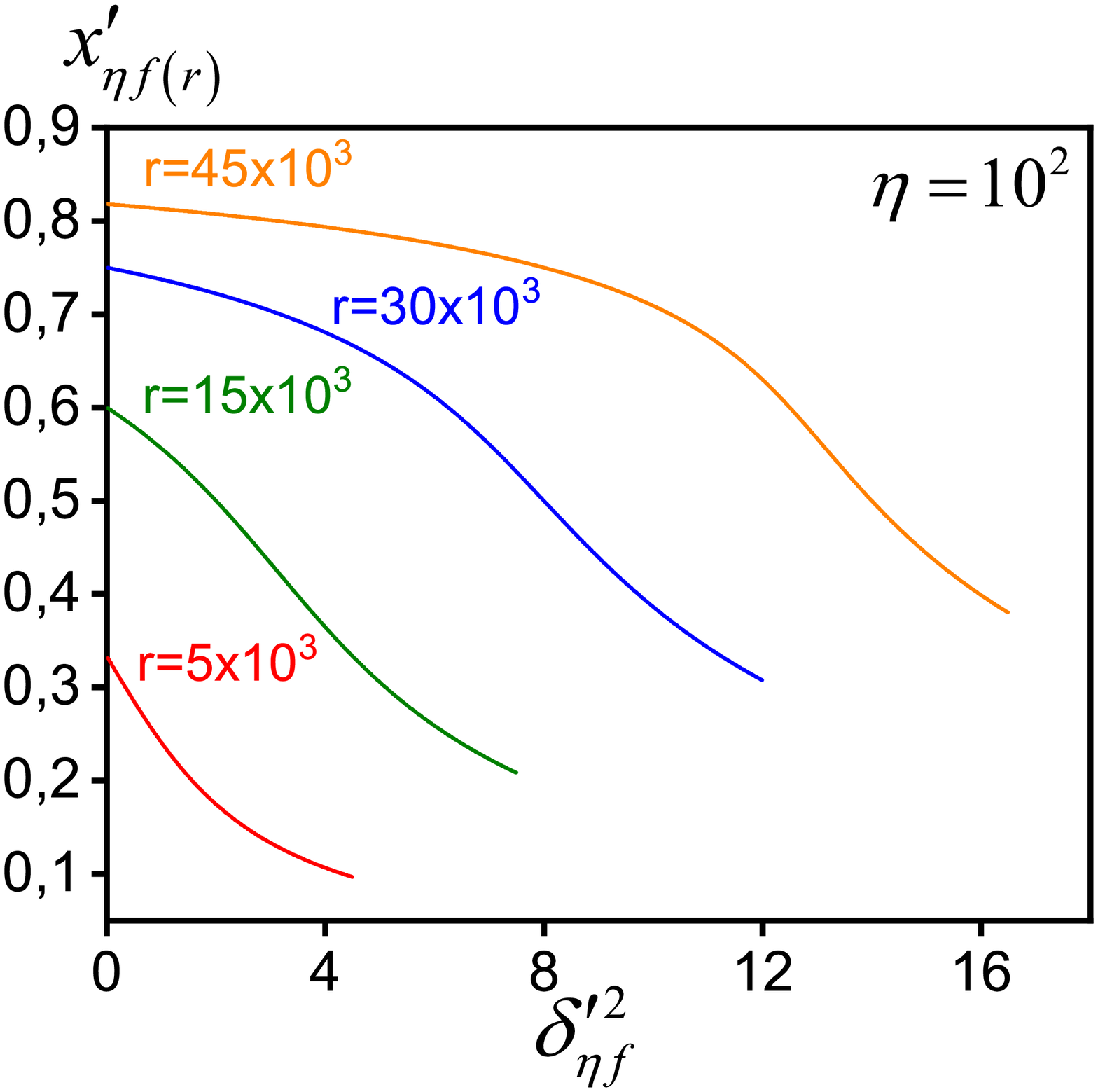}\label{<figure8a>}}
	\subfloat[]
	{\includegraphics[width=0.5\textwidth]{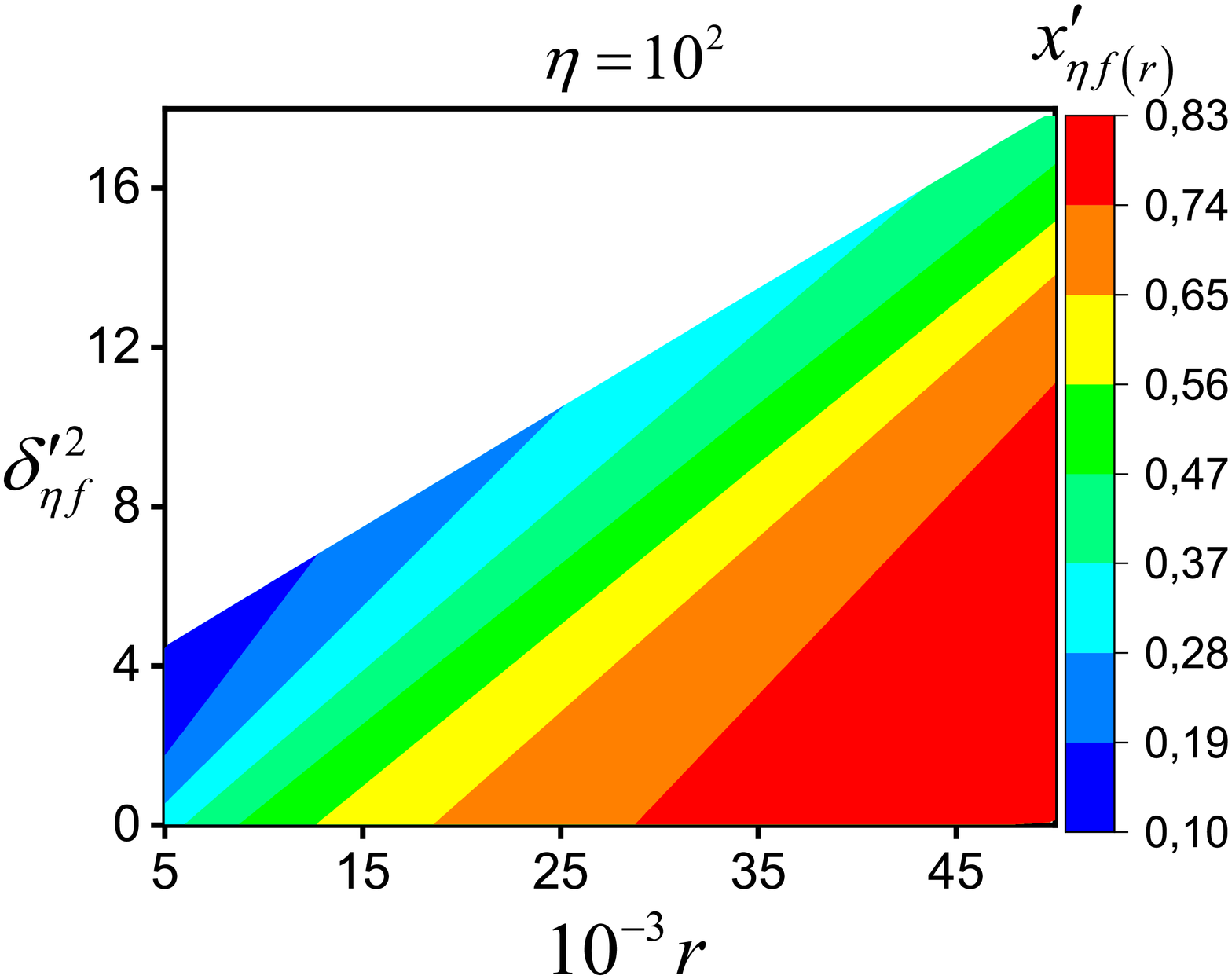}\label{<figure8b>}}
	\caption{Resonant frequency of the spontaneous photon for channel B \eqref{eq38}, \eqref{eq39} (in units of energy of the initial electron).  Fig.\ref{<figure8a>} shows the dependence of the resonant frequency ${{{x}'}_{\eta f\left( r \right)}}$ \eqref{eq39} on the square of its outgoing angle relative to the momentum of the final electron for a fixed number of absorbed photons of the wave. Fig.\ref{<figure8b>} shows the dependence of the resonant frequency ${{{x}'}_{\eta f\left( r \right)}}$ \eqref{eq39} on the square of its outgoing angle relative to the momentum of the final electron and also for different number of absorbed photons of the wave. The energy of the initial electrons ${{E}_{i}}=62.5\ \text{GeV}$, the angle between the momenta of the electrons and the laser wave ${{\theta }_{i}}=\pi $, the frequency of the wave $\omega =1\ \text{eV}$, the intensity of the laser wave $I\approx 1.861\cdot {{10}^{22}}\ \text{Wc}{{\text{m}}^{-2}}$.}
	\label{<figure8>}
\end{figure}

\begin{figure}[!h]
	\centering
	\subfloat[]
	{\includegraphics[width=0.5\textwidth]{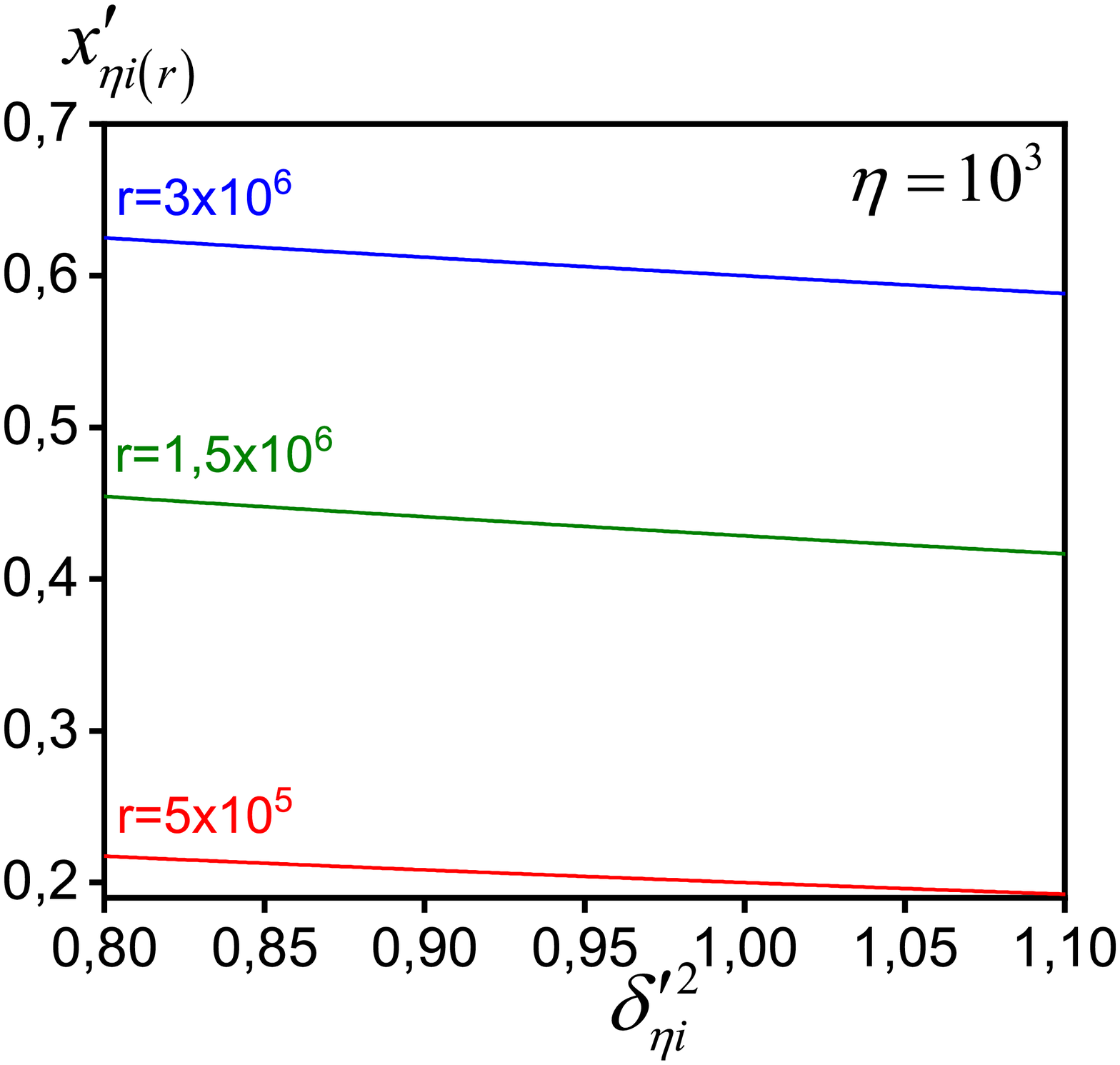}\label{<figure9a>}}
	\subfloat[]
	{\includegraphics[width=0.5\textwidth]{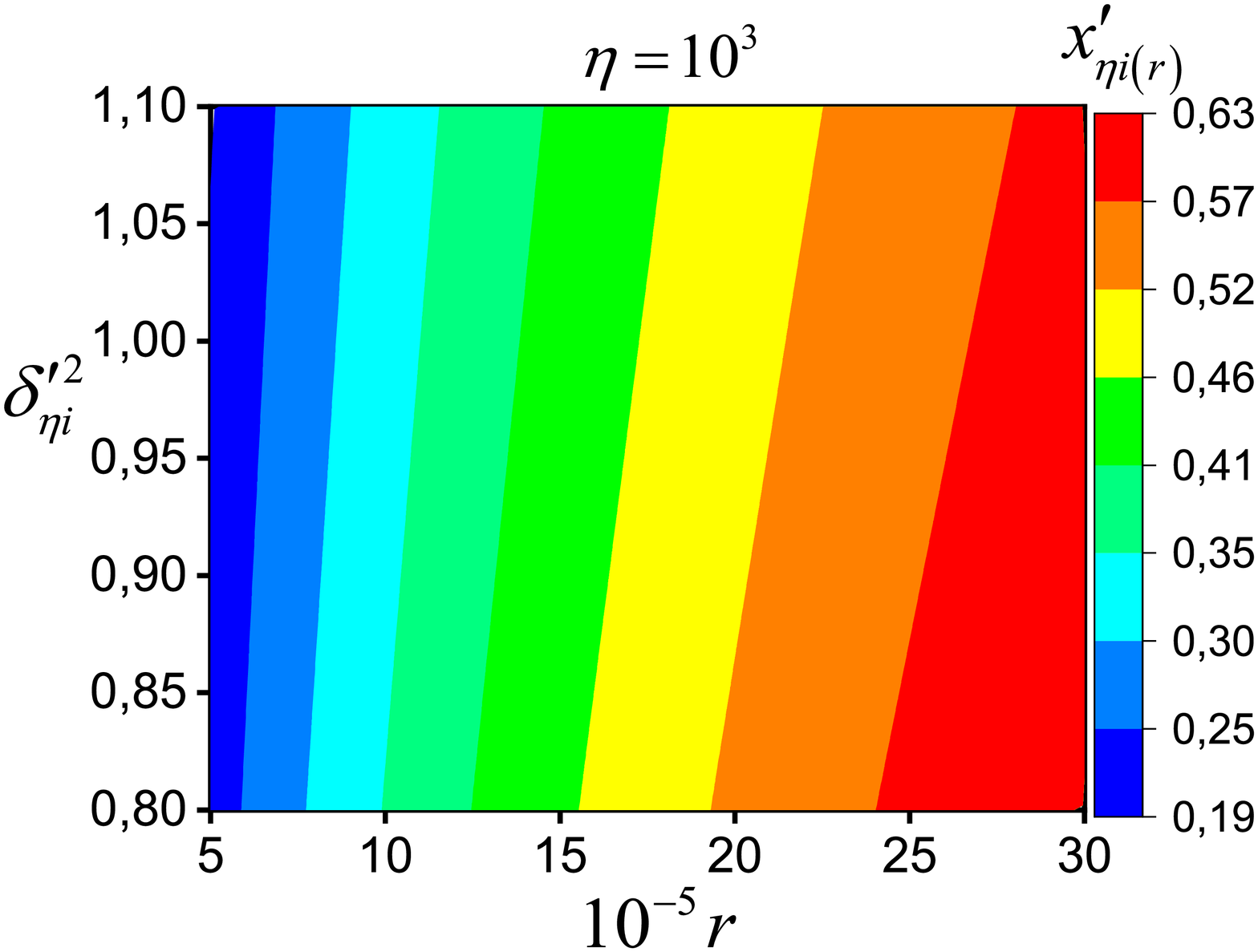}\label{<figure9b>}}
	\caption{Resonant frequency of the spontaneous photon for channel A \eqref{eq29}  (in units of energy of the initial electron).  Fig.\ref{<figure9a>} shows the dependence of the resonant frequency ${{{x}'}_{\eta i\left( r \right)}}$ \eqref{eq29} on the square of its outgoing angle relative to the momentum of the initial electron for a fixed number of absorbed photons of the wave. Fig.\ref{<figure9b>} shows the dependence of the resonant frequency ${{{x}'}_{\eta i\left( r \right)}}$ \eqref{eq29} on the square of its outgoing angle relative to the momentum of the initial electron and also for different number of absorbed photons of the wave. The energy of the initial electrons ${{E}_{i}}=62.5\ \text{GeV}$, the angle between the momenta of the electrons and the laser wave ${{\theta }_{i}}=\pi $, the frequency of the wave $\omega =1\ \text{eV}$, the intensity of the laser wave $I\approx 1.861\cdot {{10}^{24}}\ \text{Wc}{{\text{m}}^{-2}}$.}
	\label{<figure9>}
\end{figure}

\begin{figure}[!h]
	\centering
	\subfloat[]
	{\includegraphics[width=0.5\textwidth]{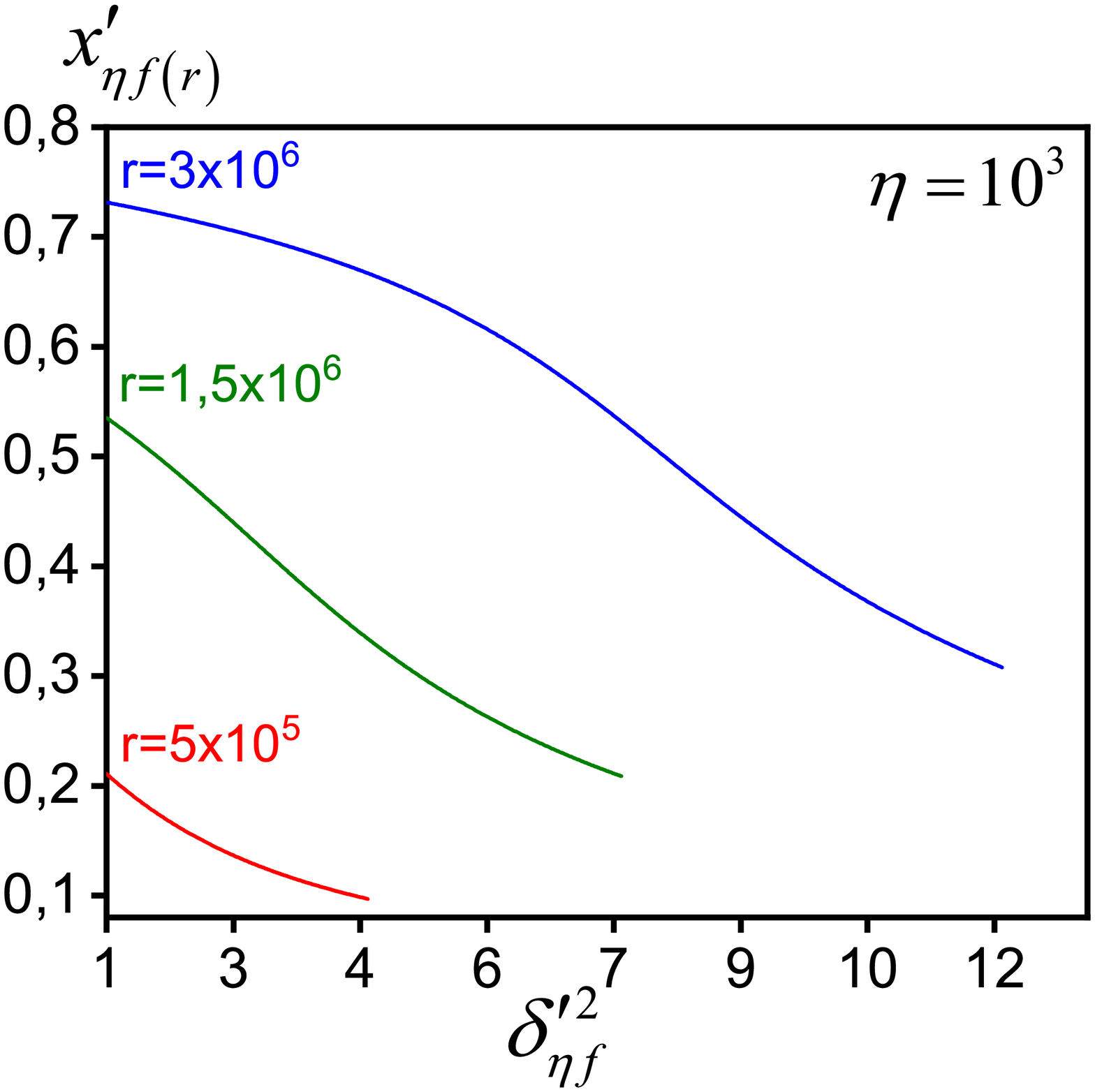}\label{<figure10a>}}
	\subfloat[]
	{\includegraphics[width=0.5\textwidth]{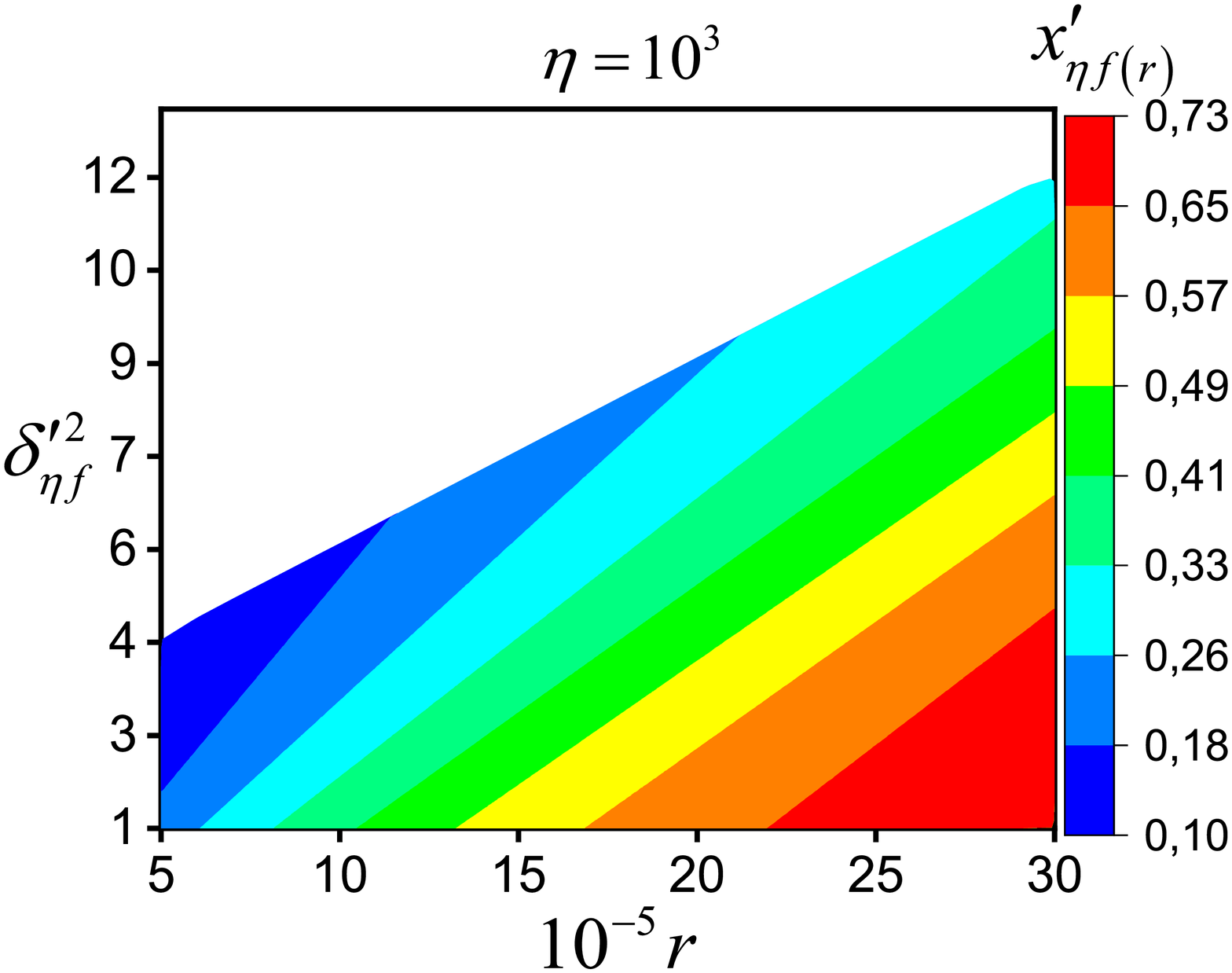}\label{<figure10b>}}
	\caption{Resonant frequency of the spontaneous photon for channel B \eqref{eq38}, \eqref{eq39} (in units of energy of the initial electron).  Fig.\ref{<figure10a>} shows the dependence of the resonant frequency ${{{x}'}_{\eta f\left( r \right)}}$ \eqref{eq39} on the square of its outgoing angle relative to the momentum of the final electron for a fixed number of absorbed photons of the wave. Fig.\ref{<figure10b>} shows the dependence of the resonant frequency ${{{x}'}_{\eta f\left( r \right)}}$ \eqref{eq39} on the square of its outgoing angle relative to the momentum of the final electron and also for different number of absorbed photons of the wave. The energy of the initial electrons ${{E}_{i}}=62.5\ \text{GeV}$, the angle between the momenta of the electrons and the laser wave ${{\theta }_{i}}=\pi $, the frequency of the wave $\omega =1\ \text{eV}$, the intensity of the laser wave $I\approx 1.861\cdot {{10}^{24}}\ \text{Wc}{{\text{m}}^{-2}}$.}
	\label{<figure10>}
\end{figure}

If the number of absorbed photons of the wave satisfies the condition
\begin{equation}\label{eq44}
r>8{{r}_{\eta }}
\end{equation}
then the nature of the solution of equation \eqref{eq36} depends on the outgoing angle of the spontaneous gamma-quantum relative to the final electron. So, for the outgoing angles in the intervals

\begin{equation}\label{eq45}
	0<{{{\delta }'}^{2}_{\eta f}}\le {{{\delta }'}^{2}_{\eta \left( r \right)-}},\quad {{{\delta }'}^{2}_{\eta \left( r \right)+}}\le {{{\delta }'}^{2}_{\eta f}}<\infty ,
\end{equation}
where

\begin{equation}\label{eq46}
{{{\delta }'}^{2}_{\eta \left( r \right)\pm }}=3\left( 1+\frac{r}{{{r}_{\eta }}} \right)+\left[ \frac{r}{8{{r}_{\eta }}}-1 \right]\left[ \frac{r}{{{r}_{\eta }}}+4\pm \sqrt{\frac{r}{{{r}_{\eta }}}\left( \frac{r}{{{r}_{\eta }}}-8 \right)} \right]
\end{equation}
there is one valid solution for the resonant frequency \eqref{eq39}-\eqref{eq41}. If the outgoing angles lie in the interval

\begin{equation}\label{eq47}
{{{\delta }'}^{2}_{\eta \left( r \right)-}}<{{{\delta }'}^{2}_{\eta f}}<{{{\delta }'}^{2}_{\eta \left( r \right)+}},
\end{equation}
then, for each outgoing angle of the spontaneous gamma-quantum, there are three possible resonant frequencies:

\begin{equation}\label{eq48}
	x'_{\eta f\left( r \right)1}=\frac{2}{3}\left[ 1+{d'_{\eta \left( r \right)}}\cos \left( \frac{{\varphi '_{\eta \left( r \right)}}}{3} \right) \right],\quad x'_{\eta f\left( r \right)2,3}=\frac{2}{3}\left[ 1+{d'_{\eta \left( r \right)}}\cos \left( \frac{\varphi '_{\eta \left( r \right)}}{3}\pm \frac{2\pi }{3} \right) \right],
\end{equation}

\begin{equation}\label{eq49}\begin{split}
& d'_{\eta \left( r \right)}=\frac{1}{\delta '_{\eta f}}\sqrt{{{{\delta }'}^{2}_{\eta f}}-3\left( 1+\frac{r}{{{r}_{\eta }}} \right)}, \\ \cos {\varphi'_{\eta \left( r \right)}}= & \delta '_{\eta f}\cdot \frac{\left( 9{r}/{{{r}_{\eta }}}\; \right)-2\left( {{{\delta }'}^{2}_{\eta f}}+9 \right)}{2{\left[ {{{\delta }'}^{2}_{\eta f}}-3\left( 1+{r}/{{{r}_{\eta }}}\; \right) \right]}^{{3}/{2}}},\quad 0\le {\varphi'_{\eta \left( r \right)}}\le \pi. 
\end{split} \end{equation}
Fig.\ref{<figure3>}-Fig.\ref{<figure10>} show the resonant frequencies of a spontaneous photon for channels A \eqref{eq29} and B \eqref{eq38}, \eqref{eq39}  as the functions of the square of its outgoing angle and a number of absorbed wave photons at different laser wave intensities from $I\approx 1.861\cdot {{10}^{18}}\ \text{Wc}{{\text{m}}^{-2}}$ to $I\approx 1.861\cdot {{10}^{24}}\ \text{Wc}{{\text{m}}^{-2}}$. From these figures, it can be seen that the resonant frequencies of spontaneous photons have energies of the order of the energy of the initial electrons only when the number of absorbed laser photons is greater than or of the order of the wave intensity (see Exps. \eqref{eq33}, \eqref{eq34}). Note that for $r<<{{\eta }^{2}}\sim I$, the resonant frequencies are small compared to the energies of the initial electrons. And in the case of $r>>{{\eta }^{2}}\sim I$, the resonant frequencies will be close to the energies of the initial electrons. However, the probability of a resonant process with such a large number of laser photons will be small (see Fig.\ref{<figure11>}-Fig.\ref{<figure18>}). Note also that for channel B, for the number of absorbed laser photons satisfying condition \eqref{eq38}, there is a maximum outgoing angle (see Exp. \eqref{eq42} and Fig.\ref{<figure4>}, Fig.\ref{<figure6>}, Fig.\ref{<figure8>}, Fig.\ref{<figure10>}).

\section{The Resonant Differential SB Cross Section in the Ultrarelativistic Energy Limit}

For the laser wave intensities \eqref{eq24} in the electron scattering amplitude on the nucleus in the wave field ${{M}_{r-l}}$ \eqref{eq11}, the second and third terms can be neglected ($\left| b_{\pm }^{0} \right|\lesssim \eta {m}/{{{E}_{i}}}<<1$, see Eq. \eqref{eq23}). As a result, this amplitude, for example, for channel A will take the form: 

\begin{equation}\label{eq50}
{{M}_{r-l}}\left( {{{\tilde{p}}}_{f}},{{{\tilde{q}}}_{i}} \right)={{\gamma }^{0}}{{L}_{r-l}}\left( {{{\tilde{p}}}_{f}},{{{\tilde{q}}}_{i}} \right),\quad {{L}_{r-l}}\left( {{{\tilde{p}}}_{f}},{{{\tilde{q}}}_{i}} \right)=\exp \left[ -i\left( r-l \right){{\chi }_{{{{\tilde{p}}}_{f}}{{{\tilde{q}}}_{i}}}} \right]{{J}_{r-l}}\left( {{\gamma }_{{{{\tilde{p}}}_{f}}{{{\tilde{q}}}_{i}}}} \right).
\end{equation}\\
In the case of channel B, substitutions must be made in Exp. \eqref{eq50}: ${{\tilde{q}}_{i}}\to {{\tilde{p}}_{i}},\ {{\tilde{p}}_{f}}\to {{\tilde{q}}_{f}}$.
It is important to emphasize that for the given initial parameters (the energy of the initial electron, the intensity and frequency of the wave, the angle between the momenta of the initial electron and the wave) for channel A, the energy of the spontaneous gamma quantum and the final electron is determined only by the outgoing angle of the spontaneous gamma quantum relative to the momentum of the initial electron (parameter ${{\delta '}^2_{\eta i}}$, see Eq. \eqref{eq29}).At the same time, for channel B, the energy of the final particles is determined by the angle between the momenta of the spontaneous gamma-quantum and the final  electron (the parameter ${{\delta '}^2_{\eta f}}$, see Exps. \eqref{eq36}, \eqref{eq39}-\eqref{eq41}, \eqref{eq48}, \eqref{eq49}). Because of this, channels A and B are distinguishable and do not interfere with each other. In addition, within the same channel (A or B), resonances with a different number of absorbed photons of the wave (a different number $r$) have different frequencies and, therefore, also do not interfere. Taking this into account, it is possible to obtain a resonant differential cross section for each reaction channel separately at fixed values of the number of photons of the wave $l$ and $r$. At the same time, in the sum by the $r$ index in expression \eqref{eq6}, it is sufficient to leave a specific term with a fixed number of absorbed photons of the wave. Using the expression for the process amplitude (see Exps. \eqref{eq4}-\eqref{eq6}, \eqref{eq12}, \eqref{eq50}), we obtain an expression for the resonant differential cross section in the case of unpolarized particles. After the standard calculations (see, for example, \cite{54}) for channel A, we get: 

\begin{equation}\label{eq51}
d{{\sigma}_{i(l,r)}}=d{{\mathbf{M}}_{i\left(r-l\right)}}\frac{8{{\pi }^{2}{m}^{2}{E}_{f}}}{{\left| \tilde{q}^{2}_{i}-m^{2}_{*} \right|}^{2}}d{\mathbf{K}_{i\left(r\right)}}.
\end{equation}
Here $d{{\mathbf{M}}_{i\left(r-l\right)}}$ is the differential cross-section of the scattering of an intermediate electron with a 4-momentum ${{\tilde{q}}_{i}}$ on the nucleus with the emission or absorption of $\left| r-l \right|$- photons of the wave \cite{41}:

\begin{equation}\label{eq52}
d{{\mathbf{M}}_{i\left(r-l\right)}}=4{{Z}^{2}}r_{e}^{2}\frac{{{m}^{2}}}{{{\mathbf{q}}^{4}}}J^{2}_{r-l}\left( {{\gamma }_{{{{\tilde{p}}}_{f}},{{{\tilde{q}}}_{i}}}} \right)\delta \left[ {{{\tilde{E}}}_{f}}-{{{\tilde{q}}}_{i0}}+\left( r-l \right)\omega  \right]{{d}^{3}}{{\tilde{p}}_{f}}.
\end{equation}
Here it is indicated:

\begin{equation}\label{eq53}
\mathbf{q}={{\mathbf{\tilde{p}}}_{f}}-{{\mathbf{\tilde{q}}}_{i}}+\left( r-l \right)\mathbf{k},
\end{equation}

\begin{equation}\label{eq54}
	{{\gamma }_{{{{\tilde{p}}}_{f}},{{{\tilde{q}}}_{i}}}}=\eta m\sqrt{-Q_{{{{\tilde{p}}}_{f}},{{{\tilde{q}}}_{i}}}^{2}},\quad Q_{{{{\tilde{p}}}_{f}},{{{\tilde{q}}}_{i}}}^{{}}=\frac{{{{\tilde{p}}}_{f}}}{\left( k{{{\tilde{p}}}_{f}} \right)}-\frac{{{{\tilde{q}}}_{i}}}{\left( k{{{\tilde{q}}}_{i}} \right)}.
\end{equation}
Function $d{\mathbf{K}_{i\left(r\right)}}$ determines the differential probability (per unit of time) of the laser-stimulated Compton effect with the absorption of $r$-photons of the wave \cite{35}:

\begin{equation}\label{eq55}
	d{\mathbf{K}_{i\left(r\right)}}=\frac{\alpha }{4\pi {\omega }'{{E}_{i}}}K\left( {{u}_{\eta i\left( r \right)}},{{v}_{\eta \left( r \right)}} \right){{d}^{3}}{k}',
\end{equation}
where

\begin{equation}\label{eq56}
	K\left( {{u}_{\eta i\left( r \right)}},{{v}_{\eta \left( r \right)}} \right)=-4J_{r}^{2}\left( {{\gamma }_{\eta i\left( r \right)}} \right)+{{\eta }^{2}}\left( 2+\frac{u_{\eta i\left( r \right)}^{2}}{1+{{u}_{\eta i\left( r \right)}}} \right)\left( J_{r+1}^{2}+J_{r-1}^{2}-2J_{r}^{2} \right),
\end{equation}

\begin{equation}\label{eq57}
	{{\gamma }_{\eta i\left( r \right)}}=2r\frac{\eta }{\sqrt{1+{{\eta }^{2}}}}\sqrt{\frac{{{u}_{\eta i\left( r \right)}}}{{{v}_{\eta \left( r \right)}}}\left( 1-\frac{{{u}_{\eta i\left( r \right)}}}{{{v}_{\eta \left( r \right)}}} \right)},
\end{equation}

\begin{equation}\label{eq58}
{{u}_{\eta i\left( r \right)}}=\frac{\left( k{k}' \right)}{\left( k{{q}_{i}} \right)}\approx \frac{{x'_{\eta i\left( r \right)}}}{\left( 1-{x'_{\eta i\left( r \right)}} \right)},\quad {{v}_{\eta \left( r \right)}}=2r\frac{\left( k{{p}_{i}} \right)}{m_{*}^{2}}=\frac{r}{{{r}_{\eta }}}.
\end{equation}
Here $\alpha $ is the fine structure constant. Given the expression for the resonant frequency in channel A \eqref{eq29}, as well as the relations \eqref{eq58}, the argument of the Bessel functions \eqref{eq57} and the ${{u}_{\eta i\left( r \right)}}$ parameter  \eqref{eq58} will take the form:

\begin{equation}\label{eq59}
	{{\gamma }_{\eta i\left( r \right)}}=2r\frac{\eta }{\sqrt{1+{{\eta }^{2}}}}\frac{{\delta'_{\eta i}}}{\left( 1+{{\delta '}^2_{\eta i}} \right)},\quad {{u}_{\eta i\left( r \right)}}={{\left( 1+{{\delta'}^2_{\eta i}} \right)}^{-1}}\frac{r}{{{r}_{\eta }}}.
\end{equation}

\begin{figure}[!h]
	\centering
	\subfloat[]
	{\includegraphics[width=0.5\textwidth]{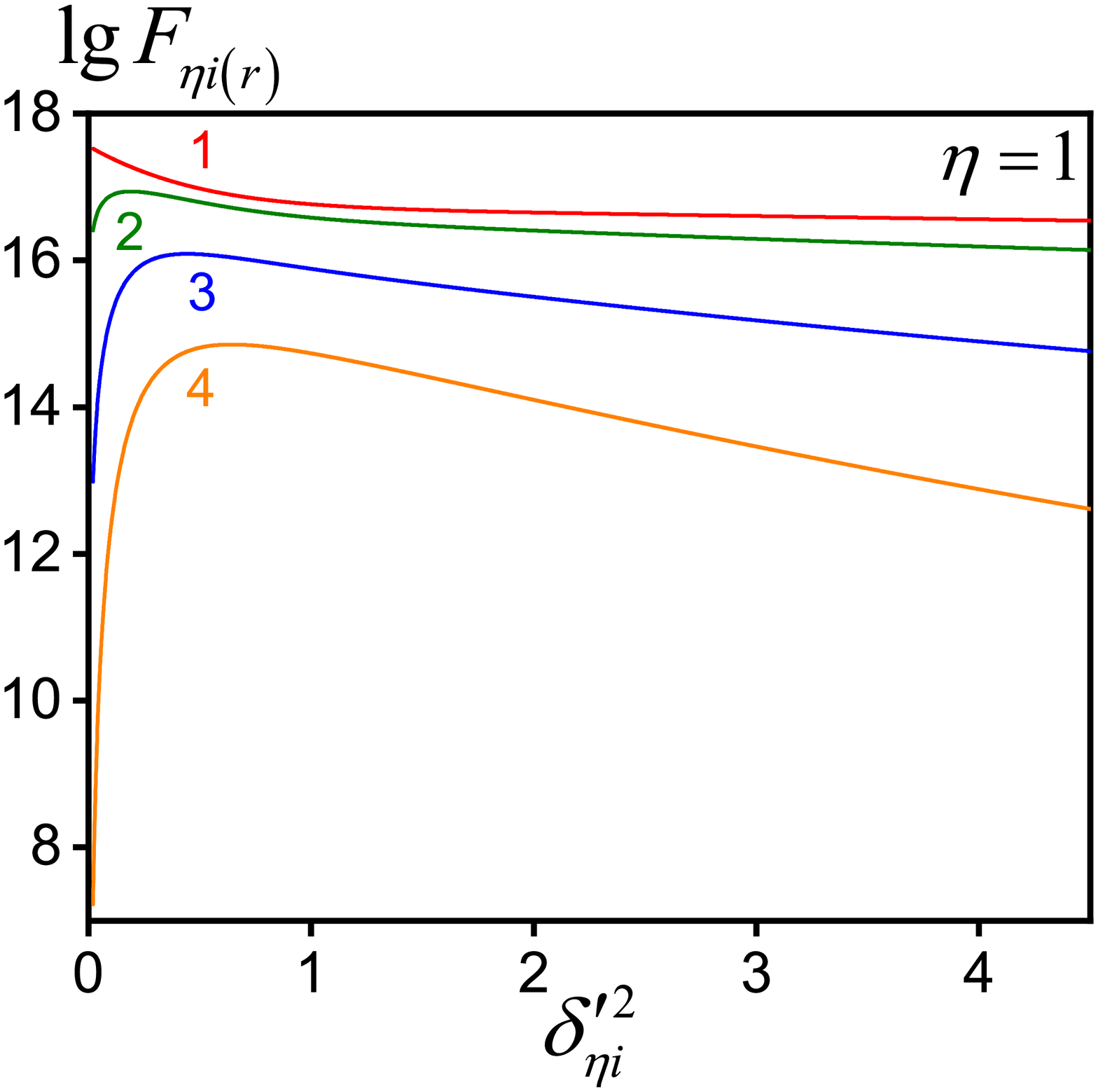}\label{<figure11a>}}
	\subfloat[]
	{\includegraphics[width=0.5\textwidth]{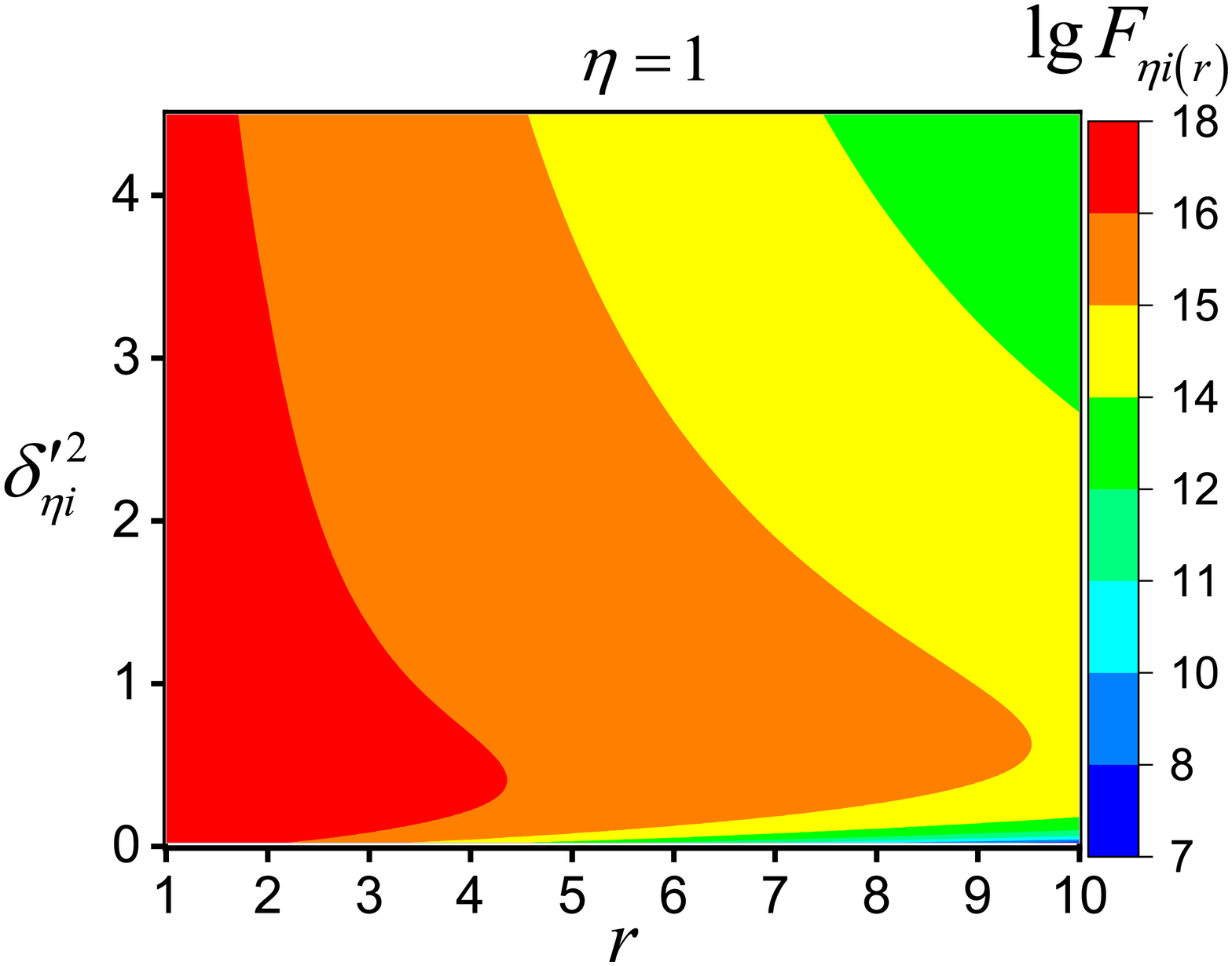}\label{<figure11b>}}
	\caption{The resonant differential cross-section for a channel A $R_{\eta i\left( r \right)}^{\max }$ \eqref{eq87}, \eqref{eq89} (in units of ${{Z}^{2}}\alpha r_{e}^{2}$).  Fig.\ref{<figure11a>} shows the dependence of the function ${{F}_{\eta i\left( r \right)}}$ \eqref{eq89} on the square of the outgoing angle of the spontaneous gamma-quantum relative to the momentum of the initial electron for a fixed number of absorbed photons of the wave: the curve 1 corresponds to $r=1$, the curve 2 corresponds to $r=2$, the curve 3 corresponds to $r=5$, the curve 4 corresponds to $r=10$.  Fig.\ref{<figure11b>} shows the dependence of the function ${{F}_{\eta i\left( r \right)}}$ \eqref{eq89} on the square of the outgoing angle of the spontaneous gamma-quantum relative to the momentum of the initial electron and also for different number of absorbed photons of the wave. The energy of the initial electrons ${{E}_{i}}=62.5\ \text{GeV}$, the angle between the momenta of the electrons and the laser wave ${{\theta }_{i}}=\pi $, the frequency of the wave $\omega =1\ \text{eV}$, the intensity of the laser wave $I\approx 1.861\cdot {{10}^{18}}\ \text{Wc}{{\text{m}}^{-2}}$.}
	\label{<figure11>}
\end{figure}

\begin{figure}[!h]
	\centering
	\subfloat[]
	{\includegraphics[width=0.5\textwidth]{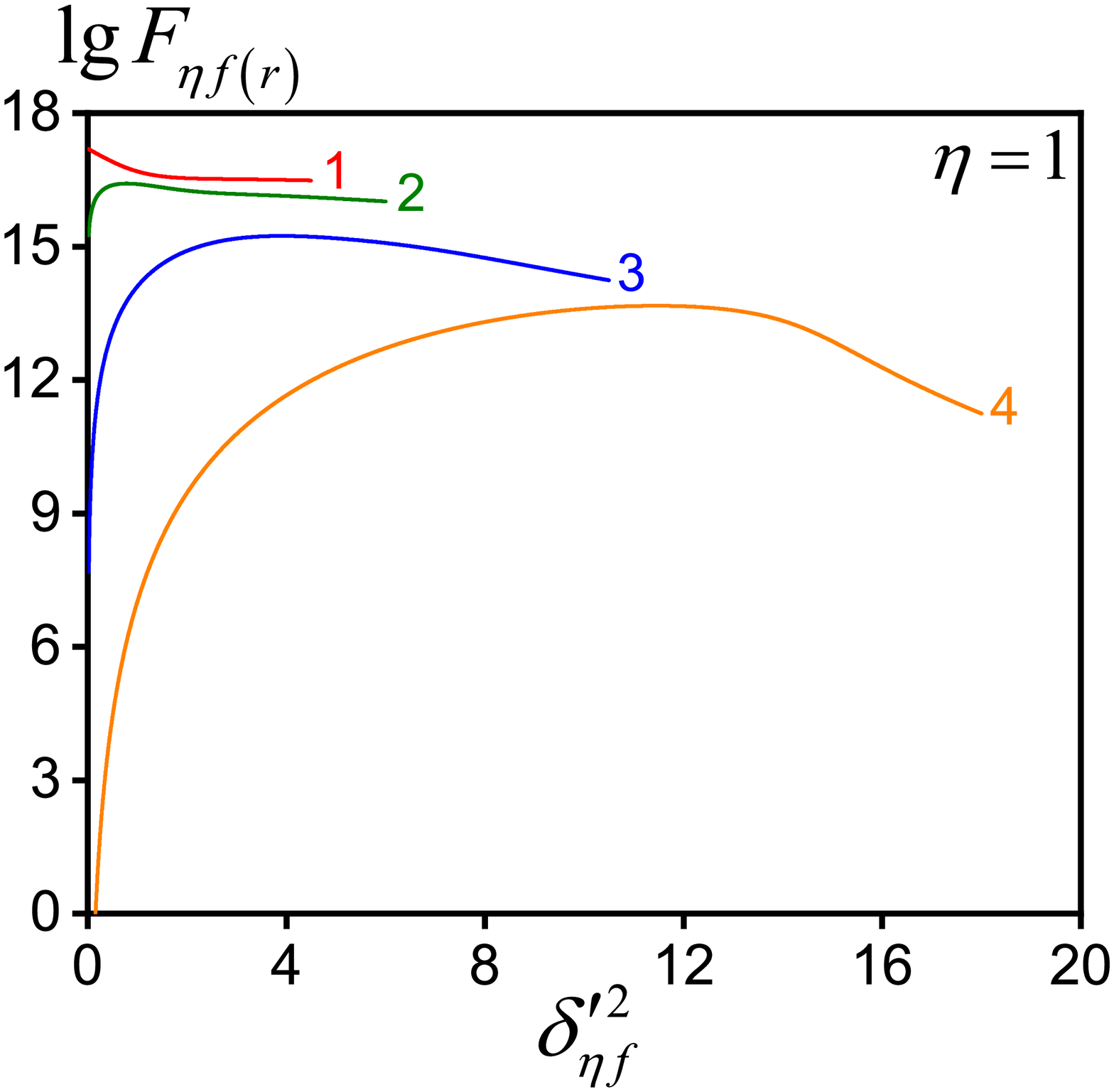}\label{<figure12a>}}
	\subfloat[]
	{\includegraphics[width=0.5\textwidth]{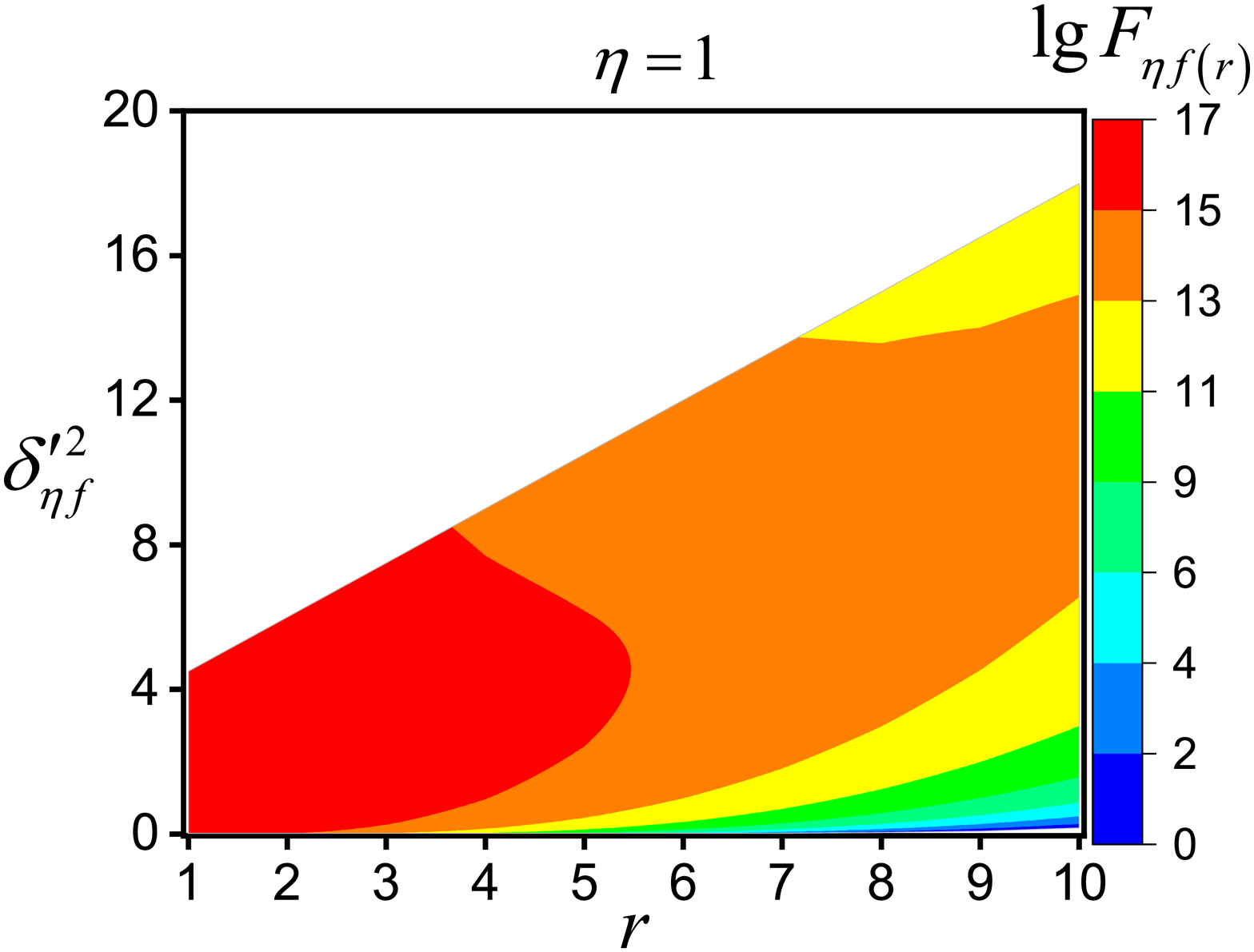}\label{<figure12b>}}
	\caption{The resonant differential cross-section for a channel B $R_{\eta f\left( r \right)}^{\max }$ \eqref{eq88}, \eqref{eq90} (in units of ${{Z}^{2}}\alpha r_{e}^{2}$).  Fig.\ref{<figure12a>} shows the dependence of the function ${{F}_{\eta f\left( r \right)}}$ \eqref{eq90} on the square of the outgoing angle of the spontaneous gamma-quantum relative to the momentum of the final electron for a fixed number of absorbed photons of the wave: the curve 1 corresponds to $r=1$, the curve 2 corresponds to $r=2$, the curve 3 corresponds to $r=5$, the curve 4 corresponds to $r=10$.  Fig.\ref{<figure12b>} shows the dependence of the function ${{F}_{\eta f\left( r \right)}}$ \eqref{eq90} on the square of the outgoing angle of the spontaneous gamma-quantum relative to the momentum of the final electron and also for different number of absorbed photons of the wave. The energy of the initial electrons ${{E}_{i}}=62.5\ \text{GeV}$, the angle between the momenta of the electrons and the laser wave ${{\theta }_{i}}=\pi $, the frequency of the wave $\omega =1\ \text{eV}$, the intensity of the laser wave $I\approx 1.861\cdot {{10}^{18}}\ \text{Wc}{{\text{m}}^{-2}}$.}
	\label{<figure12>}
\end{figure}

\begin{figure}[!h]
	\centering
	\subfloat[]
	{\includegraphics[width=0.5\textwidth]{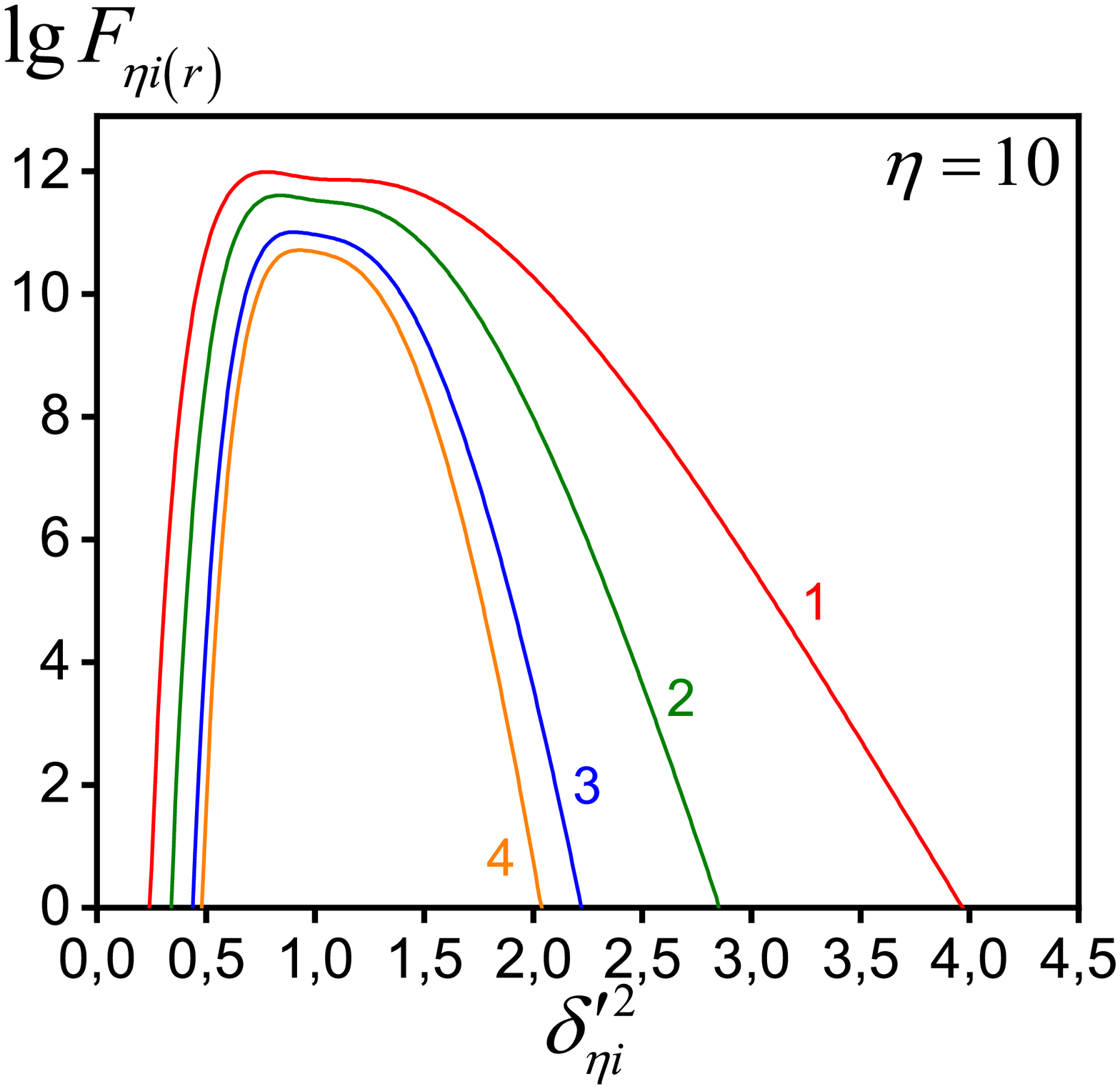}\label{<figure13a>}}
	\subfloat[]
	{\includegraphics[width=0.5\textwidth]{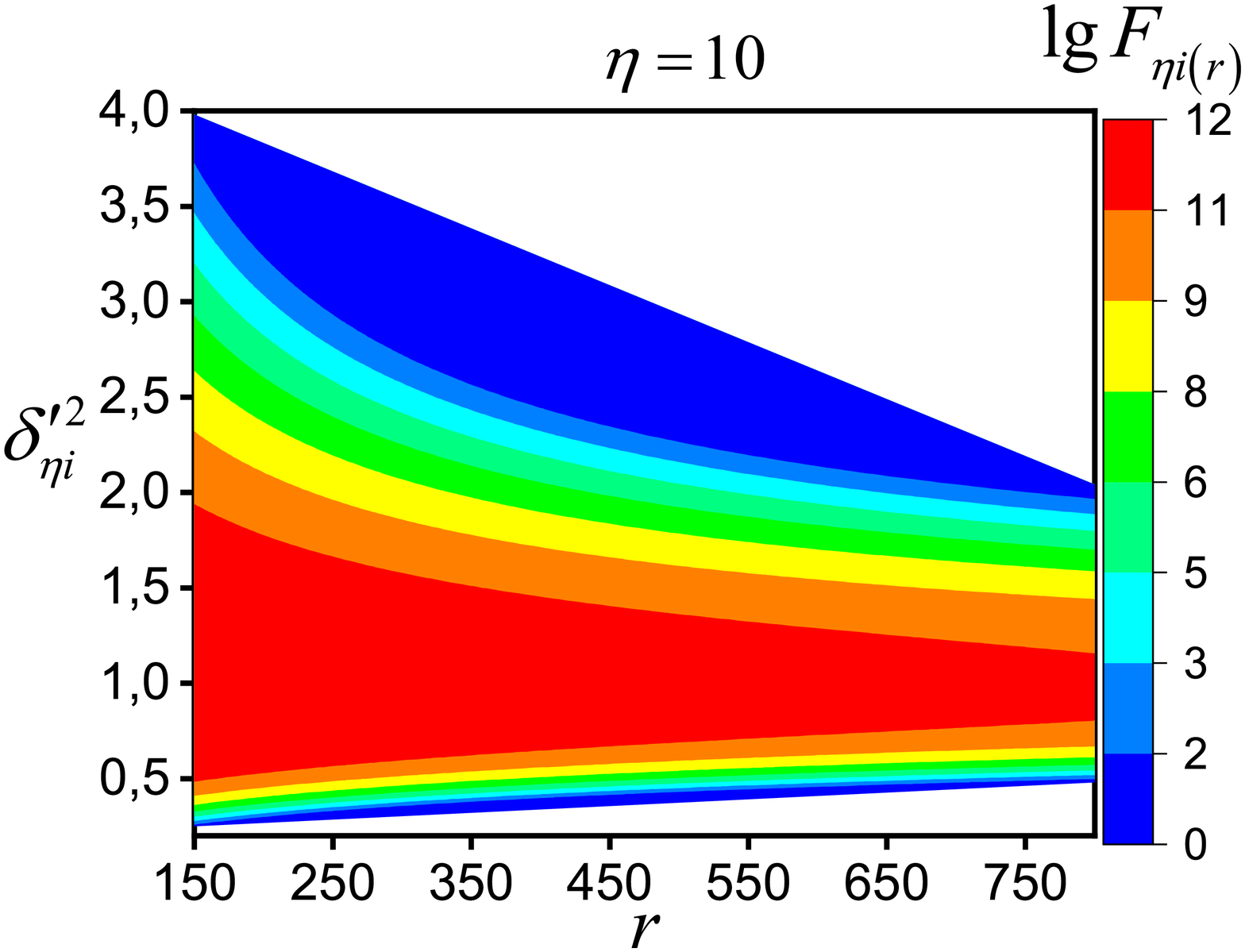}\label{<figure13b>}}
	\caption{The resonant differential cross-section for a channel A $R_{\eta i\left( r \right)}^{\max }$ \eqref{eq87}, \eqref{eq89} (in units of ${{Z}^{2}}\alpha r_{e}^{2}$).  Fig.\ref{<figure13a>} shows the dependence of the function ${{F}_{\eta i\left( r \right)}}$ \eqref{eq89} on the square of the outgoing angle of the spontaneous gamma-quantum relative to the momentum of the initial electron for a fixed number of absorbed photons of the wave: the curve 1 corresponds to $r=150$, the curve 2 corresponds to $r=300$, the curve 3 corresponds to $r=600$, the curve 4 corresponds to $r=800$.  Fig.\ref{<figure13b>} shows the dependence of the function ${{F}_{\eta i\left( r \right)}}$ \eqref{eq89} on the square of the outgoing angle of the spontaneous gamma-quantum relative to the momentum of the initial electron and also for different number of absorbed photons of the wave. The energy of the initial electrons ${{E}_{i}}=62.5\ \text{GeV}$, the angle between the momenta of the electrons and the laser wave ${{\theta }_{i}}=\pi $, the frequency of the wave $\omega =1\ \text{eV}$, the intensity of the laser wave $I\approx 1.861\cdot {{10}^{20}}\ \text{Wc}{{\text{m}}^{-2}}$.}
	\label{<figure13>}
\end{figure}

\begin{figure}[!h]
	\centering
	\subfloat[]
	{\includegraphics[width=0.5\textwidth]{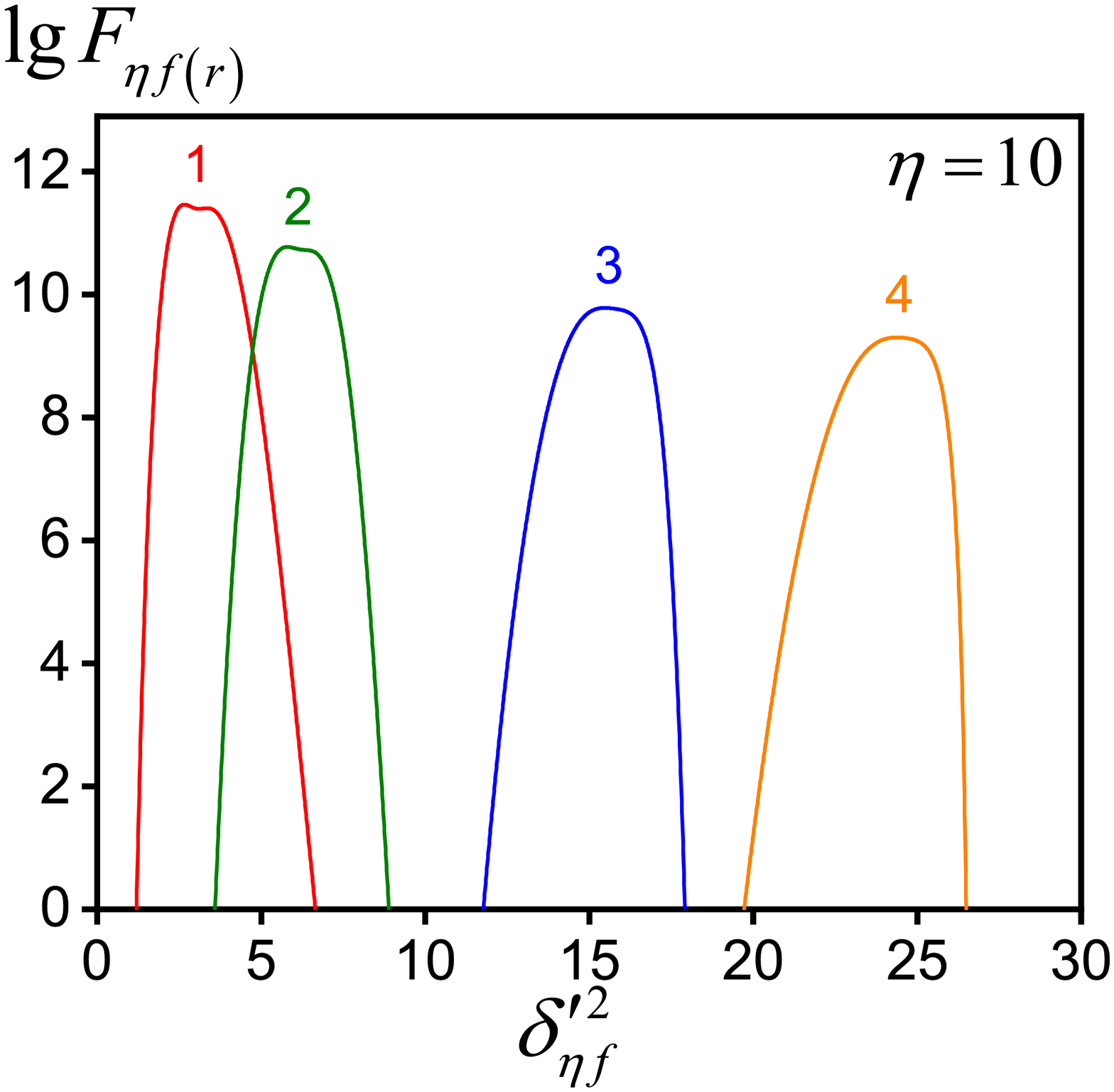}\label{<figure14a>}}
	\subfloat[]
	{\includegraphics[width=0.5\textwidth]{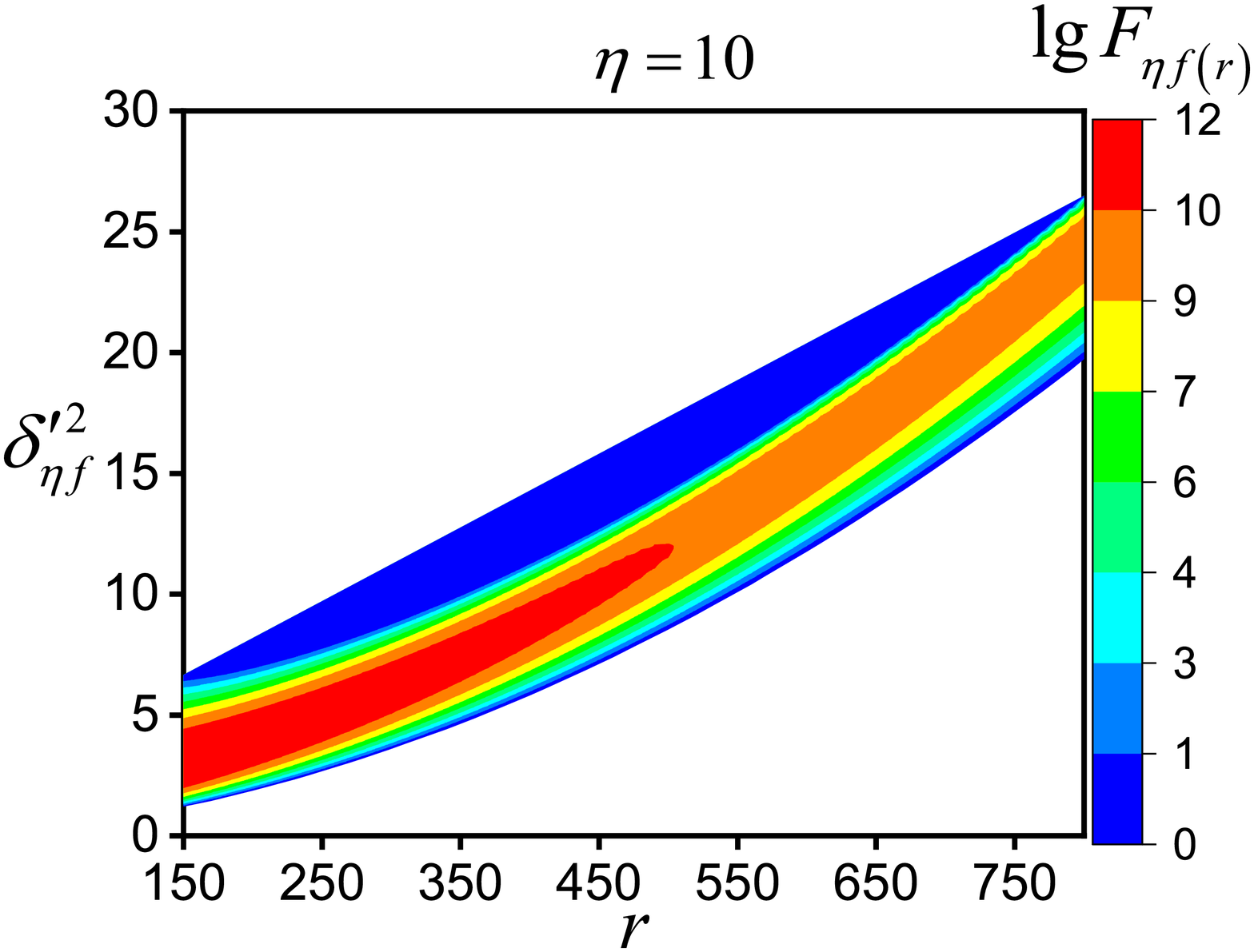}\label{<figure14b>}}
	\caption{The resonant differential cross-section for a channel B $R_{\eta f\left( r \right)}^{\max }$ \eqref{eq88}, \eqref{eq90} (in units of ${{Z}^{2}}\alpha r_{e}^{2}$).  Fig.\ref{<figure14a>} shows the dependence of the function ${{F}_{\eta f\left( r \right)}}$ \eqref{eq90} on the square of the outgoing angle of the spontaneous gamma-quantum relative to the momentum of the final electron for a fixed number of absorbed photons of the wave: the curve 1 corresponds to $r=150$, the curve 2 corresponds to $r=300$, the curve 3 corresponds to $r=600$, the curve 4 corresponds to $r=800$.  Fig.\ref{<figure14b>} shows the dependence of the function ${{F}_{\eta f\left( r \right)}}$ \eqref{eq90} on the square of the outgoing angle of the spontaneous gamma-quantum relative to the momentum of the final electron and also for different number of absorbed photons of the wave. The energy of the initial electrons ${{E}_{i}}=62.5\ \text{GeV}$, the angle between the momenta of the electrons and the laser wave ${{\theta }_{i}}=\pi $, the frequency of the wave $\omega =1\ \text{eV}$, the intensity of the laser wave $I\approx 1.861\cdot {{10}^{20}}\ \text{Wc}{{\text{m}}^{-2}}$.}
	\label{<figure14>}
\end{figure}

Note that, due to condition \eqref{eq24}, in the cross section \eqref{eq52}, it is possible to put \\ ${{d}^{3}}{{\tilde{p}}_{f}}\approx {{d}^{3}}{{p}_{f}}\approx E_{f}^{2}d{{E}_{f}}d{{\Omega }_{f}}$ and easily carry out the integration with respect to the energy of the final electron.

The elimination of the resonant infinity in channels A and B can be achieved by an imaginary addition to the effective mass of the intermediate electron \cite{53}. So, for channel A, we have:

\begin{equation}\label{eq60}
	{{m}_{*}}\to {{\mu }_{*}}={{m}_{*}}+i{{\Gamma }_{\eta i\left( r \right)}},\quad {{\Gamma }_{\eta i\left( r \right)}}=\frac{{{{\tilde{q}}}_{i0}}}{2{{m}_{*}}}W\left( {{r}_{\eta }} \right).
\end{equation}
Here ${W\left( {{r}_{\eta }} \right)}$ is the total probability (per unit time) of the laser-stimulated Compton effect on an intermediate electron with the 4-momentum ${{\tilde{q}}_{i}}$ \cite{35}.

\begin{equation}\label{eq61}
	W\left( {{r}_{\eta }} \right)=\frac{\alpha m^2}{4E_i}\mathbf{K} \left( r_\eta  \right),
\end{equation}

\begin{equation}\label{eq62}
\mathbf{K} \left(r_\eta \right)=\sum_{s=1}^{\infty} \mathbf{K}_{s} \left(r_\eta \right),\quad \mathbf{K}_s \left(r_\eta \right)=\int\limits_0^{s/r_\eta} \frac{du}{\left(1+u\right)^2} K \left(u, \frac{s}{r_\eta} \right).
\end{equation}
Given the relations \eqref{eq61}, \eqref{eq62}, the resonant width \eqref{eq60} will take the form:

\begin{equation}\label{eq63}
	{{\Gamma }_{\eta i\left( r \right)}}=\alpha m\frac{\left( 1-{x'_{\eta i\left( r \right)}} \right)}{8\sqrt{1+{{\eta }^{2}}}}\mathbf{K} \left( {{r}_{\eta }} \right).
\end{equation}
Here the function $K\left( u,{r}/{{{r}_{\eta }}}\; \right)$ is obtained from the expressions \eqref{eq56}-\eqref{eq58} by replacing: ${{u}_{\eta i\left( r \right)}}\to u,\ r\to s$. Taking into account the relations \eqref{eq60}-\eqref{eq63}, the resonant denominator can be represented as:      

\begin{equation}\label{eq64}
\left| \tilde{q}^{2}_{i}-\mu^{2}_{*} \right|^2=m^4x'^2_{\eta i\left( r \right)}\left[ \left( 1+\eta^2 \right)^2 \left( {\delta'}^2_{\eta i}-{\delta'}^2_{\eta i\left( r \right)} \right)^2+\Upsilon^2_{\eta i\left( r \right)} \right].
\end{equation}
Here ${{\Upsilon }_{\eta i\left( r \right)}}$ is the angular resonant width.

\begin{equation}\label{eq65}
	\Upsilon_{\eta i\left( r \right)}=\frac{2\Gamma _{\eta i\left(r \right)}}{m_*x'_{\eta i\left( r \right)}}=\alpha \frac{\left(1-x'_{\eta i\left(r \right)}\right)}{4x'_{\eta i\left( r \right)}} \mathbf{K} \left( r_\eta \right).
\end{equation}
In expression \eqref{eq64}, the ${\delta '}^2_{\eta i\left( r \right)}$ parameter is related to the resonant frequency by the ratio \eqref{eq29}, and the ${\delta'}^2_{\eta i}$ parameter  changes independently of the frequency of the spontaneous gamma-quantum. Note that the angular resonance width ${{\Upsilon }_{\eta i\left( r \right)}}$ \eqref{eq65} increases significantly with increasing wave intensity. This is due to the growth of the function $\mathbf{K} \left( {{r}_{\eta }} \right)$ \eqref{eq62}. So, for $\eta \sim 1$ a function $\mathbf{K} \left( {{r}_{\eta }} \right)\sim 1$, and for $\eta \sim {{10}^{3}}$ a function $\mathbf{K} \left( {{r}_{\eta }} \right)\sim {{10}^{2}}$. It should also be noted that in this paper we have limited the intensity of the laser field $F\lesssim {{10}^{14}}\ {\text{V}}/{\text{cm}}\;$ (see condition \eqref{eq24} and the text after this formula). Because of this, the angular resonance width ${{\Upsilon }_{\eta i\left( r \right)}}$ will be significantly greater than the corresponding radiation corrections \cite{33}. 
Given this, as well as Exps. \eqref{eq20}, \eqref{eq29}, \eqref{eq36}, \eqref{eq24}, \eqref{eq51}, \eqref{eq64} after simple calculations \cite{54}, we obtain expressions for the resonant differential SB cross section for channels A and B in the following form:
\begin{equation}\label{eq66} \begin{split}
	\frac{d\sigma_{\eta i\left(l,r\right)}}{dx'_{\eta i\left( r \right)}d{{\delta }'}^{2}_{\eta i}}=\left( Z^2 \alpha r^2_e \right)\frac{J^2_{r-l}\left( \alpha_{\eta i\left(r \right)} \right)}{g^4_A} & \frac{\left(2\pi \right)^2 \left( 1-x'_{\eta i\left( r \right)} \right)^3} {\left[ \left( 1+\eta^2 \right)^2 \left({{\delta }'}^{2}_{\eta i}-\delta'^2_{\eta i\left( r \right)} \right)+\Upsilon^2_{\eta i\left( r \right)} \right]x'_{\eta i\left( r \right)}} \times {} \\ & {} \times  K\left( u_{\eta i\left( r \right)},\frac{r}{r_\eta} \right)d \delta'^2_{\eta f}d \varphi'_{-},
\end{split} \end{equation}

\begin{equation}\label{eq67} \begin{split}
	\frac{d\sigma_{\eta f\left( l,r \right)}}{dx'_{\eta f\left( r \right)}d{{\delta }'}^{2}_{\eta f}}=\left( Z^2\alpha r^2_e \right)K\left( u_{\eta f\left( r \right)},\frac{r}{r_{\eta }} \right) & \frac{\left(2\pi  \right)^2\left(1-x'_{\eta f\left( r \right)} \right)^{-1}}{\left[ \left( 1+\eta^2 \right)^2 \left( {{\delta }'}^{2}_{\eta f}-{{\delta }'}^{2}_{\eta f\left( r \right)} \right)^2+\Upsilon^2_{\eta f\left( r \right)} \right]x'_{\eta f\left( r \right)}} \times {} \\ & {} \times  \frac{J^2_{r-l}\left( \alpha_{\eta f\left( r \right)} \right)}{g^4_B}d{{\delta }'}^{2}_{\eta i} d\varphi'_{-},
\end{split} \end{equation}
where
\begin{equation}\label{eq68}
g_{A}^{2}=g_{\eta 0}^{2}+\frac{m_{*}^{2}}{2E_{i}^{2}}{{g}_{\eta i\left( l,r \right)}},\quad g_{B}^{2}=g_{\eta 0}^{2}+\frac{m_{*}^{2}}{2E_{i}^{2}}{{g}_{\eta f\left( l,r \right)}}.
\end{equation}
Here ${\varphi '_{-}}$ is the angle between the planes $\left( \mathbf{{k}'},{{\mathbf{p}}_{i}} \right)$ and $\left( \mathbf{{k}'},{{\mathbf{p}}_{f}} \right)$, the ${\delta '}^2_{\eta f\left( r \right)}$ parameter  is related to the resonant frequency by the ratio \eqref{eq36}, and the ${\delta '}^2_{\eta f}$ parameter  changes independently of the frequency of the spontaneous gamma-quantum . The relativistic-invariant parameter ${u_{\eta f\left( r \right)}}$ and the resonant width for channel B have the form:

\begin{equation}\label{eq69}
	u_{\eta f\left( r \right)}=\frac{\left( k{k}' \right)}{\left( k{{p}_{f}} \right)}\approx \frac{{x'_{\eta f\left( r \right)}}}{\left( 1-{x'_{\eta f\left( r \right)}} \right)},
\end{equation}

\begin{equation}\label{eq70}
	\Upsilon _{f\left( r \right)}^{{}}=\frac{\alpha }{4{x'_{\eta f\left( r \right)}}\left( 1-{x'_{\eta f\left( r \right)}} \right)}\mathbf{K} \left( {{r}_{\eta }} \right).
\end{equation}
The $K\left( {{u}_{\eta f\left( r \right)}},{r}/{{{r}_{\eta }}}\; \right)$ function is obtained from the corresponding expression for channel A (see Eqs. \eqref{eq56}-\eqref{eq57}), if a replacement is made in the latter ${{u}_{\eta i\left( r \right)}}\to {{u}_{\eta f\left( r \right)}}$. The $g_{\eta 0}^{2}$ and ${{g}_{\eta i\left( l,r \right)}},\ {{g}_{\eta f\left( l,r \right)}}$ functions  \eqref{eq68} determine the square of the momentum transmitted to the nucleus, taking into account the relativistic corrections of the order ${m_{*}^{2}}/{E_{i}^{2}}\;$ for channels A and B. So, for channel A, we get:
\begin{equation}\label{eq71}
g_{\eta 0}^2={\delta '}^2_{\eta i}+\tilde{\delta '}^2_{\eta f}-2{\delta'_{\eta i}}{\tilde \delta'_{\eta f}}\cos {\varphi '_-},\quad {\tilde \delta'_{\eta f}}=\frac{{{E}_{f}}}{{{E}_{i}}}{\delta'_{\eta f}}\approx \left( 1-{x'_{\eta f\left( r \right)}} \right){\delta'_{\eta f}},
\end{equation}

\begin{equation}\label{eq72}
	{{g}_{\eta i\left( l,r \right)}}=g_{\eta i\left( l,r \right)}^{\left( 0 \right)}+\frac{1}{\left( 1+{{\eta }^{2}} \right)}g_{\eta i\left( l,r \right)}^{\left( 1 \right)}+\frac{1}{{{\left( 1+{{\eta }^{2}} \right)}^{2}}}g_{\eta i\left( r \right)}^{\left( 2 \right)},
\end{equation}

\begin{equation}\label{eq73}\begin{split}
	g_{\eta i\left( l,r \right)}^{\left( 0 \right)}=\left( 1-{x'_{\eta i\left( r \right)}} \right)+\frac{1}{\left( 1-{x'_{\eta i\left( r \right)}} \right)}+&\frac{{x'_{\eta i\left( r \right)}}{{\delta }'}^4_{\eta i}}{6}\left[ \frac{1}{{{\left( 1-{x'_{\eta i\left( r \right)}} \right)}^{3}}}-1 \right]+ {} \\ & {} + \frac{{{{{\beta }'}}_{\eta \left( l \right)}}}{\left( n{{n}_{i}} \right)}\left[ {{{{\beta }'}}_{\eta \left( l \right)}}-\frac{{x'_{\eta i\left( r \right)}}\left( 1+{x'_{\eta i\left( r \right)}} \right)}{{{\left( 1-{x'_{\eta i\left( r \right)}} \right)}^{2}}}{{\delta }'}^{2}_{\eta i} \right],
\end{split} \end{equation}

\begin{equation}\label{eq74}\begin{split}
	&g_{\eta i\left( l,r \right)}^{\left( 1 \right)}=\frac{{x'_{\eta i\left( r \right)}}}{\left( 1-{x'_{\eta i\left( r \right)}} \right)}\left[ {{{{\beta }'}}_{\eta \left( l \right)}}+\frac{{{{{x}'}}_{\eta i\left( r \right)}}}{\left( 1-{x'_{\eta i\left( r \right)}} \right)}{{\delta }'}^{2}_{\eta i} \right],{} \\ & {} g_{\eta i\left( r \right)}^{\left( 2 \right)}=\frac{{x'_{\eta i\left( r \right)}}}{2}\left[ 1+\frac{{x'_{\eta i\left( r \right)}}{{\left( 2-{x'_{\eta i\left( r \right)}} \right)}^{2}}-1}{{{\left( 1-{x'_{\eta i\left( r \right)}} \right)}^{3}}} \right],
\end{split} \end{equation}

\begin{equation}\label{eq75}
	\beta'_{\eta \left( l \right)}=\left( \frac{{{\eta }^{2}}}{1+{{\eta }^{2}}} \right)\frac{{x'_{\eta i\left( r \right)}}}{\left( 1-{x'_{\eta i\left( r \right)}} \right)}-\frac{l}{{{r}_{\eta }}}.
\end{equation}

Note that for channel B, the ${{g}_{\eta f\left( l,r \right)}}$ functions are obtained from ${{g}_{\eta i\left( l,r \right)}}$ functions \eqref{eq72}-\eqref{eq75} by replacing: ${x'_{\eta i\left( r \right)}}\to {x'_{\eta f\left( r \right)}},\ {\delta'_{\eta i}}\to {\tilde \delta'_{\eta f}}$. The arguments of the Bessel functions that determine the process of radiation or absorption of  photons of the laser field during the scattering of an intermediate electron on the nucleus for channels A and B have the form:

\begin{equation}\label{eq76}
	{{\alpha }_{\eta i\left( r \right)}}\approx \frac{2{{r}_{\eta }}}{\left( 1-{x'_{i\left( r \right)}} \right)}\frac{\eta }{\sqrt{1+{{\eta }^{2}}}}\sqrt{g_{\eta 0}^{2}},\quad {{\alpha }_{\eta f\left( r \right)}}\approx 2{{r}_{\eta }}\frac{\eta }{\sqrt{1+{{\eta }^{2}}}}\sqrt{g_{\eta 0}^{2}}.
\end{equation}

\begin{figure}[!h]
	\centering
	\subfloat[]
	{\includegraphics[width=0.5\textwidth]{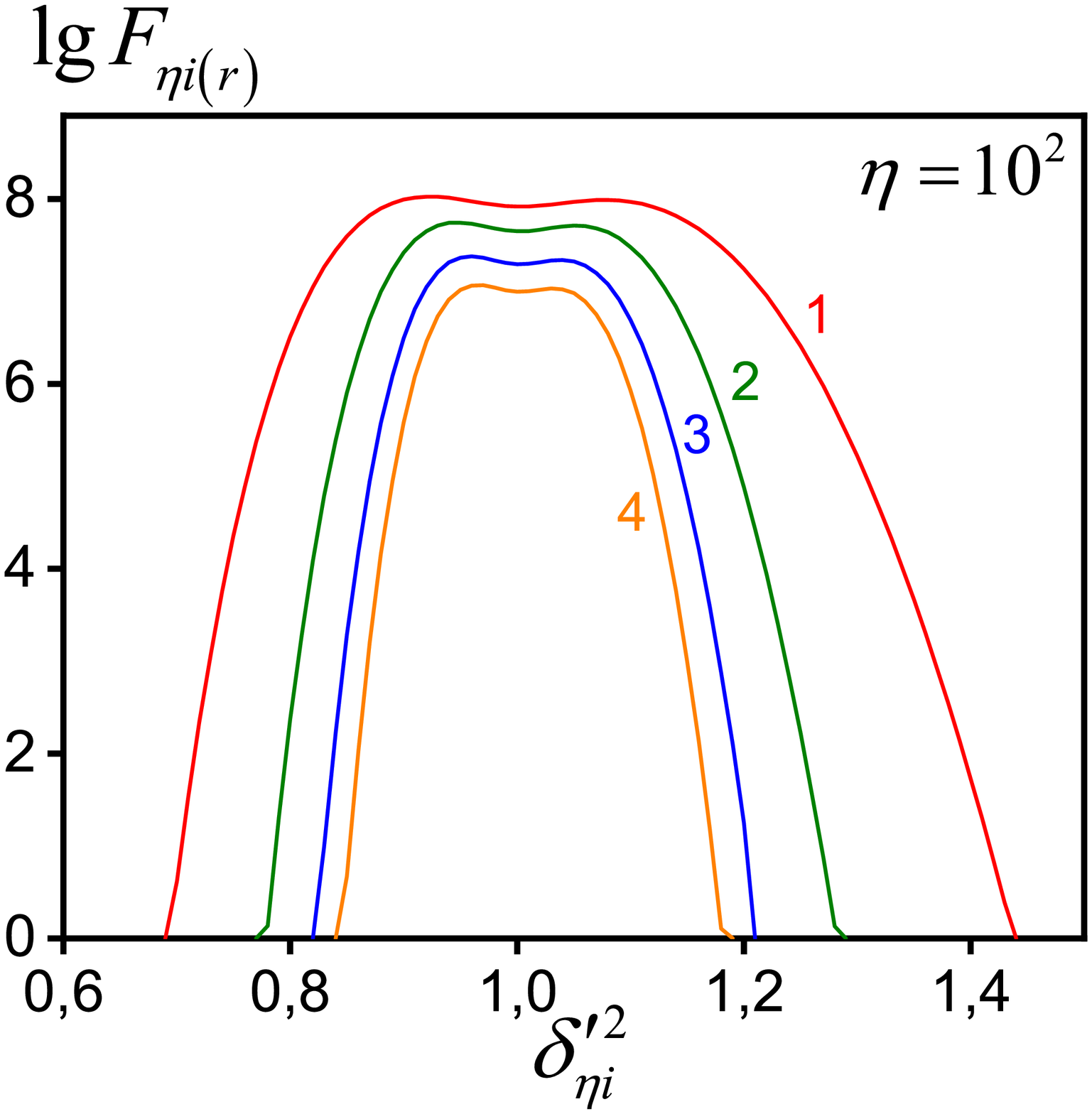}\label{<figure15a>}}
	\subfloat[]
	{\includegraphics[width=0.5\textwidth]{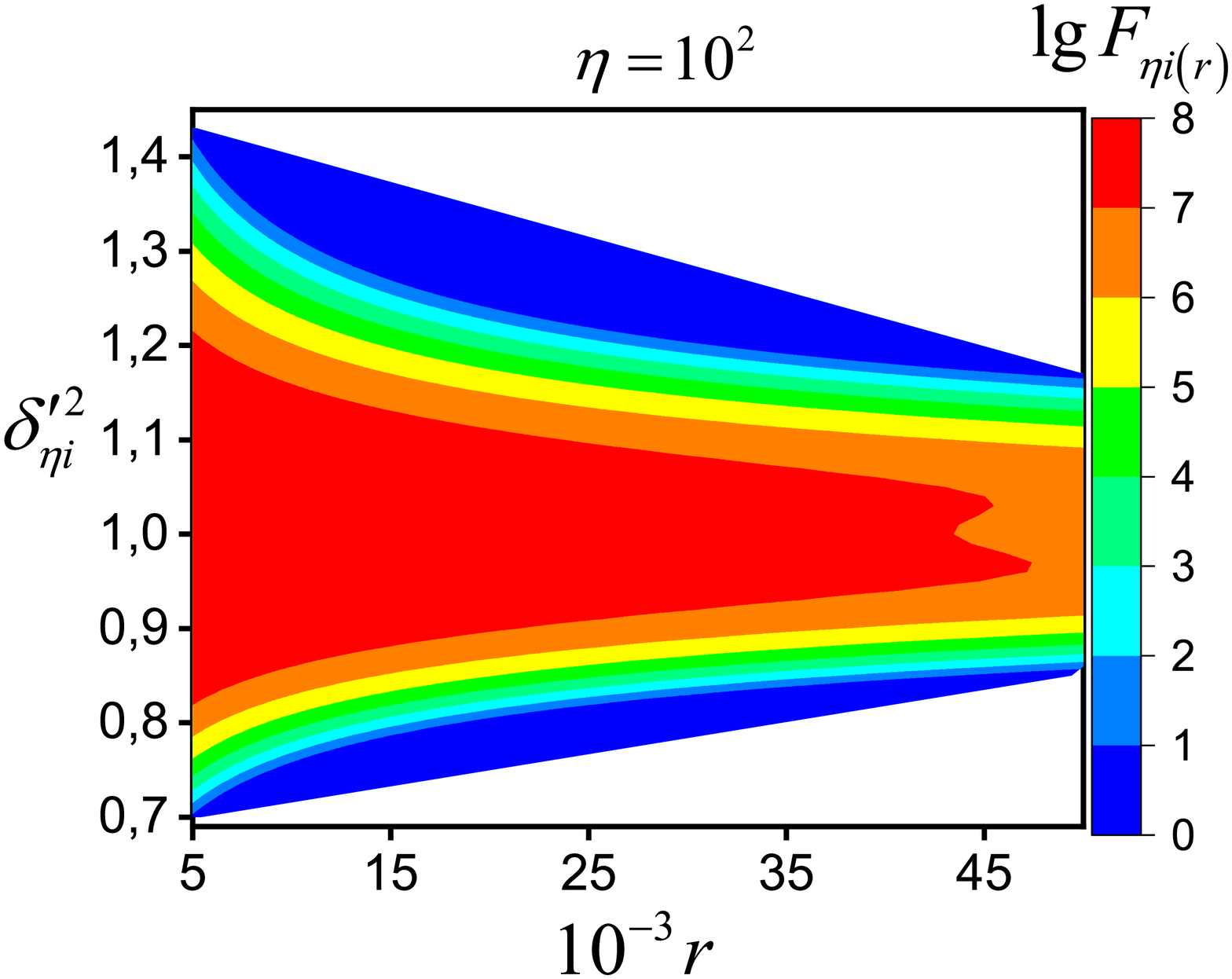}\label{<figure15b>}}
	\caption{The resonant differential cross-section for a channel A $R_{\eta i\left( r \right)}^{\max }$ \eqref{eq87}, \eqref{eq89} (in units of ${{Z}^{2}}\alpha r_{e}^{2}$).  Fig.\ref{<figure15a>} shows the dependence of the function ${{F}_{\eta i\left( r \right)}}$ \eqref{eq89} on the square of the outgoing angle of the spontaneous gamma-quantum relative to the momentum of the initial electron for a fixed number of absorbed photons of the wave: the curve 1 corresponds to $r=5\cdot {{10}^{3}}$, the curve 2 corresponds to $r=1.5\cdot {{10}^{4}}$, the curve 3 corresponds to $r=3\cdot {{10}^{4}}$, the curve 4 corresponds to $r=4.5\cdot {{10}^{4}}$.  Fig.\ref{<figure15b>} shows the dependence of the function ${{F}_{\eta i\left( r \right)}}$ \eqref{eq89} on the square of the outgoing angle of the spontaneous gamma-quantum relative to the momentum of the initial electron and also for different number of absorbed photons of the wave. The energy of the initial electrons ${{E}_{i}}=62.5\ \text{GeV}$, the angle between the momenta of the electrons and the laser wave ${{\theta }_{i}}=\pi $, the frequency of the wave $\omega =1\ \text{eV}$, the intensity of the laser wave $I\approx 1.861\cdot {{10}^{22}}\ \text{Wc}{{\text{m}}^{-2}}$.}
	\label{<figure15>}
\end{figure}

\begin{figure}[!h]
	\centering
	\subfloat[]
	{\includegraphics[width=0.5\textwidth]{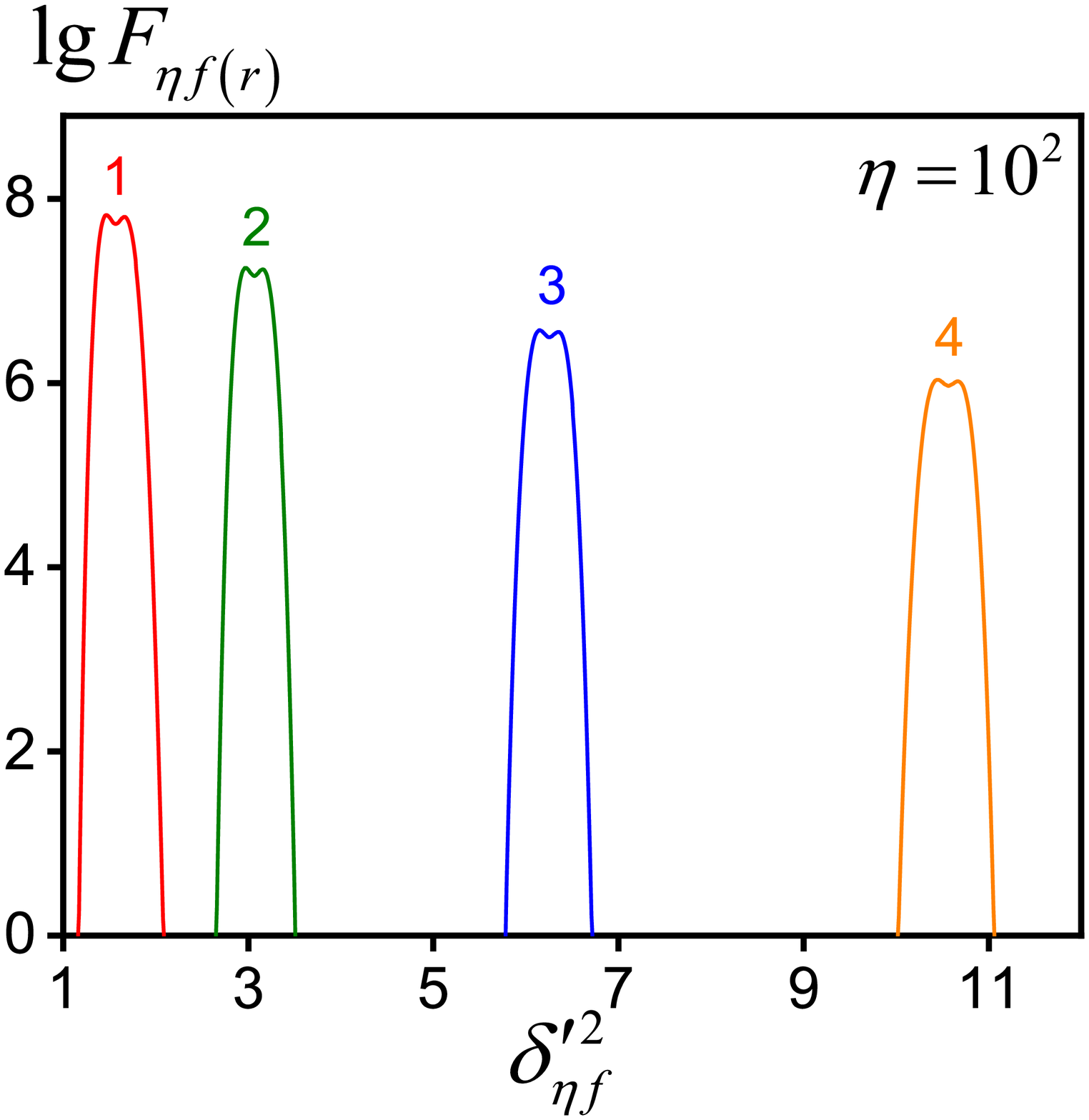}\label{<figure16a>}}
	\subfloat[]
	{\includegraphics[width=0.5\textwidth]{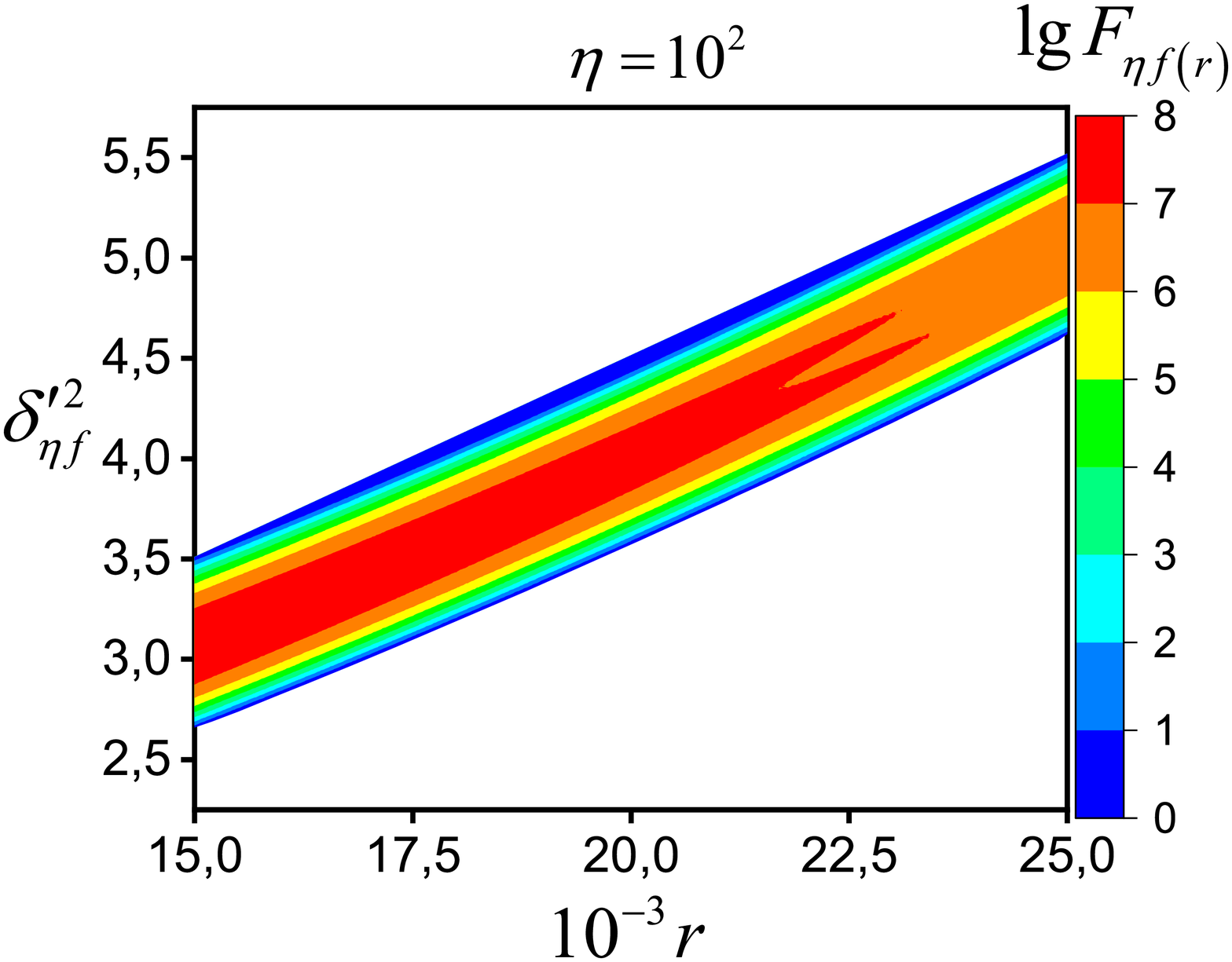}\label{<figure16b>}}
	\caption{The resonant differential cross-section for a channel B $R_{\eta f\left( r \right)}^{\max }$ \eqref{eq88}, \eqref{eq90} (in units of ${{Z}^{2}}\alpha r_{e}^{2}$).  Fig.\ref{<figure16a>} shows the dependence of the function ${{F}_{\eta f\left( r \right)}}$ \eqref{eq90} on the square of the outgoing angle of the spontaneous gamma-quantum relative to the momentum of the final electron for a fixed number of absorbed photons of the wave: the curve 1 corresponds to $r=5\cdot {{10}^{3}}$, the curve 2 corresponds to $r=1.5\cdot {{10}^{4}}$, the curve 3 corresponds to $r=3\cdot {{10}^{4}}$, the curve 4 corresponds to $r=4.5\cdot {{10}^{4}}$.  Fig.\ref{<figure16b>} shows the dependence of the function ${{F}_{\eta f\left( r \right)}}$ \eqref{eq90} on the square of the outgoing angle of the spontaneous gamma-quantum relative to the momentum of the final electron and also for different number of absorbed photons of the wave. The energy of the initial electrons ${{E}_{i}}=62.5\ \text{GeV}$, the angle between the momenta of the electrons and the laser wave ${{\theta }_{i}}=\pi $, the frequency of the wave $\omega =1\ \text{eV}$, the intensity of the laser wave $I\approx 1.861\cdot {{10}^{22}}\ \text{Wc}{{\text{m}}^{-2}}$.}
	\label{<figure16>}
\end{figure}

\begin{figure}[!h]
	\centering
	\subfloat[]
	{\includegraphics[width=0.5\textwidth]{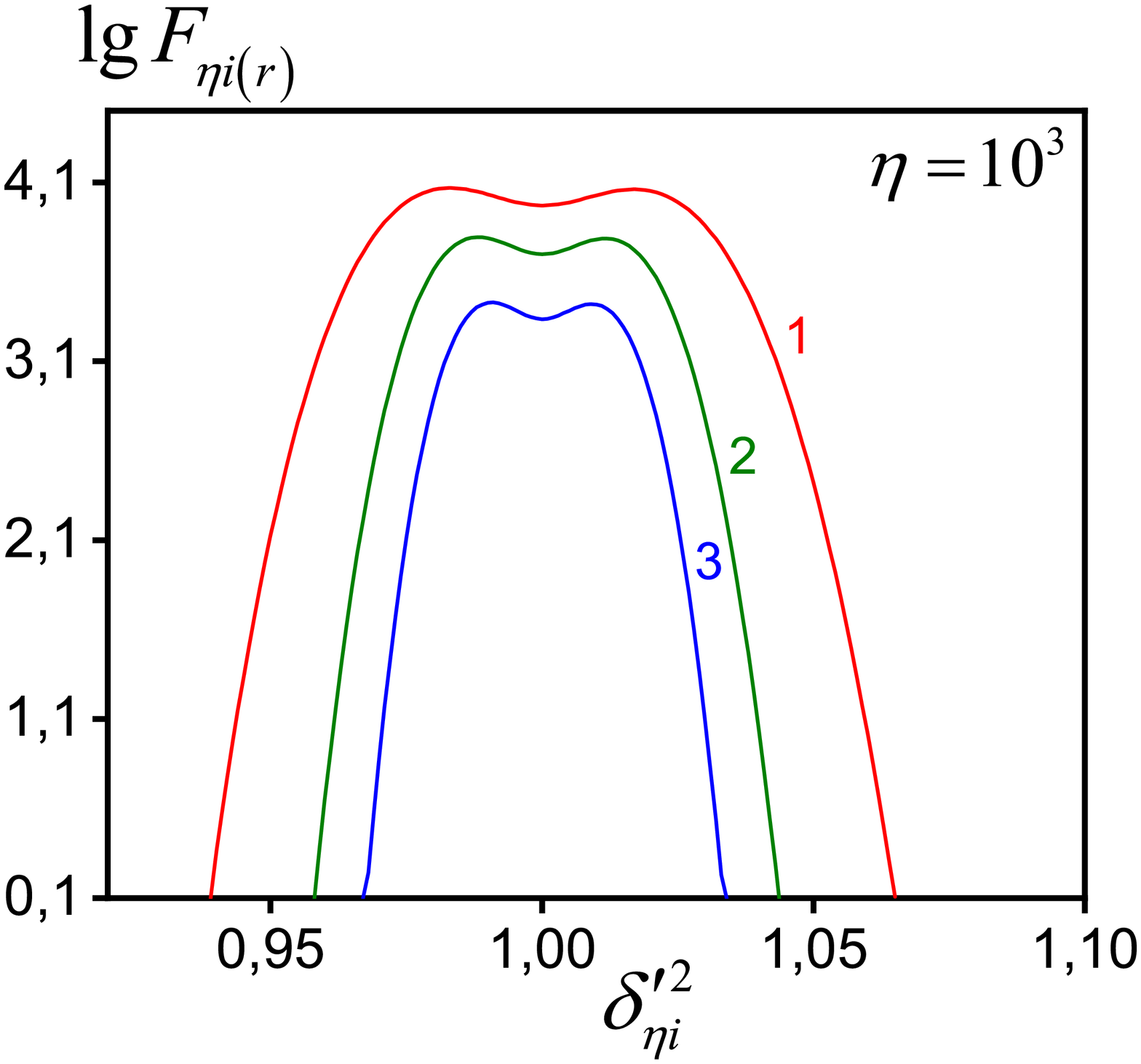}\label{<figure17a>}}
	\subfloat[]
	{\includegraphics[width=0.5\textwidth]{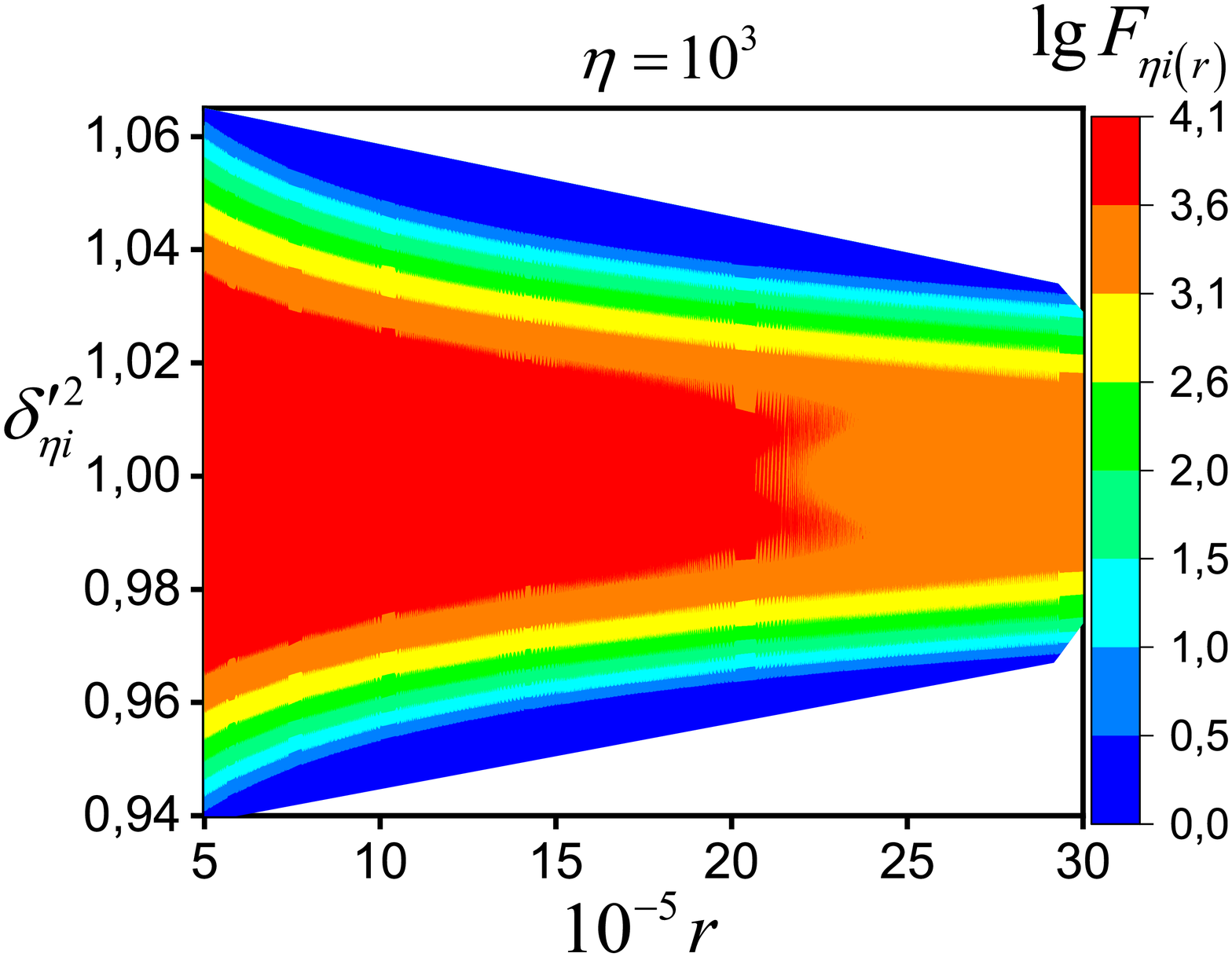}\label{<figure17b>}}
	\caption{The resonant differential cross-section for a channel A $R_{\eta i\left( r \right)}^{\max }$ \eqref{eq87}, \eqref{eq89} (in units of ${{Z}^{2}}\alpha r_{e}^{2}$).  Fig.\ref{<figure17a>} shows the dependence of the function ${{F}_{\eta i\left( r \right)}}$ \eqref{eq89} on the square of the outgoing angle of the spontaneous gamma-quantum relative to the momentum of the initial electron for a fixed number of absorbed photons of the wave: the curve 1 corresponds to $r=5\cdot {{10}^{5}}$, the curve 2 corresponds to $r=1.5\cdot {{10}^{6}}$, the curve 3 corresponds to $r=3\cdot {{10}^{6}}$. Fig.\ref{<figure17b>} shows the dependence of the function ${{F}_{\eta i\left( r \right)}}$ \eqref{eq89} on the square of the outgoing angle of the spontaneous gamma-quantum relative to the momentum of the initial electron and also for different number of absorbed photons of the wave. The energy of the initial electrons ${{E}_{i}}=62.5\ \text{GeV}$, the angle between the momenta of the electrons and the laser wave ${{\theta }_{i}}=\pi $, the frequency of the wave $\omega =1\ \text{eV}$, the intensity of the laser wave $I\approx 1.861\cdot {{10}^{24}}\ \text{Wc}{{\text{m}}^{-2}}$.}
	\label{<figure17>}
\end{figure}

\begin{figure}[!h]
	\centering
	\subfloat[]
	{\includegraphics[width=0.5\textwidth]{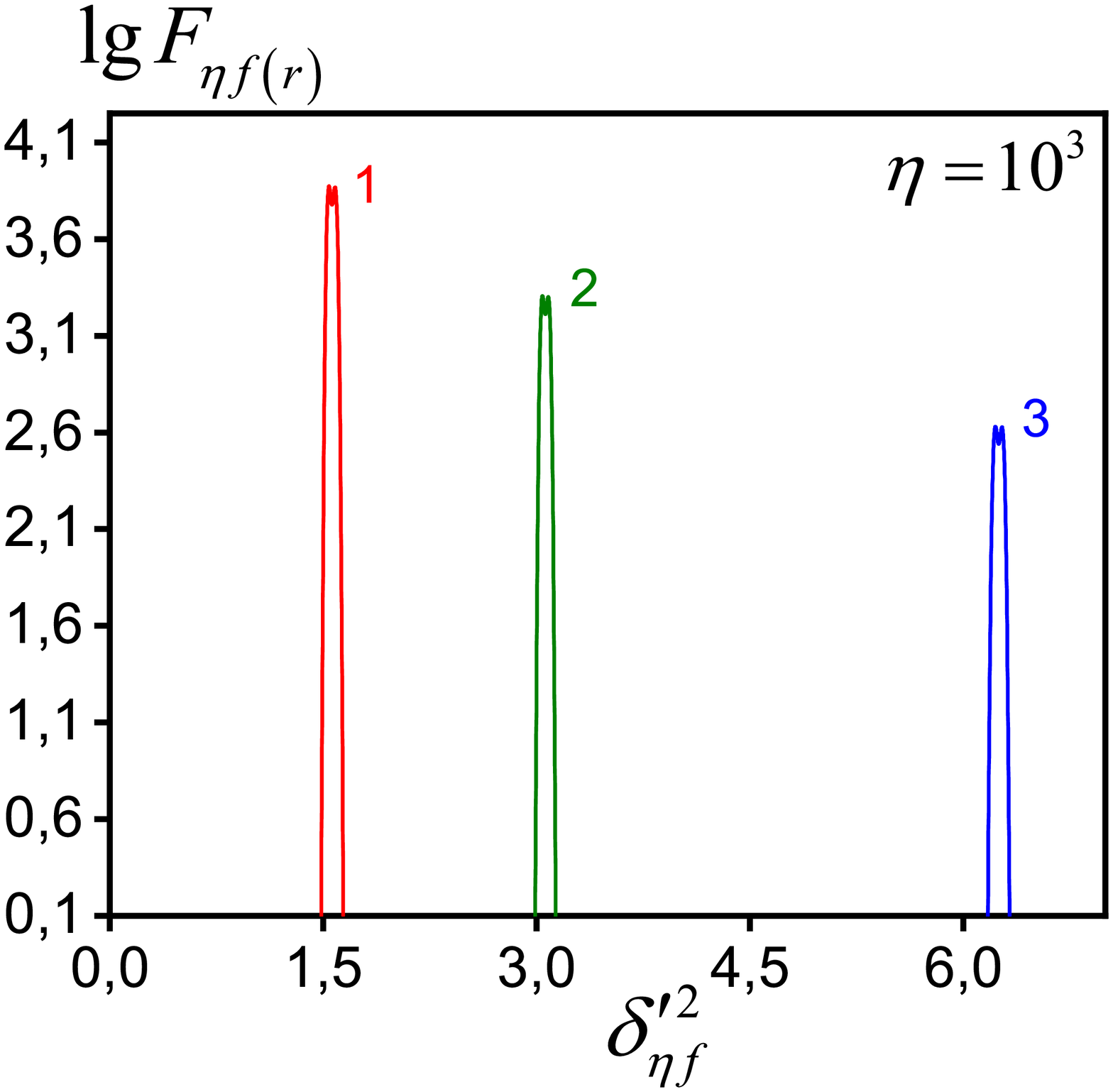}\label{<figure18a>}}
	\subfloat[]
	{\includegraphics[width=0.5\textwidth]{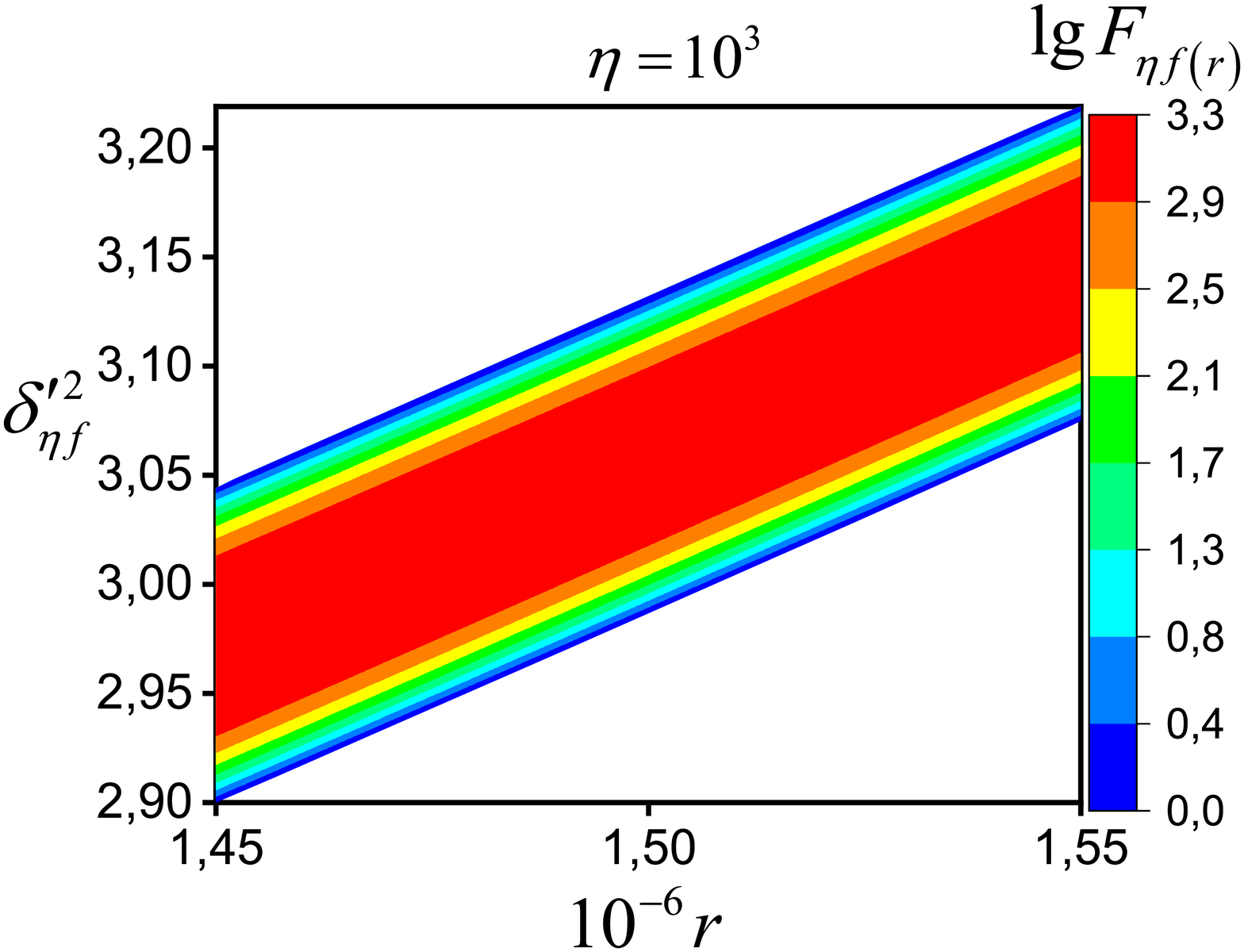}\label{<figure18b>}}
	\caption{The resonant differential cross-section for a channel B $R_{\eta f\left( r \right)}^{\max }$ \eqref{eq88}, \eqref{eq90} (in units of ${{Z}^{2}}\alpha r_{e}^{2}$).  Fig.\ref{<figure18a>} shows the dependence of the function ${{F}_{\eta f\left( r \right)}}$ \eqref{eq90} on the square of the outgoing angle of the spontaneous gamma-quantum relative to the momentum of the final electron for a fixed number of absorbed photons of the wave: the curve 1 corresponds to $r=5\cdot {{10}^{5}}$, the curve 2 corresponds to $r=1.5\cdot {{10}^{6}}$, the curve 3 corresponds to $r=3\cdot {{10}^{6}}$. Fig.\ref{<figure18b>} shows the dependence of the function ${{F}_{\eta f\left( r \right)}}$ \eqref{eq90} on the square of the outgoing angle of the spontaneous gamma-quantum relative to the momentum of the final electron and also for different number of absorbed photons of the wave. The energy of the initial electrons ${{E}_{i}}=62.5\ \text{GeV}$, the angle between the momenta of the electrons and the laser wave ${{\theta }_{i}}=\pi $, the frequency of the wave $\omega =1\ \text{eV}$, the intensity of the laser wave $I\approx 1.861\cdot {{10}^{24}}\ \text{Wc}{{\text{m}}^{-2}}$.}
	\label{<figure18>}
\end{figure}

It is important to emphasize that in the differential cross-sections \eqref{eq66} and \eqref{eq67}, small corrections proportional to the ${{\left( {{{m}_{*}}}/{{{E}_{i}}}\; \right)}^{2}}<<1$ value are introduced in the denominators . Note that these corrections make a dominant contribution to the corresponding differential cross-sections under the conditions
\begin{equation}\label{eq77}
	{{\varphi }_{-}}\lesssim \frac{{{m}_{*}}}{{{E}_{i}}},\quad \left| {\delta'_{\eta i}}-{{\tilde \delta '}_{\eta f}} \right|\lesssim \frac{{{m}_{*}}}{{{E}_{i}}}<<1.
\end{equation}
In this case, the function $g_{\eta 0}^{2}\to 0$ and the corresponding differential cross-sections \eqref{eq66} and \eqref{eq67} have sharp peaks. Note that the nature of these peaks is due to the small transmitted momenta in the ultrarelativistic limit of electron energies (see \cite{54}). The resonant cross-sections can be integrated near these peaks \eqref{eq77} by the saddle points method. So, for channel A, the differential cross-section \eqref{eq66} can be represented as:
\begin{equation}\label{eq78}
	\frac{d{{\sigma }_{\eta i\left( l,r \right)}}}{d{x'_{\eta i\left( r \right)}}d{{\delta }'}^{2}_{\eta i}}=\frac{{{\left( 2\pi  \right)}^{2}}{{Z}^{2}}\alpha r_{e}^{2}\left( 1-{x'_{\eta i\left( r \right)}} \right)K\left( {{u}_{\eta i\left( r \right)}},{{\varepsilon }_{\eta \left( r \right)}} \right)}{\left[ {{\left( 1+{{\eta }^{2}} \right)}^{2}}{{\left( {{\delta }'}^{2}_{\eta i}-{{\delta }'}^{2}_{\eta i\left( r \right)} \right)}^{2}}+\Upsilon _{\eta i\left( r \right)}^{2} \right]{x'_{\eta i\left( r \right)}}}{{C}_{\eta i\left( l,r \right)}},
\end{equation}
where
\begin{equation}\label{eq79}
	{{C}_{\eta i\left( l,r \right)}}=\int\limits_{0}^{\pi }{d{\varphi '_-}}\int\limits_{0}^{\infty }{J_{r-l}^{2}\left( {{\alpha }_{\eta i\left( r \right)}} \right){{\left[ g_{\eta 0}^{2}+\frac{m_{*}^{2}}{2E_{i}^{2}}{{g}_{\eta i\left( l,r \right)}} \right]}^{-2}}}d\tilde{{\delta }'}^{2}_{\eta f}.
\end{equation}
The integral in Exp. \eqref{eq78} has a sharp maximum in the area of angles \eqref{eq77}. At the same time, the function $J_{r-l}^{2}\left( {{\alpha }_{\eta i\left( r \right)}} \right)\lesssim 1$ and we are allowed to take it out of the integral. As a result, we get:
\begin{equation}\label{eq80}
	C_{\eta i\left( l,r \right)}\approx \left. J_{r-l}^{2}\left( {\alpha }_{\eta i\left( r \right)} \right) \right|_{\begin{smallmatrix} 
				{\varphi'_-}=0, \\ 
				{{\tilde \delta '}_{f}}={\delta '_{i}} 
	\end{smallmatrix}}\cdot \int\limits_{0}^{\pi }{d{\varphi '_-}}\int\limits_{0}^{\infty }{d\tilde{{\delta }'}^{2}_{\eta f}}\exp \left[ {{f}_{\eta }}\left( {\varphi '_-},{\tilde \delta '_{\eta f}} \right) \right],
\end{equation}
where
\begin{equation}\label{eq81}
	{{f}_{\eta }}\left( {\varphi '_{-}},{{\tilde \delta '}_{\eta f}} \right)=-2\ln \left( g_{\eta 0}^{2}+{{a}_{\eta i}} \right),\quad {{a}_{\eta i}}=\frac{m_{*}^{2}}{2E_{i}^{2}}{{g}_{\eta i\left( l,r \right)}}.
\end{equation}
We expense the ${{f}_{\eta }}$ function \eqref{eq81} into a Taylor series near the points ${\varphi '_-}=0,\ \ {\tilde \delta'_f}={\delta'_i}$:
\begin{equation}\label{eq82}
	{{f}_{\eta }}\left( {\varphi '_-},{\tilde \delta '_{\eta f}} \right)\approx -2\ln \left( {{a}_{\eta i}} \right)-\frac{2{{\delta }'}^{2}_{\eta i}}{\left( {{a}_{\eta i}} \right)}\cdot {\varphi '^2_-}-\frac{1}{2{{a}_{\eta i}}{{\delta }'}^{2}_{\eta i}}\cdot {{\left( \tilde{{\delta }'}^{2}_{f}-{{\delta }'}^{2}_{i} \right)}^{2}}.
\end{equation}
Substituting Exp. \eqref{eq82} into Eq. \eqref{eq80} and performing simple integrations, we finally get:
\begin{equation}\label{eq83}
	{C_{\eta i\left( l,r \right)}}\approx \frac{\pi }{{{g}_{\eta i\left( l,r \right)}}}\left( \frac{E_{i}^{2}}{m_{*}^{2}} \right)J_{r-l}^{2}\left( 0 \right)=\frac{\pi }{{{g}_{\eta i\left( l,r \right)}}}\left( \frac{E_{i}^{2}}{m_{*}^{2}} \right)\quad \left( l=r \right).
\end{equation}

Here it is taken into account that $J_{r-l}^{2}\left( 0 \right)=1$ at $l=r$. This means that the process of scattering of the intermediate electron on the nucleus, basically, takes place without the absorption or emission of photons of the wave. Similar calculations can be made for channel B by performing integrations in the area \eqref{eq77}  by $d{{\varphi }_{-}}$ and $d{{\delta }'}^{2}_{\eta i}$. Finally, the resonant cross-sections for channels A and B will take the form:
\begin{equation}\label{eq84}
	\frac{d{{\sigma }_{\eta i\left( r \right)}}}{d{x'_{\eta i\left( r \right)}}d{{\delta }'}^{2}_{\eta i}}=\left( {{Z}^{2}}\alpha r_{e}^{2} \right)\frac{E_{i}^{2}}{m_{*}^{2}}\frac{4{{\pi }^{3}}\left( 1-{x'_{\eta i\left( r \right)}} \right)}{{{g}_{\eta i\left( r \right)}}{x'_{\eta i\left( r \right)}}\left[ {{\left( 1+{{\eta }^{2}} \right)}^{2}}{{\left( {{\delta }'}^{2}_{\eta i}-{{\delta }'}^{2}_{\eta i\left( r \right)} \right)}^{2}}+\Upsilon _{\eta i\left( r \right)}^{2} \right]}K\left( {{u}_{\eta i\left( r \right)}},\frac{r}{{{r}_{\eta }}} \right),
\end{equation}

\begin{equation}\label{eq85}
\frac{d{{\sigma }_{\eta f\left( r \right)}}}{d{x'_{\eta f\left( r \right)}}d{{\delta }'}^{2}_{\eta f}}=\left( {{Z}^{2}}\alpha r_{e}^{2} \right)\frac{E_{i}^{2}}{m_{*}^{2}}\frac{4{{\pi }^{3}}{{\left( 1-{x'_{\eta f\left( r \right)}} \right)}^{-1}}}{{{g}_{\eta f\left( r \right)}}{x'_{\eta f\left( r \right)}}\left[ {{\left( 1+{{\eta }^{2}} \right)}^{2}}{{\left( {{\delta }'}^{2}_{\eta f}-{{\delta }'}^{2}_{\eta f\left( r \right)} \right)}^{2}}+\Upsilon _{\eta f\left( r \right)}^{2} \right]}K\left( {{u}_{\eta f\left( r \right)}},\frac{r}{{{r}_{\eta }}} \right).
\end{equation}
Here, the ${{g}_{\eta i\left( r \right)}}$ and ${{g}_{\eta f\left( r \right)}}$ functions are obtained from the ${{g}_{\eta i\left( l,r \right)}}$ and ${{g}_{\eta f\left( r \right)}}$ functions \eqref{eq72}-\eqref{eq75}, respectively, under the condition $l=r$. When the conditions are met
\begin{equation}\label{eq86}
\left( {{\delta }'}^{2}_{\eta i}-{{\delta }'}^{2}_{\eta i\left( r \right)} \right)^{2}<<\frac{\Upsilon _{\eta i\left( r \right)}^{2}}{\left(1+{\eta }^{2} \right)^{2}},\quad \left( {{\delta }'}^{2}_{\eta f}-{{\delta }'}^{2}_{\eta f\left( r \right)} \right)^{2}<<\frac{\Upsilon _{\eta f\left( r \right)}^{2}}{\left( 1+{{\eta }^{2}} \right)^{2}}
\end{equation}
we obtain the maximum resonant differential cross-sections for channels A and B:
\begin{equation}\label{eq87}
R_{\eta i\left( r \right)}^{\max }=\frac{d\sigma _{\eta i\left( r \right)}^{\max }}{d{{{{x}'}}_{\eta i\left( r \right)}}d{{\delta }'}^{2}_{\eta i}}=\left( {{Z}^{2}}\alpha r_{e}^{2} \right){{F}_{\eta i\left( r \right)}},
\end{equation}

\begin{equation}\label{eq88}
	R_{\eta f\left( r \right)}^{\max }=\frac{d\sigma _{\eta f\left( r \right)}^{\max }}{d{x'_{\eta f\left( r \right)}}d{{\delta }'}^{2}_{\eta f}}=\left( {{Z}^{2}}\alpha r_{e}^{2} \right){{F}_{\eta f\left( r \right)}}.
\end{equation}
Here, the ${{F}_{\eta i\left( r \right)}}$ and ${{F}_{\eta f\left( r \right)}}$ functions determine the spectral-angular distribution of the resonant SB cross-section for channels A and B:
\begin{equation}\label{eq89}
{{F}_{\eta i\left( r \right)}}={{b}_{\eta i}}\frac{{x'_{\eta i\left( r \right)}}}{\left( 1-{x'_{\eta i\left( r \right)}} \right){{g}_{\eta i\left( r \right)}}{{\mathbf{K} }^{2}}\left( {{r}_{\eta }} \right)}K\left( {{u}_{\eta i\left( r \right)}},\frac{r}{{{r}_{\eta }}} \right),
\end{equation}

\begin{equation}\label{eq90}
{{F}_{\eta f\left( r \right)}}={{b}_{\eta i}}\frac{{x'_{\eta f\left( r \right)}}\left( 1-{x'_{\eta f\left( r \right)}} \right)}{{{g}_{\eta f\left( r \right)}}{{\mathbf{K} }^{2}}\left( {{r}_{\eta }} \right)}K\left( {{u}_{\eta f\left( r \right)}},\frac{r}{{{r}_{\eta }}} \right),
\end{equation}
where
\begin{equation}\label{eq91}
	{{b}_{\eta i}}=\pi {{\left( \frac{8\pi {{E}_{i}}}{\alpha {{m}_{*}}} \right)}^{2}}.
\end{equation}

Figures \ref{<figure11>}-\ref{<figure18>} show the resonant differential cross sections (for channels A (89) and B (90)) as the functions of the square of its outgoing angle and a number of absorbed wave photons at different laser wave intensities from $I\approx 1.861\cdot {{10}^{18}}\ \text{Wc}{{\text{m}}^{-2}}$ to $I\approx 1.861\cdot {{10}^{24}}\ \text{Wc}{{\text{m}}^{-2}}$. 

It should be noted that the graphs of the resonant differential cross sections for each laser wave intensity are given for the same values of the spontaneous photon outgoing angles and the number of absorbed photons of the wave as for the resonant frequencies (see Figs.\ref{<figure3>}-\ref{<figure10>}).

 It can be seen from Fig.\ref{<figure11>} (channel A) and Fig.\ref{<figure12>} (channel B) that in the region of medium fields $\left( \eta =1 \right)$, the resonant differential cross-sections with simultaneous registration of the frequency and outgoing angle of a spontaneous gamma quantum have a maximum value when one laser photon is absorbed and is equal to the value $F_{\eta i\left( 1 \right)}^{\max }\approx {{10}^{17.53}}$, $F_{\eta f\left( 1 \right)}^{\max }\approx {{10}^{17.19}}$ at zero outgoing angle $\left({\delta'}^{2}_{\eta i,\eta f}\approx 0 \right)$. For the number of absorbed laser photons $r\ge 2$, maxima appear in the angular distribution of resonant differential cross sections, which shift to the region of large outgoing angles with an increase in the number of absorbed photons. At the same time, the value of the resonant differential cross-section decreases (see curves 2, 3, 4 in Fig.\ref{<figure11a>} and Fig.\ref{<figure12a>}, as well as Fig.\ref{<figure11b>} and Fig.\ref{<figure12b>}). So, for channel A, in the maximum of distributions ${{F}_{\eta i\left( r \right)}}$, we have: $F_{\eta i\left( 2 \right)}^{\max }\approx {{10}^{16.94}}$ at ${{\delta }'}^{2}_{\eta i}\approx 0.19$, $F_{\eta i\left( 5 \right)}^{\max }\approx {{10}^{16.09}}$ at ${{\delta }'}^{2}_{\eta i}\approx 0.44$, $F_{\eta i\left( 10 \right)}^{\max }\approx {{10}^{14.86}}$ at ${{\delta }'}^{2}_{\eta i}\approx 0.64$. At the same time, for channel B, the corresponding values of the maximum differential cross-section ${{F}_{\eta f\left( r \right)}}$ are less than for channel A. Indeed, $F_{\eta f\left( 2 \right)}^{\max }\approx {{10}^{16.42}}$ at ${{\delta }'}^{2}_{\eta i}\approx 0.19$, $F_{\eta i\left( 5 \right)}^{\max }\approx {{10}^{16.09}}$ at ${{\delta }'}^{2}_{\eta i}\approx 0.44$, $F_{\eta i\left( 10 \right)}^{\max }\approx {{10}^{14.86}}$ at ${{\delta }'}^{2}_{\eta i}\approx 0.64$. See also tables \ref{tab1} - table \ref{tab4}.

\begin{table}[!h]
\begin{center}
	\caption{The values of the resonant frequency and the square of the outgoing angle of the spontaneous gamma-quantum for the maximum value of the resonant differential cross-section (see Fig.\ref{<figure3a>}, Fig.\ref{<figure11a>} and Fig.\ref{<figure4a>}, Fig,\ref{<figure12a>}). The intensity of the laser wave and the energy of the initial electrons are $I\approx 1.861\cdot {10}^{18} \text{W}{{\text{cm}}^{-2}}$ and ${{E}_{i}}=62.5\ \text{GeV}$.}
	\label{tab1}
	\setlength{\extrarowheight}{2mm} 
	\begin{tabular}{|c|c|c|c|c|c|c|}
		\hline
		$r$ & ${\delta}^{2} _{\eta i}$ & ${{{\omega }'}_{\eta i\left( r \right)}},\ \text{GeV}$ & $F_{\eta i\left( r \right)}^{\max }$ & ${\delta}^{2} _{\eta f}$ & ${{{\omega }'}_{\eta f\left( r \right)}},\ \text{GeV}$ & $F_{\eta f\left( r \right)}^{\max }$\\[2mm]
		\hline
		1 & 0.02 & 20.558 & ${{10}^{17.52}}$  & 0.02 & 20.709 & ${{10}^{17.19}}$\\ \hline
		2 & 0.19 & 28.538 & ${{10}^{16.94}}$  & 0.79 & 27.870 & ${{10}^{16.42}}$\\ \hline
		5 & 0.44 & 39.656 & ${{10}^{16.09}}$  & 3.91 & 38.186 & ${{10}^{15.24}}$\\ \hline
		10 & 0.64 & 47.063 & ${{10}^{14.85}}$  & 11.46 & 45.896 & ${{10}^{13.67}}$\\ \hline
	\end{tabular}
\setlength{\extrarowheight}{0mm} 
\end{center}
\end{table}

\begin{table}[!h]
	\begin{center}
				\caption{The values of the resonant frequency and the square of the outgoing angle of the spontaneous gamma-quantum for the maximum value of the resonant differential cross-section (see Fig.\ref{<figure5a>}, Fig.\ref{<figure13a>} and Fig.\ref{<figure6a>}, Fig,\ref{<figure14a>}). The intensity of the laser wave and the energy of the initial electrons are $I\approx 1.861\cdot {10}^{20} \text{W}{{\text{cm}}^{-2}}$ and ${{E}_{i}}=62.5\ \text{GeV}$.}
		\label{tab2}
		\setlength{\extrarowheight}{2mm} 
		\begin{tabular}{|c|c|c|c|c|c|c|}
			\hline
			$r$ & ${\delta}^{2} _{\eta i}$ & ${{{\omega }'}_{\eta i\left( r \right)}},\ \text{GeV}$ & $F_{\eta i\left( r \right)}^{\max }$ & ${\delta}^{2} _{\eta f}$ & ${{{\omega }'}_{\eta f\left( r \right)}},\ \text{GeV}$ & $F_{\eta f\left( r \right)}^{\max }$\\[2mm]
			\hline
			150 & 0.78 & 28.427 & ${{10}^{12.09}}$  & 2.66 & 28.275 & ${{10}^{11.56}}$\\ \hline
			300 & 0.84 & 38.592 & ${{10}^{11.71}}$  & 5.80 & 38.432 & ${{10}^{10.87}}$\\ \hline
			600 & 0.90 & 47.354 & ${{10}^{11.11}}$  & 15.47 & 47.168 & ${{10}^{9.88}}$\\ \hline
			800 & 0.93 & 50.264 & ${{10}^{10.81}}$  & 24.39 & 50.108 & ${{10}^{9.40}}$\\ \hline
		\end{tabular}
		\setlength{\extrarowheight}{0mm} 
	\end{center}
\end{table}

\begin{table}[!h]
	\begin{center}
				\caption{The values of the resonant frequency and the square of the outgoing angle of the spontaneous gamma-quantum for the maximum value of the resonant differential cross-section (see Fig.\ref{<figure7a>}, Fig.\ref{<figure15a>} and Fig.\ref{<figure8a>}, Fig,\ref{<figure16a>}). The intensity of the laser wave and the energy of the initial electrons are $I\approx 1.861\cdot {10}^{22} \text{W}{{\text{cm}}^{-2}}$ and ${{E}_{i}}=62.5\ \text{GeV}$.}
		\label{tab3}
		\setlength{\extrarowheight}{2mm} 
		\begin{tabular}{|c|c|c|c|c|c|c|}
			\hline
			$r$ & ${\delta}^{2} _{\eta i}$ & ${{{\omega }'}_{\eta i\left( r \right)}},\ \text{GeV}$ & $F_{\eta i\left( r \right)}^{\max }$ & ${\delta}^{2} _{\eta f}$ & ${{{\omega }'}_{\eta f\left( r \right)}},\ \text{GeV}$ & $F_{\eta f\left( r \right)}^{\max }$\\[2mm]
			\hline
\multirow{3}{*}{$5\cdot 10^3$}  & 0.92300 & 12.911 & $10^{8.12}$ & 1.46 & 12.876 & $10^{7.92}$ \\ \cline{2-7}
						    	& 1.00326 & 12.498 & $10^{8.02}$ & 1.56 & 12.468 & $10^{7.82}$  \\ \cline{2-7}
								& 1.07758 & 12.111 & $10^{8.09}$ & 1.66 & 12.078 & $10^{7.90}$  \\ \hline
								
\multirow{3}{*}{$1.5\cdot 10^4$}& 0.94722 & 27.171 & $10^{7.84}$ & 2.97 & 27.186 & $10^{7.35}$ \\ \cline{2-7}
								& 1.00239 & 26.783 & $10^{7.75}$ & 3.06 & 26.776 & $10^{7.26}$  \\ \cline{2-7}
								& 1.05171 & 26.406 & $10^{7.81}$ & 3.15 & 26.386 & $10^{7.33}$  \\ \hline
								
\multirow{3}{*}{$3\cdot 10^4$}  & 0.95901 & 37.800 & $10^{7.47}$ & 6.15 & 37.800 & $10^{6.67}$ \\ \cline{2-7}
								& 1.00245 & 37.498 & $10^{7.39}$ & 6.25 & 37.490 & $10^{6.59}$  \\ \cline{2-7}
								& 1.03887 & 37.200 & $10^{7.44}$ & 6.35 & 37.201 & $10^{6.65}$  \\ \hline
								
\multirow{3}{*}{$4.5\cdot 10^4$}& 0.96520 & 43.468 & $10^{7.16}$ & 10.45 & 43.494 & $10^{6.13}$ \\ \cline{2-7}
								& 1.00247 & 43.267 & $10^{7.09}$ & 10.56 & 43.261 & $10^{6.07}$  \\ \cline{2-7}
								& 1.03235 & 43.068 & $10^{7.13}$ & 10.67 & 43.046 & $10^{6.12}$  \\ \hline
		\end{tabular}
		\setlength{\extrarowheight}{0mm} 
	\end{center}
\end{table}

\begin{table}[!h]
	\begin{center}
				\caption{The values of the resonant frequency and the square of the outgoing angle of the spontaneous gamma-quantum for the maximum value of the resonant differential cross-section (see Fig.\ref{<figure9a>}, Fig.\ref{<figure17a>} and Fig.\ref{<figure10a>}, Fig.\ref{<figure18a>}). The intensity of the laser wave and the energy of the initial electrons are $I\approx 1.861\cdot {10}^{24} \text{W}{{\text{cm}}^{-2}}$ and ${{E}_{i}}=62.5\ \text{GeV}$.}
		\label{tab4}
		\setlength{\extrarowheight}{2mm} 
		\begin{tabular}{|c|c|c|c|c|c|c|}
			\hline
			$r$ & ${\delta}^{2} _{\eta i}$ & ${{{\omega }'}_{\eta i\left( r \right)}},\ \text{GeV}$ & $F_{\eta i\left( r \right)}^{\max }$ & ${\delta}^{2} _{\eta f}$ & ${{{\omega }'}_{\eta f\left( r \right)}},\ \text{GeV}$ & $F_{\eta f\left( r \right)}^{\max }$\\[2mm]
			\hline
\multirow{3}{*}{$5\cdot 10^5$}  & 0.98310 & 12.585 & $10^{4.068}$ & 1.54 & 12.586 & $10^{3.872}$ \\ \cline{2-7}
								& 1.00014 & 12.499 & $10^{3.971}$ & 1.56 & 12.497 & $10^{3.780}$  \\ \cline{2-7}
								& 1.01694 & 12.415 & $10^{4.059}$ & 1.58 & 12.414 & $10^{3.861}$  \\ \hline
			
\multirow{3}{*}{$1.5\cdot 10^6$}& 0.98831 & 26.877 & $10^{3.793}$ & 3.04 & 26.875 & $10^{3.305}$ \\ \cline{2-7}
								& 1.00010 & 26.785 & $10^{3.698}$ & 3.06 & 26.783 & $10^{3.212}$  \\ \cline{2-7}
								& 1.01166 & 26.693 & $10^{3.785}$ & 3.08 & 26.695 & $10^{3.302}$  \\ \hline
			
\multirow{3}{*}{$3\cdot 10^6$}  & 0.99077 & 37.567 & $10^{3.426}$ & 6.23 & 37.568 & $10^{2.630}$ \\ \cline{2-7}
								& 1.00009 & 37.499 & $10^{3.335}$ & 6.25 & 37.499 & $10^{2.540}$  \\ \cline{2-7}
								& 1.00916 & 37.432 & $10^{3.420}$ & 6.27 & 37.430 & $10^{2.626}$  \\ \hline
		\end{tabular}
		\setlength{\extrarowheight}{0mm} 
	\end{center}
\end{table}

\newpage

\section{Conclusion}

We considered Oleinik resonances for resonant spontaneous bremsstrahlung of ultrarelativistic electrons with energies less than or on the order of $100\ \text{GeV}$ in the nucleus field and a strong light field up to the intensities $I\sim {{10}^{24}}\ \text{Wc}{{\text{m}}^{\text{-2}}}$.       
           
•	All the particles (initial and final) fly in a narrow cone. It is important to note that the resonant frequency of a spontaneous gamma-quantum significantly depends on its outgoing angle, as well as on the number of absorbed photons of the wave. At the same time, for channel A, the resonant frequency depends on the outgoing angle relative to the initial electron momentum, and for channel B - on the outgoing angle relative to the final electron momentum.

•	In this process, there is a characteristic parameter ${{r}_{\eta }}$ that is determined by the parameters of the laser installation and is proportional to the intensity of the wave. This parameter determines the number of absorbed laser photons. The resonant frequencies will be of the order of the energy of the initial electrons only if the number of absorbed laser photons is greater than or of the order of a given characteristic parameter.

•	The resonant differential cross-section with simultaneous registration of the frequency and the outgoing angle of the spontaneous gamma-quantum has a maximum value of about $R_{\eta i\left( r \right)}^{\max }\sim {{10}^{18}}$ in units  ${{Z}^{2}}\alpha r_{e}^{2}$. for average laser wave intensities of about $I\sim {{10}^{18}}\ \text{Wc}{{\text{m}}^{\text{-2}}}$. As the wave intensity increases, the value of the resonant differential cross-section decreases. This is due to an increase in the number of absorbed laser photons, as well as an increase in the resonance width. So, for a wave intensity of the order of $I\sim {{10}^{22}}\ \text{Wc}{{\text{m}}^{\text{-2}}}$, the resonant differential cross section has an order of magnitude of $R_{\eta i\left( r \right)}^{\max }\sim {{10}^{8}}$ in units  ${{Z}^{2}}\alpha r_{e}^{2}$. And for a wave intensity of $I\sim {{10}^{24}}\ \text{Wc}{{\text{m}}^{\text{-2}}}$, the resonant section has an order of magnitude of $R_{\eta i\left( r \right)}^{\max }\sim {{10}^{4}}$ in units  ${{Z}^{2}}\alpha r_{e}^{2}$.

This theoretical research predicts a number of new physical effects and can be tested in international research project ELI (Extreme Light Infrastructure, Czech Republic).


\begin{thebibliography}{99}

\bibitem{1} F. V. Bunkin and M. V. Fedorov, Zh. Eksp. Teor. Fiz. 49, 1215 (1965) [Sov. Phys. JETP 22, 844 (1966)].

\bibitem{2} V. P. Oleinik, Zh. Eksp. Teor. Fiz. 52, 1049 (1967) [Sov. Phys. JETP 25, 697 (1967)].

\bibitem{3} V. P. Oleinik, Zh. Eksp. Teor. Fiz. 53, 1997 (1967) [Sov. Phys. JETP 26, 1132 (1968)].

\bibitem{4} N. B. Narozhny and M. S. Fofanov, Zh. Eksp. Teor. Fiz. 110, 26 (1996) [Sov. Phys. JETP 83, 14 (1996)].

\bibitem{5} R. V. Karapetian and M. V. Fedorov, Zh. Eksp. Teor. Fiz. 75, 816 (1978).

\bibitem{6} F. Zhou and L. Rosenberg, Phys. Rev. A 48, 505 (1993).

\bibitem{7} V. P. Krainov and S. P. Roshchupkin, Zh. Eksp. Teor. Fiz. 84, 1302 (1983).

\bibitem{8} I. V. Lebedev, Opt. Spectrosk. 32, 120 (1972).

\bibitem{9} A. V. Borisov, V. C. Zhukovskii, and P. A. Eminov, Sov. Phys. JETP 51, 267 (1980).

\bibitem{10} S. P. Roshchupkin, Yad. Fiz. 41, 1244 (1985).

\bibitem{11} S. P. Roshchupkin, Laser Phys. 12, 498 (2002).

\bibitem{12} M. Dondera and V. Florescu, Radiat. Phys. Chem. 75, 1380 (2006).

\bibitem{13} A. Florescu and V. Florescu, Phys. Rev. A 61, 033406 (2000).

\bibitem{14} A. N. Zheltukhin, A. V. Flegel, M. V. Frolov, N. L. Manakov, and A. F. Starace, Phys. Rev. A 89, 023407 (2014).

\bibitem{15} A. V. Flegel, M. V. Frolov, N. L. Manakov, A. F. Starace, and A. N. Zheltukhin, Phys. Rev. A 87, 013404 (2013).

\bibitem{16} A. N. Zheltukhin, A. V. Flegel, M. V. Frolov, N. L. Manakov, and A. F. Starace, J. Phys. B 48, 075202 (2015).

\bibitem{17} A. Li, J. Wang, N. Ren, P. Wang, W. Zhu, X. Li, R. Hoehn, and S. Kais, J. Appl. Phys. 114, 124904 (2013).

\bibitem{18} E. L$\mathrm{\ddot{o}}$tstedt, U. D. Jentschura, and C. H. Keitel, Phys. Rev. Lett. 98, 043002 (2007).

\bibitem{19} S. Schnez, E. L$\mathrm{\ddot{o}}$tstedt, U. D. Jentschura, and C. H. Keitel, Phys. Rev. A. 75, 053412 (2007).

\bibitem{20} A. A. Lebed' and S. P. Roshchupkin, Eur. Phys. J. D. 53, 113 (2009).

\bibitem{21} A. Lebed' and S. Roshchupkin, Laser Phys. Lett. 6, 472 (2009).

\bibitem{22} A. A. Lebed' and S. P. Roshchupkin, Phys. Rev. A 81, 033413 (2010).

\bibitem{23} S. P. Roshchupkin and O. B. Lysenko, Laser Phys. 9, 494 (1999).

\bibitem{24} S. P. Roshchupkin and O. B. Lysenko, JETP 89, 647 (1999).

\bibitem{25} A. A. Lebed', E. A. Padusenko, S. P. Roshchupkin, and V. V. Dubov, Phys. Rev. A 94, 013424 (2016).

\bibitem{26} A. A. Lebed', E. A. Padusenko, S. P. Roshchupkin, and V. V. Dubov, Phys. Rev. A 97, 043404 (2018).

\bibitem{27} A. Dubov, V. V. Dubov and S. P. Roshchupkin, Laser Phys. Lett. 17, 045301 (2020).

\bibitem{28} A. Dubov, V. V. Dubov and S. P. Roshchupkin, Universe 6, 143 (2020).

\bibitem{29} S. P. Roshchupkin, A. Dubov and V. V. Dubov, Laser Phys. Lett. 18, 045301 (2021).

\bibitem{30} P. A. Krachkov, A. Di. Piazza, and A. I. Milstein, Phys. Lett. B 797. 134814 (2019).

\bibitem{31} S. P. Roshchupkin, V. A. Tsybul'nik, and A. N. Chmirev, Las. Phys. 10, 1256 (2000).

\bibitem{32}  R. Kanya, Y. Morimoto, and K. Yamanouchi, Phys. Rev. Lett. 105, 123202 (2010).

\bibitem{33}  A. Hartin, International Journal of Modern Physics A 33, 1830011 (2018).

\bibitem{34} S. P. Roshchupkin, V. V. Dubov, A. V. Dubov, D. V. Doroshenko, N. R. Larin, G. K. Sizykh, and V. D. Serov, in High Power Lasers and Applications, Vol. 11777, edited by J. Hein, T. J. Butcher, P. Bakule, C. L. Haefner, G. Korn, and L. O. Silva, International Society for Optics and Photonics (SPIE, 2021) pp. 40 – 55. 

\bibitem{35} V. I. Ritus and A. I. Nikishov, Quantum electrodynamics phenomena in the intense field, in {\it Trudy FIAN}, edited by V. L. Ginzburg (Nauka, Moscow, 1979), Vol. 111.

\bibitem{36} F. Ehlotzky, A. Jaro$\mathrm{\acute{n}}$, and J. Z. Kami$\mathrm{\acute{n}}$ski, Phys. Rep. 297, 63 (1998).

\bibitem{37} F. Ehlotzky, K. Krajewska, and J. Z. Kami$\mathrm{\acute{n}}$ski, Rep. Prog. Phys. 72, 046401 (2009).

\bibitem{38} S. P. Roshchupkin, A. A. Lebed', and E. A. Padusenko, Las. Phys. 22, 1513 (2012).

\bibitem{39} S. P. Roshchupkin, Las. Phys. 6, 837 (1996).

\bibitem{40} S. P. Roshchupkin, A. A. Lebed', E. A. Padusenko, and A. I. Voroshilo, Las. Phys. 22, 1113 (2012).

\bibitem{41} M. V. Fedorov, {\it An Electron in a Strong Light Field} (Nauka, Moscow, 1991).

\bibitem{42} S. P. Roshchupkin and A. I. Voroshilo, {\it Resonant and Coherent Effects of Quantum Electrodynamics in the Light Field} (Naukova Dumka, Kiev, 2008).

\bibitem{43} S. P. Roshchupkin and A. A. Lebed', {\it Effects of Quantum Electrodynamics in the Strong Pulsed Laser Fields} (Naukova Dumka, Kiev, 2013).

\bibitem{44} S. P. Roshchupkin, A. A. Lebed', E. A. Padusenko, and A. I. Voroshilo, Resonant effects of quantum electrodynamics in the pulsed light field, in {\it Quantum Optics and Laser Experiments}, edited by S. Lyagushyn (Intech, Rijeka, Croatia, 2012), Chap. 6, pp. 107-156.

\bibitem{45}  C. Bula, K. T. McDonald, E. J. Prebys, C. Bamber, S. Boege, T. Kotseroglou, A. C. Melissinos, D. D. Meyerhofer, W. Ragg, D. L. Burke $et \: al.$, Phys. Rev. Lett. 76, 3116 (1996).

\bibitem{46} D. L. Burke, R. C. Field, G. Horton-Smith, J. E. Spencer, D. Walz, S. C. Berridge, W. M. Bugg, K. Shmakov, A. W. Weidemann, C. Bula $et \: al.$, Phys. Rev. Lett. 79, 1626 (1997).

\bibitem{47} G. A. Mourou, T. Tajima, and S. V. Bulanov, Rev. Mod. Phys. 78, 309 (2006).

\bibitem{48}   V. Bagnoud, B. Aurand, A. Blazevic, S. Borneis, C. Bruske, B. Ecker, U. Eisenbarth, J. Fils, A. Frank, E. Gaul $et \: al.$, Appl. Phys. B 100, 137 (2010).

\bibitem{49}  A. Di. Piazza, C. M$\mathrm{\ddot{u}}$ller, K. Z. Hatsagortsyan, and C. H. Keitel, Rev. Mod. Phys. 84, 1177 (2012).

\bibitem{50} D. M. Volkov, Zeit. Phys. 94, 250 (1935).

\bibitem{51} J. Schwinger, Phys. Rev. 82, 664 (1951).

\bibitem{52} L. S. Brown and T. W. B. Kibblie, Phys. Rev. 133, A705 (1964).

\bibitem{53} G. Breit and E. Wigner, Phys. Rev. 49, 519 (1936).

\bibitem{54} V. B. Berestetskii, E. M. Lifshitz, and L. P. Pitaevskii, {\it Quantum Electrodynamics} (Nauka, Moscow, 1980).


\end{thebibliography}
\end{document}